\documentclass[journal]{IEEEtran}

\makeatletter

\newcommand{\Rmnum}[1]{\expandafter\@slowromancap\romannumeral #1@}
\makeatother

\usepackage{verbatim}
\usepackage{cite}
\ifCLASSINFOpdf
   \usepackage[pdftex]{graphicx}
\else
 \usepackage[dvips]{graphicx}
\fi

\usepackage[cmex10]{amsmath}
\usepackage{latexsym}
\usepackage{amssymb}

\usepackage{algorithm}
\usepackage{algorithmic}

\usepackage[dvips]{graphicx}
\usepackage[tight,footnotesize]{subfigure}
\usepackage{sidecap}

\usepackage{caption}

\begin{document}
%
\title{Throughput Optimizing Localized Link Scheduling for Multihop Wireless Networks Under Physical Interference Model}
%
%
%

\author{Yaqin~Zhou,
       Xiang-Yang~Li,~\IEEEmembership{Senior Member,~IEEE,}
        Min~Liu,~\IEEEmembership{Member,~IEEE},
        Xufei~Mao,~\IEEEmembership{Member,~IEEE},
   Shaojie~Tang,~\IEEEmembership{Member,~IEEE},
     Zhongcheng~Li,~\IEEEmembership{Member,~IEEE}
\thanks{Yaqin~Zhou, Min~Liu, and Zhongcheng~Li are with Research Center of Network Technology, Institute of Computing Technology, Chinese Academy of Sciences, Beijing, China.}
\thanks{Xiang-Yang~Li and Shaojie~Tang are with Department of Computer Science, Illinois Institute of Technology, Chicago, IL, USA.}
\thanks{Xufei~Mao is with TNLIST, School of Software, Tsinghua University, Beijing, China.}
}

%
%

\markboth{}%
{Shell \MakeLowercase{\textit{et al.}}:  Throughput Optimizing Distributed Link Scheduling for Multihop Wireless Networks Under Physical Interference Model}
%



\maketitle

\begin{abstract}
We study throughput-optimum localized link scheduling in  wireless networks. The majority of results on link scheduling assume binary interference models that simplify interference constraints in actual wireless communication. While the physical interference model reflects the physical reality more precisely, the problem becomes notoriously harder under the physical interference model. There have been just a few existing results on link scheduling under the physical interference model, and even fewer on more practical distributed or localized scheduling.  In this paper, we tackle the challenges of localized link scheduling posed by the complex physical interference constraints. By cooperating the partition and shifting strategies into the pick-and-compare scheme, we present a class of localized scheduling algorithms with provable throughput guarantee subject to physical interference constraints. The algorithm in the linear power setting is the first localized algorithm that achieves at least a constant fraction of the optimal capacity region subject to  physical interference constraints. The algorithm in the uniform power setting is  the first localized algorithm with a logarithmic approximation ratio to the optimal solution. Our extensive simulation results demonstrate correctness and performance efficiency of our algorithms.
\end{abstract}

\begin{IEEEkeywords}
localized link scheduling, physical interference model, maximum weighted independent set of links (MWISL), capacity region.
\end{IEEEkeywords}

\IEEEpeerreviewmaketitle

\section{Introduction}

\IEEEPARstart{A}{s} a fundamental problem in wireless networks, link scheduling is crucial for improving  networking performances through maximizing throughput and fairness. It has recently regained much interest from networking research community because of wide deployment of multihop wireless networks, \emph{e.g.}, wireless sensor networks for monitoring physical or environment \cite{li2009underground} \cite{liu2010passive} through collection of sensing data \cite{S:effcient}.
Generally, link scheduling involves determination of which links should transmit at what times, what modulation and coding schemes to use, and at what transmission power levels should communication take place\cite{S:GMS}. In addition to its great significance in wireless networks, developing an efficient scheduling algorithm is extremely difficult due to the intrinsically complex interference among simultaneously transmitting links in the network.

The link scheduling problem has been studied with different optimization objectives, \emph{e.g.}, throughput-optimum scheduling, minimum length scheduling. Our study mainly focuses on maximizing throughput in multihop wireless networks. Nevertheless,  a common primary issue here is to find a set of conflict-free links with regard to the interference models.

Interference models matter significantly in scheduling algorithm design and analysis. It is well known that the time complexity of a throughput-optimum scheduling that tries to find a maximum weighted independent set of links, is polynomial under the $1$-hop interference model in switched wired networks, whereas it becomes NP-hard under more general interference models in wireless networks\cite{S:GMS}. Here the weight associated with every link is its corresponding queue length. Motivated by providing efficient solutions for the NP-hard problem, numerous low complexity scheduling algorithms with suboptimal throughput guarantee \cite{S:MWM2}, \cite{S:GMS}, \cite{S:constant1}, \cite{S:constant2}, \cite{S:constant3}, \cite{S:constant4}, \cite{S:pick1}, \cite{S:pick2}, \cite{S:pick3}, \cite{S:MS}, have been proposed in literature.

Despite of the numerous significant results gained for the problem, most of them assume simple binary interference models, \emph{e.g.}, hop-based, range-based, and protocol interference models \cite{S:phy6}. Under this category of interference model, a set of links are conflict-free if they are pairwise conflict-free. Conflict of transmissions on two distinct links is predetermined independently of the concurrent transmissions of other links. Thus, interference relationships based on these models can be represented by conflict graphs, and we can leverage classic graph-theoretical tools for solutions. However, in actual wireless communication, interference constraints among concurrent transmissions are not local and pairwise, but global and additive. Conflict of distinct transmissions is determined by the cumulative interference from all concurrent transmissions, which is often depicted by the  physical interference model, \emph{e.g.},  the Signal-to-Interference-plus-Noise Ratio (SINR) interference model. But due to the simplification of binary interference models on physical reality, the corresponding results no longer hold under the more realistic physical interference model. Recent results \cite{S:phy13}, \cite{S:phy14} have shown that throughput differs significantly under the two kinds of models. Moreover, the global and additive nature of the physical interference model drives previous traditional techniques based on conflict graphs inapplicable or trivial. Consequently, designing and analyzing scheduling algorithms under the physical interference model becomes especially challenging.

Some recent research results \cite{S:phy1}, \cite{S:phy6}, \cite{S:phy2},\cite{multicast}, \cite{S:phy5}, \cite{S:phy8}, \cite{S:phy12}, \cite{S:phy15}, \cite{S:phy9}  have addressed a few challenges related to the link scheduling problem under the physical interference model. To the best of our knowledge, however, all of these, with throughput maximization or other optimization objectives such as a minimum length schedule \cite{S:phy6}, just focus on centralized implementation.  Distributed or even localized scheduling under the physical interference model is  more demanding out of practical relevance.

Though  \cite{S:phy17}, \cite{S:phy7}, \cite{brar-icdcs08} consider distributed implementation of centralized algorithms, they fail to provide an effective localized scheduling algorithm with satisfactory theoretical guarantee \cite{S:phy17}\cite{S:phy7} and require global propagation of messages.
A scheduling algorithm without theoretical guarantee may cause arbitrarily bad throughput performance, and  global propagation of messages throughout network is inefficient in terms of time complexity\cite{peleg87}, especially for  large-scale networks.
This motivates us to develop localized link scheduling algorithms with provable throughput performance. Here by a localized scheduling algorithm we mean  that each node only needs information within constant distance to make scheduling decisions, while a distributed algorithm may inexplicitly need information far away.  Since just local information is available for each node to collaborate on schedulings with globally coupled interference constraint, it poses significant challenges to designing  efficient localized scheduling algorithms with theoretical throughput guarantee.

\indent In this paper we tackle these challenges of practical localized scheduling for throughput maximization under the realistic physical interference model with the commonly-used linear and uniform power assignment. For the primary challenge of decoupling the global interference constraints, on observing that distance dominates the interference, we partition the links into disjoint local link sets with a certain distance away from each other. By this way, we bound the interference from  all the other local link sets with a constant so that independent scheduling inside each local link set is possible. Then we shift the partitions to ensure every link get scheduled.  Using this we  implement localized schedulings. Furthermore, we novelly combine these two strategies into the pick-and-compare scheme by which we can prove that our algorithms obtain a theoretical capacity guarantee.

 We summarize our main contributions as follows.

\begin{enumerate}
    \item We propose a localized link scheduling algorithm under the linear power setting with provable throughput guarantee. We successfully decouple the global interference constraints for the implementation of localized scheduling, by proving that there exists a constant distance between any two disjoint local link sets to ensure independent link scheduling inside a local link set under the linear power setting. We then prove that our algorithm can achieve a constant fraction of capacity region.
  \item We further revise the algorithm to work under the uniform power setting, and give a detailed analysis on the achievable capacity region.  While under the uniform power setting, we prove that if the size of the scheduling set for local link sets is upper bounded by a constant, the solution to decouple the global interference constraints is also applicable. We prove that our revised algorithm can achieve $O(\log |V|)$ fraction of capacity region where $|V|$ is the number of nodes.
  \item We conduct extensive simulations to evaluate our algorithms. The simulation results demonstrate the correctness of our algorithms and corroborate our theoretical analysis.  The comparisons with two centralized algorithms in terms of average throughput performance further show performance efficiency of our algorithms.
\end{enumerate}

The remainder of the paper is organized as follows. In Section \Rmnum{2}, we define the exact system models for our problem. In Section \Rmnum{3}, we describe the basic ideas of our solutions, and the proposed localized scheduling algorithm with theoretical performance analysis. In Section \Rmnum{4}, we revise our solutions for the uniform power setting, and propose a localized algorithm with a logarithmic approximation ratio. We evaluate our solutions for both power settings in Section \Rmnum{5}. In Section \Rmnum{6}, we review related works on link scheduling.
In Section \Rmnum{7} we discuss some related issues on proposed algorithms.
We conclude this paper in Section \Rmnum{8}.

\section{Models and Assumptions}

\subsection{Network Communication Model}
We model a wireless network by a two-tuple $(V,E)$, where $V$  denotes the set of nodes and $E$  denotes the set of links. Each directed link $l=(u,v)\in E$  represents a communication request from a sender  $u$ to a receiver $v$. Let $\Vert l\Vert$  or $\Vert uv\Vert$  denote the length of link $l$. We assume each node knows  its own location, which is necessary in the partition and shifting process.

\subsection{Interference model}
 Under the physical interference model, a \emph{feasible schedule} is defined as an independent set of  links (ISL), each satisfying
{\small{\[
SINR_{uv} \buildrel \Delta \over = \frac{P_{u} \cdot \eta \cdot \left\|uv\right\|^{-\kappa} }{\sum\nolimits_{w \in T_u} {P_{w} \cdot
\eta \cdot \left\|wv\right\|^{-\kappa} +\xi } }\ge \sigma _{,}
\]}}
where $\xi $ denotes the ambient noise, $\sigma $  denotes certain threshold, and $T_u$ denotes the set of simultaneous transmitters with $u$.  It assumes path gain $\eta \cdot \left\|uv\right\|^{-\kappa} \le 1 $, where the constant $\kappa > 2 $ is path-loss exponent, and $\eta $ is the reference loss factor.

We consider the  following  two  transmission power settings.
\begin{enumerate}[\IEEEsetlabelwidth{12)}]
\item \emph{Linear power setting:} a sender $u$ transmits to a receiver $v$ always at
the power
$P_{u} =c\cdot \left\| uv \right\|^\beta \le P$  where $c$ and $\beta$ are both constant satisfying $c>0, 0 <\beta < \kappa,$ and $P$ is the maximum
transmission power.
\item \emph{Uniform power setting:} all links always transmit at the same power $P_{u}=P$.
\end{enumerate}
We use $P_l$ and $P_u$ alternatively to denote the transmitting power of link $l=(u,v)$.

We also assume all links  having a length less than the maximum transmission radius ${\sqrt[{\kappa}]{\frac{\eta P}{\sigma \xi}}}$. The distance between $u$ and $v$ satisfies $ r \le \left\| uv \right\|\le R$, where $r$ and $R$ respectively denotes the shortest link length and the longest link length. We suppose that $r$ and $R$  are known by each node.

\vspace*{-1\baselineskip}
\subsection{Traffic models and scheduling}
The maximum throughput scheduling is often studied in the following models. It assumes time-slotted wireless systems, and single-hop flows with stationary stochastic packet arrival process at an average arrival rate $\lambda _l$.
 The vector $ A(t)=\{A_l(t)\}$ denotes the number of packets arriving at each link in time slot $t$.
Every packet arrival process $A_l(t)$ is assumed to be i.i.d over time. We also assume all packet arrival process $A_l(t)$ have bounded second moments and they are bound by $A_{\max}$, i.e., $A_l(t) \leq A_{\max}, \forall l \in E$.
Let a vector $\{0,1\}^{|E|}$  denote a schedule $ S(t)$ at each time slot $t$, where $S_l(t)=1$ if link $l$ is active in time slot $t$ and $S_l(t)=0$ otherwise. Packets departure transmitters of activated links at the end of time slots.
Then, the \emph{queue length} (it is also referred to as \emph{weight} or \emph{backlog}) vector $ Q(t)=\{Q_l(t)\}$ evolves according to the following dynamics:
\[
     Q(t+1) = \max\{ 0,  Q(t)- S(t)\}+  A(t+1).
\]

 The throughput performance of link scheduling algorithms is measured by a set of supportable arrival rate vectors, named \emph{capacity region} or \emph{ throughput capacity}. That is, a scheduling policy is \emph{stable}, if for any arrival rate vector in its capacity region\cite{S:pick3},
\[    \lim_{ t\rightarrow \infty} \mathbb{E}[{ Q(t)}] < \infty.\]

A \textit{throughput-optimal} scheduling algorithm can achieve the optimal capacity region. It is well known that the policy of finding a maximum independent set of links to schedule regarding to the underlying interference models is throughput-optimal\cite{S:MWM1}. Unfortunately, the problem of finding a MWISL itself is NP-hard generally\cite{sharma2006complexity}. Thus we have to rely on approximation or heuristic methods to develop suboptimal scheduling algorithms running in polynomial time.
A suboptimal scheduling policy can just achieve a fraction of the optimal capacity region. The fractionally guaranteed capacity region is depicted by efficiency ratio $\gamma$ \cite{S:GMS}.

A suboptimal scheduling policy with efficiency ratio $\gamma $
must find a $\gamma $-approximation scheduling at every time slot $t$ to
achieve $\gamma$ times of the optimal capacity region \cite{lin2005info}. It remains difficult to achieve in a decentralized manner. The  pick-and-compare approach proposed in \cite{S:MWM2} enables that we just need to find a $\gamma $-approximation scheduling with a  constant positive probability. The basic pick-and-compare \cite{S:MWM2} works as follows. At every time slot, it generates a feasible schedule that has a constant probability of achieving the optimal capacity region. If the weight of this new solution is greater than the current solution, it replaces the current one. Using this approach achieves the optimal capacity region. The proposition below further extends this approach to suboptimal cases.

\textit{Proposition 1: (\cite{S:pick1}) Given any }$\gamma \in (0,1]$\textit{, suppose that an algorithm has a probability at least }$\delta >0$\textit{ of generating an independent set }$\mathcal{X}(t)$\textit{ of links with weight at least }$\gamma $\textit{ times the weight of the optimal. Then, capacity }$\gamma \cdot \Lambda $\textit{ can be achieved by switching links to the new independent set when its weight is larger than the previous one(otherwise, previous set of links will be kept for scheduling). The algorithm should generate the new scheduling} $S(t)$ \textit{ from the old scheduling }$S(t-1)$\textit{ and current queue length }$Q(t)$\textit{.}

In the rest of the paper we look for solutions to generate localized link
schedulings of approximately optimal weight, with  a certain constant probability at least.

\section{The Algorithm in the Linear Power Setting}

In this section, we focus on the design of localized scheduling algorithm for the linear power setting. We firstly demonstrate the basic ideas of the algorithm design before involved in the details of implementation.
\vspace*{-1\baselineskip}

\begin{table*}[t]\setlength{\tabcolsep}{3pt}
\begin{center}
\caption{Summary of notations}
\label{notations}
\begin{tabular}{c|c|c|c|c|c}
  \hline
 $J$ & side length of sub-square & $L_{ij}$ & link set of sub-square$(i,j)$ & $Y_{ij}$ & link set of super-subSquare$(i,j)$\\
    \hline
 $K$ & side length of super-subSquare & $\mathcal{X}_{ij}(t)$ & new scheduling for $L_{ij}$ at $t$ & $S_{ij}(t)$ & scheduling for $Y_{ij}$ at $t$\\
  \hline
 $Z_i$ & local link set   & $OPT_{ij}^*(t)$ & local optimal MWISL for $L_{ij}$ & $S^{\ast}(t)$ & global optimal MWISL at $t$\\
 \hline
 $R$ & longest link length   & $S_{ij}^{\ast}(t)$ & intersection of $L_{ij}$ and $S^{\ast}(t)$
 & $S(t)$ & global scheduling at time slot t\\

  \hline
  $d$ & side length of cell & $\|uv\|$ & link length & $r_S(l)$ & relative interference $l$ get from link set $S$\\
  \hline
  $a_S(l)$ & affectness $l$ get from link set $S$ & $Q(t)$ & queue length vector &$W(S)$ &weight of link set S \\
  \hline
  $I_S^l$ & interference  link $l$ suffered from link set $S$ &
  $I_{max}^l$ & the maximum interference $l$ can bear & $I_{max}$ & the minimum of $I_{max}^l$\\

  \hline
\end{tabular}
\end{center}\vspace{-10pt}
\end{table*}

\subsection{Basic idea}
The basis of our idea is to create a set of disjoint local link sets in which the scheduling can be done independently without violating the global interference constraint. The decoupling of the global interference constraint is based on the fact that distance dominates the interference between two distinct links. That is, if a transmitting link is placed a certain distance away from all the other transmitting links, the total interference it receives may get bounded. We prove later that this assumption holds actually.
Based on this, then we employ the partition strategies to divide the network graph into disjoint local areas such that each local area is separated away by a certain distance to enable independent local computation of schedulings inside every
local area. Links lying outside local areas will keep silent to ensure
separation of local areas. As links lying outside local areas cannot remain
unscheduled all the time or it will induce network instability, we use the
shifting strategy to change partitions at every time slot to make sure that every
backlogged link will be scheduled. These locally computed scheduling link sets
compose a new global schedule $\mathcal{X}(t)$ at every time slot $t$.

In light of the pick-and-compare scheme, we choose a more weighted schedule, denoted as $S(t)$,
between a new generated schedule $\mathcal{X}(t)$ and the last-time schedule
$S(t-1)$ using $Q(t)$. Meanwhile, by Proposition 1, if we guarantee
that
\begin{equation}
\mathbb{P}(S(t)\cdot Q(t)\ge \gamma S^\ast (t)\cdot Q(t))\ge \delta
\end{equation}
for some constant $\gamma >0,\mbox{ }\delta >0$, the queue length vector $Q(t)$
will eventually converge to a stable state. Then we can get a constant approximation ratio scheduling policy for the optimal.
\vspace*{-0.5\baselineskip}

\subsection{Detailed description}
Now we describe the details of our method.
\setlength{\textfloatsep}{3pt plus 1pt minus 1pt}
\begin{figure}
\centering
\subfigure[{\small{Partition$(K,a_t,b_t )$. Here the gray area is the local area the
link set of which participate in computing the new schedule; the links in the
white area keep silent.}}]{\includegraphics[width=4.2cm]{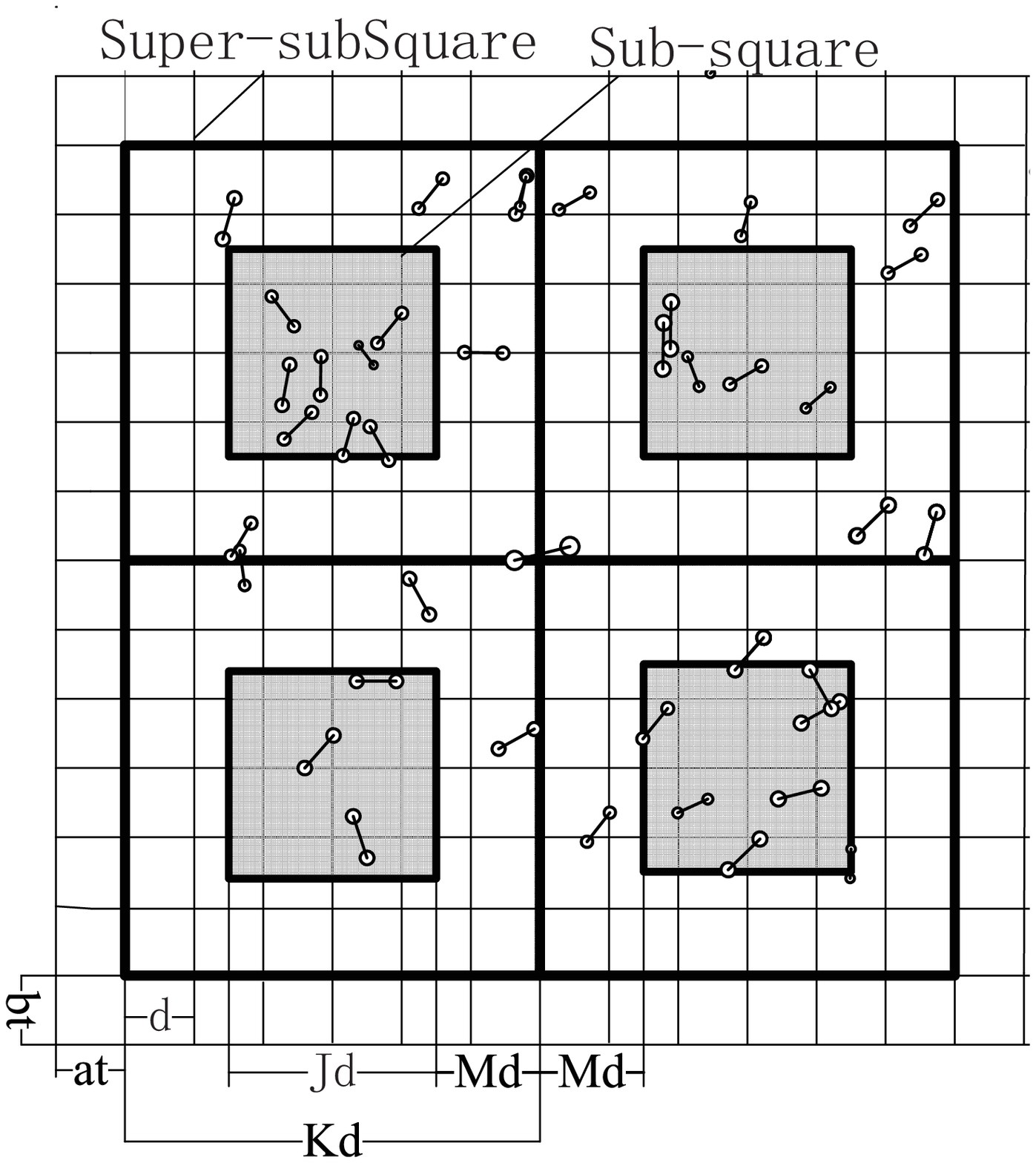}}
\hfill
\centering
\subfigure[{\small{Partition$(K,a_{(t+1)},b_t )$. Change the partition through shifting to the right by one cell, ensuring that links in the white area in previous partitions have opportunity to be scheduled. }}]{\includegraphics[width=4.2cm]{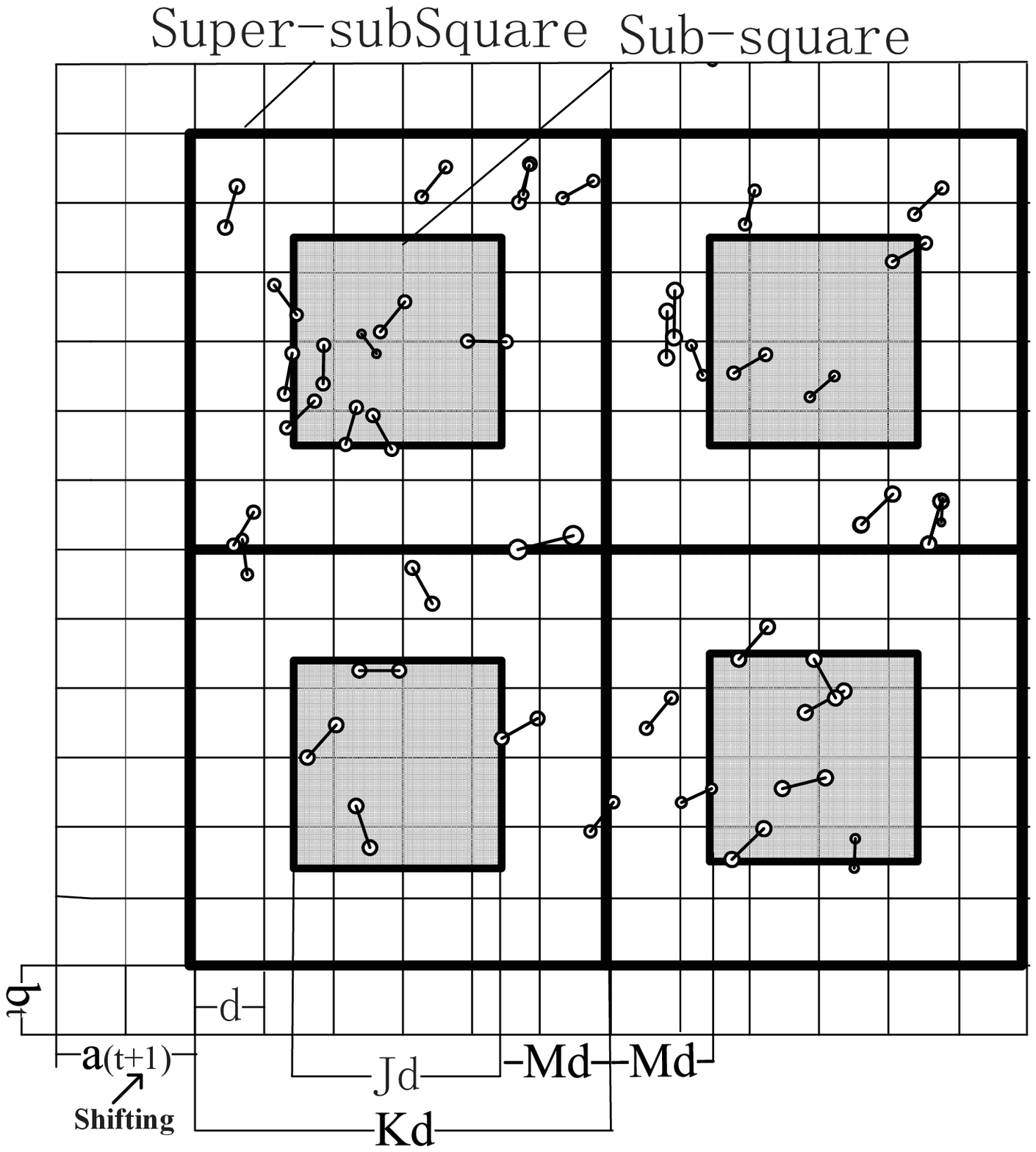}}

\caption{The partition and shifting process. The partition changes at different time slots by shifting vertically or horizontally by a step of one cell-length. }
\end{figure}

We first give a brief description of the partition and shifting strategies \cite{S:pick2}, as illustrated in Fig. 1.
We partition the plane into cells with side length $d =R$, using horizontal lines $x=i$ and vertical lines $y=j$ for all integers $i$ and $j$. A vertical strip with index $i$ is $\left\{ {(x,y)\vert i<x\le i+1} \right\}$. Similarly, we define the horizontal strip $j$. Let $\mbox{cell}(i,j)$ denote the intersection area of a vertical strip $i$ and a horizontal strip $j$.
A super-subSquare$(i,j)$ is the set of cells: $ \{cell(x,y)\vert x\in [i\ast K+a_t, (i+1)\ast K+a_t ), y\in [j\ast K+b_t, (j+1)\ast K+b_t )\} \mbox{,~}$
and a sub-square$(i,j)$ inside it is the set of cells:$\{cell(x,y)\vert x\in [i\ast K+a_t +M,(i+1)\ast K+a_t -M), y\in [j\ast K+b_t +M,(j+1)\ast K+b_t -M)\}.$
The corresponding link set $Y_{ij}$ (or $L_{ij}) $ consists of links with both ends inside super-subSquare$(i,j)$ (or sub-square$(i,j)$).
Let constant $K=2M+J$, $0\le a_t,b_t <K$. $a_t,\mbox{ }b_t $ are adjustable variables, referred to as the
\emph{horizontal} and \emph{vertical} \emph{shifting} respectively. $M$ is a constant, relating to the distance for independent  scheduling in local areas. We will formally define it later in Lemma 2.  We define the above process as \emph{Partition$(K,a_t,b_t )$}. By separately adjusting $a_t$, $b_t $  by a step of single cell, we can get $K^2$ different partitions for a plane totally.

Note that herein nodes do not shift, it looks like that a virtually partitioned cover having the same size with the original topology shifts vertically or horizontally. Initially, the virtual cover and the original topology of the network coincide completely. When the virtual cover shifts, nodes located in the edge area will certainly get uncovered if the distance between the edge of topology and the nearest nodes is less than $Kd$. To tackle this, we just need to assume a larger virtual cover, e.g., the distance between the edge of topology and the nearest nodes is equal or greater than $Kd$.

Then at each time slot nodes cooperate to compute a globally feasible scheduling as follows.

\textbf{Time slot } $t=0$: Every node first decides in which cell it
resides by a partition using $(a_0,b_0 )=(0,0)$; then it
participates in the process of computing a local scheduling $S_{ij} (0)$ for
the sub-square$(i,j)$ it belongs to. Let the solution $S(0) $ of time
slot $0$ be the union of the local solutions $S_{ij} (0)$ for all sub-squares.

\textbf{Time slot} $t \geq 1$: Every node decides in which cell
it resides by a partition starting from $(a_t, b_t )$. The shifting strategy for $a_t$ and $b_t$ works as follows. We let $a_t=t \mod K;$ and $b_t = (b_t+1) \mod K $ if $ a_t = 0$, or it keeps unchanged. Each node then participates in computing the new local scheduling, denoted as $\mathcal{X}_{ij} (t)$,
for its sub-square$(i,j)$ using the weight $Q(t)$. Let $S_{ij}(t-1)$ be the set of links from $S(t-1)$(the global solution at time slot $t-1)$ falling in the super-subSquare$(i,j)$ instead of sub-square$(i,j)$. If $S_{ij} (t-1)\cdot Q(t)>\mathcal{X}_{ij} (t)\cdot Q(t)$, let $S_{ij} (t)=S_{ij}(t-1)$, else $S_{ij} (t)=\mathcal{X}_{ij} (t)$, the global solution is the union
of $S_{ij} (t)$ from all super-subSquares.

The pseudo-codes are given in Algorithm 1. Since every node knows the locality from which it will collect information, it then participates the corresponding local computation, and at last it sends (if it is a coordinator) or receives (if not) the results. It may require multihop propagation to collect and broadcast the needed information inside super-subSquares. The details are omitted here. Please refer to \cite{brar-icdcs08} for detailed implementation.
The coordinator computes a new local scheduling link set by enumeration  in constant time $2^{|L_{ij}|}$, because the size of interference-free links for a sub-square$(i,j)$ is bounded by a constant (we claim it in Lemma 1).

\begin{algorithm}[htpb]
\caption{Distributed Scheduling by node $v$ under the linear power setting}
\label{alg1}
\begin{algorithmic}[1]
\STATE state $=$ White; active $= $ No; Coordinator $=$ No;
\STATE {Calculate which cell node $v$ resides in regarding to the current partition$(K, a_t, b_t)$;}
        \IF{ $v$ is the closest node to the center of super-subSquare}
            \STATE {Coordinator $=$ Yes;}
        \ENDIF
        \IF {Coordinator $=$ Yes}
            \STATE Collect $Q(t) $ and $S_{ij}(t-1)$.
            \STATE Compute $\mathcal{X}_{ij}(t)$ in sub-square$(i,j)$;
            \IF {$S_{ij}(t-1) \cdot Q(t) > \mathcal{X}_{ij}(t) \cdot Q(t) $}
                \STATE $S_{ij}(t)=S_{ij}(t-1)$;
                \ELSE
                    \STATE $S_{ij}(t)=\mathcal{X}_{ij}(t)$;
            \ENDIF
            \STATE Broadcast RESULT$(S_{ij}(t))$ in super-subSquare$(i,j)$;
        \ENDIF
        \IF{ state $=$ White}
            \IF{ receive message RESULT$(S_{ij}(t))$}
                \IF{$v \in S_{ij}(t)$ }
                    \STATE state $=$ Red; active $=$ Yes;
                \ELSE
                    \STATE state $=$ Black; active $=$ No;
                \ENDIF
            \ENDIF
        \ENDIF
     \end{algorithmic}
\end{algorithm}

\subsection{Theoretical analysis and proof}
We will then prove that our algorithm is correct and has a constant approximation ratio to the optimal capacity region.

Given a network $(V,E)$, supposing  $\cup {Z_{i}}$ is a set of disjoint local link sets inside for scheduling, where $Z_i \in E$ and $Z_i \cap Z_j = \phi \mbox{~if~} i \neq j, $
for any link $l\in Z_{i} $, if $l$ is activated, then
{\small{\[
I^{l}=I_{in}^l +I_{out}^l
\vspace*{-0.5\baselineskip}\]}}
where $I^{l} $ denotes cumulative interference from all other activated links in the network, $I_{in}^l $ denotes the total interference from simultaneously transmitting
links inside $Z_{i} $ and $I_{out}^l $ denotes the total interference from transmissions outside.

Therefore, we can do independent scheduling inside $Z_i$ without consideration of $I_{out}^l$ from concurrent transmissions outside $Z_i$, if $I_{out}^l $ gets bounded by a constant, \emph{i.e.},
\vspace*{-0.25\baselineskip}
{\small{
\[ I_{in}^l \le (1-\varepsilon) \cdot I_{\max}^l \mbox{, }
 I_{out}^l \le \varepsilon \cdot I_{\max } \mbox{, } 0< \varepsilon <1
 \mbox{, } l\in Z_{i}  \mbox{. }
 \vspace*{-0.5\baselineskip}\]}}
We let $I_{\max } $ denote the maximum interference that the longest links in $E$ can
tolerant during a successful transmission, and $I_{_{\max } }^l $ represent the maximum interference
that  an activated link $l$ can tolerant during a successful transmission.

We will give a formal statement that  $I_{out}^l$ of each activated link $l$ is indeed bounded by $\varepsilon I_{\max}$ in Lemma 2. However, to prove Lemma 2, we shall claim the following Lemma 1 in advance.

\newtheorem{lemma}{Lemma}
\begin{lemma}
In the linear power setting, the number of interference-free links for a local
link set $Z_i $ inside a square  with a size length $JR$ is bounded by a constant. Let $OPT_{i}$ refer to the optimal MWISL for $Z_i$. That is,
{\small{
\[\left| {OPT_{i}}
\right|\le \frac{(\sqrt 2 JR)^\kappa }{(1-\varepsilon )}\left[
{\frac{1}{\sigma }-\frac{\xi \cdot r^{\beta -\kappa
}}{c\eta }} \right]+1,\]}}
\vspace*{-0.5\baselineskip}
\label{lemma1}
\end{lemma}

\begin{IEEEproof}
This proof is available in Appendix A.
\end{IEEEproof}
We let $|OPT_{i}|_{ub}$ denote an upper bound of the size of the local optimal MWISL for $Z_i$, then we have,

\begin{lemma}
In a given network $(V,E)$ under the physical interference model in the linear power setting, if the Euclidean distance between any two disjoint local link sets is at least $M\times R $, then activated links in each local link set suffer a bounded cumulative interference from all other activated link sets, \emph{i.e.}, for each activated link $l$ in local link set $Z_i $,
{\small{\[
I_{out}^l \le \varepsilon \cdot I_{\max } \mbox{, } 0<\varepsilon<1  \mbox{,}
\vspace*{-0.2\baselineskip}
\]}}
where M is a constant, satisfying
{\small{$ M \ge \biggl[ \frac{2 \pi c \eta R^{\beta-\kappa} \cdot |OPT_{i}|_{ub}}{(\kappa-2) \varepsilon I_{\max}} \biggr]^{\frac{1}{\kappa}}_{\mbox{.}}$}}
\vspace*{-0.5\baselineskip}
\label{lemma1}
\end{lemma}
\begin{IEEEproof}
This proof is available in Appendix B.
\end{IEEEproof}

To analyze the theoretical performance of our method, we first review the following definitions.

\textit{Definition 2: (affectness \cite{S:phy8}) The relative interference of link }$l^* $\textit{ on }$l $\textit{ is the increase caused by }$l^* $\textit{ in the inverse of the SINR at }$l $\textit{, namely }$r_{l^* } (l )=I_{l^* }^{l } /(P_{l } \eta \|uv\|^{-\kappa})$\textit{. For convenience, define }$r_{l} (l )=0$\textit{. Let }$c_l =\frac{\sigma }{1-\sigma \xi /(P_{l } \eta \|uv\|^{-\kappa}) }$\textit{ be a constant that indicates the extent to which the ambient noise approaches the required signal at receiver }$t_v $\textit{. The affectness of link }$l $\textit{ caused by a set }$S$\textit{ of links, is the sum of relative interference of the links in }$S$\textit{ on }$l
$\textit{, scaled by }$c_l $\textit{, or}
{\small{$a_S (l )=c_l \cdot \sum\limits_{l^* \in S} {r_{l^* } (l )}.$}}

\textit{Definition 3: (}$p\mbox{-signal}$\textit{ set \cite{S:phy8}) We define a }$p\mbox{-signal}$\textit{ set to be one where the affectness of any link is at most }$1 \mathord{\left/ {\vphantom {1 p}}
\right. \kern-\nulldelimiterspace} p$\textit{. Clearly, any ISL is a }$\mbox{1-signal}$\textit{ set.}

\begin{lemma}
(\cite{S:phy8}) There is a polynomial-time protocol that takes a $p\mbox{-signal}$ set
and refines into a $p'\mbox{-signal}$ set, for $p'>p$, increasing the number of slots by a factor of at
most $4(\frac{p'}{p})^2$.
\label{lemma3}
\end{lemma}

That is, if a $p'\mbox{-signal}$ set of links can be scheduled simultaneously in a single slot, then we can  use some algorithm to partition a $p\mbox{-signal}$ into  at most $4(\frac{p'}{p})^2$ slots such that each set of links in every slot is a $p'\mbox{-signal}$ set.
This lemma implies that a $p\mbox{-signal}$ set can be refined into at most $4(\frac{p'}{p})^2$ $p'\mbox{-signal}$ set through a polynomial-time algorithm, e.g., a first-fit algorithm.

Next in Lemma 4 we present a constant approximation relationship
between each locally computed scheduling link set and its counterpart of the global optimal
scheduling set.
\begin{lemma}
The weight of $\mathcal{X}_{ij} (t)$ has a constant approximation ratio to the weight of the intersection set by the local link set $L_{ij} $ and the global optimal MWISL $S^\ast (t)$.
\label{lemma4}
\end{lemma}

\begin{IEEEproof}
Though the total interference that every link $l$ can tolerate at a
successful transmission becomes $(1-\varepsilon )I_{\max }^l $, we show that
it does little influence on local computation of optimal link scheduling set
for $L_{ij}$.

Normally any ISL is a $1$-signal set. Nevertheless, in order to keep independence of
sub-squares, the locally computed ISL should be a $p$-signal set, where $p$ is
bigger than $1$. That is, for the affectness of a normal ISL it holds that:

 \begin{equation}
 a_S (l ) = c_l \cdot \sum\limits_{l^* \in S} {r_{l^*} (l )}
 \le \frac{\sigma }{1-\sigma \xi /P_{l } }\cdot \frac{I^{l }_{\max }
}{P_{l } } \le 1,
 \end{equation}
whereas the affectness of a locally computed ISL for sub-square must satisfy that,

 \begin{equation}
a_{S_{ij(t)} } (l ) = c_l \cdot \sum\limits_{l^* \in S_{ij(t)} } {r_{l^*} (l )}
    \le \frac{\sigma \cdot (1-\varepsilon )}{1-\sigma \xi /P_{l } }\cdot \frac{I_{\max }^{ l } }{P_{l } }
 \le 1-\varepsilon .
 \end{equation}

Therefore, by Lemma 3, any $p\mbox{-signal}$ set can be refined into at most
$4(\frac{p'}{p})^2$  $p'\mbox{-signal}$ sets, where $p'>p$. So a normal MWISL
can be refined into $\frac{4}{(1-\varepsilon )^2}$  $\frac{1}{(1-\varepsilon
)}$-Signal link sets at most. Since the $\frac{1}{(1-\varepsilon )}$-Signal
link set returned by enumeration is most weighted, so the locally computed link
scheduling sets $\mathcal{X}_{ij} (t)$ has a weight
\begin{equation}
W(\mathcal{X}_{ij} (t))\ge \frac{(1-\varepsilon)^2}{4}W(OPT_{ij} (t)).
\end{equation}

Let $S_{ij}^\ast (t)=S^\ast (t)\cap L_{ij} $ denote the intersection by
$L_{ij} $ and $S^\ast (t)$, where $S^\ast (t)$ is the global
optimal MWISL at time slot $t$. Let $W(S)$ denote the summed weight of links in
a link set $S$. It is obvious that
\begin{equation}
W(S_{ij}^\ast (t))\le W(OPT_{ij} (t))\le \frac{4}{(1-\varepsilon)^2}W(\mathcal{X}_{ij} (t))_{.}
\end{equation}

Thus it proves the lemma.
\end{IEEEproof}

\newtheorem{theorem}{Theorem}
\begin{theorem}
$S(t)=\cup S_{ij} (t)$ computed by our algorithm is an independent link set under the physical interference model in the linear power setting. The weight of $S(t)$, \emph{i.e.}, $W(S(t))$, is a constant approximation of the weight of the global optimal MWISL with probability of at least $1/K^2$.
\label{theorem1}
\end{theorem}
\begin{IEEEproof}
The proof consists of two phases. We first prove that $S(t)=\cup S_{ij} (t)$ is an independent set. We next derive the approximation bound  $S(t)$ achieves.

\textbf{Phase I}: We rely on induction to infer that at every time slot $S(t)$ is a union of disjoint activated local link sets that are separated by at least $M$ cells from each other. Then we have that $S(t)$ is an independent link set under the physical interference model by Lemma 2. The following are the details.

For any link $l\in S(t)$, assuming $l\in
S_{ij} (t)$, the total interference $l$ suffers from all the other
simultaneously transmitting links in $S(t)$ is denoted by $I_{S(t)}^l $.

At time slot $0$, every local activated link set $S_{ij} (0)= \mathcal{X}_{ij}(0)$ is kept $2M$
cells away from each other, so $S(0)$ is an independent set by Lemma 2.

At time slot $1$, either $S_{ij} (0)$ or $\mathcal{X}_{ij} (1)$ is chosen to be a part of
$S(1)$. For those super-subSquares whose $S_{ij} (1) = S_{i'j'}(0) \cap Y_{ij} (0)$, their distance is kept at least $2M$ cells away. For those super-subSquares whose $S_{ij} (1)=\mathcal{X}_{ij} (1)$, their distance is also kept at least $2M$ cells away. And the distance between the two kinds of link set is at least $2M-1$ cells away. So $S(1)$ is an independent set.

\begin{figure}
\centering
\includegraphics[width=6cm]{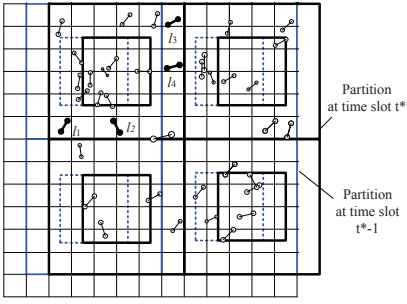}
\caption{Partitions at time slot $t^*-1$ and time slot $t^*$}
\end{figure}

At some time slot $t^\ast $, $t^\ast >1$, for some super-subSquares,
$S_{ij} (t^\ast )$ consists of disjoint subsets from several
different $S_{i'j'} (t^\ast -1)$  which fall into $Y_{ij} (t^\ast )$,
\emph{i.e.},
{\small{\begin{equation}
S_{ij} (t^\ast ) = S(t^\ast -1)\cap Y_{ij} (t^\ast )
=\bigcup\limits_{i'j'} {\left\{ {S_{i'j'} (t^\ast -1)\cap Y_{ij} (t^\ast )}\right\}}.
\end{equation}
}}
  For instance, as illustrated in Fig. 2,  link sets $\{l_1,l_2\}$ and $\{l_3,l_4\}$ respectively get scheduled in two different super-subSquares (\emph{i.e.}, super-subSquare$(1,2)$ and super-subSquare$(2,2)$) at time slot $t^*-1$, the links $\{l_1,l_2,l_3,l_4\}$ are then get scheduled in the same super-subSquare$(1,2)$ at time slot $t^*$.

We then let
\begin{equation}
\Phi _{i'j'}^{ij} (t^\ast )=S_{i'j'} (t^\ast -1)\cap Y_{ij} (t^\ast)
\end{equation}
 for brevity. Each $S_{i'j'} (t^\ast -1)$ is kept at least $M$ cells away from each other,
so is each $\Phi _{i'j'}^{ij} $.

Clearly, $S(t^\ast )$ can be divided into two separated subsets, one formed by some subsets of $S(t^\ast -1)$, the other formed by newly computed $\mathcal{X}_{ij} (t^\ast )$, \emph{i.e.},
\vspace*{-0.5\baselineskip}
{\small{
\setlength{\arraycolsep}{0.0em}
\begin{eqnarray}
 S(t^\ast ) &=& \bigcup\limits_{ij} {S_{ij} (t^\ast )} \\
 &=& \left\{{\bigcup\limits_{pq} {\bigcup\limits_{i'j'} {\Phi _{i'j'}^{pq}
(t^\ast )} } } \right\}\bigcup \left\{ {\bigcup\limits_{\substack{mn,\\mn\ne pq}} {\mathcal{X}_{mn}
(t^\ast )} } \right\}_{.}
\vspace*{-3\baselineskip}
\end{eqnarray}
}}

Since $\bigcup\limits_{pq} {\bigcup\limits_{i'j'} {\Phi _{i'j'}^{pq} (t^\ast)} } $ is a subset of $S(t^\ast -1)$, it is composed by disjoint subsets with a mutual distance of  $M$ cells at least. The distance between any distinct $\mathcal{X}_{mn}(t^\ast )$ is no less than $2M$ cells. Then we consider the distance between a disjoint subset of $\bigcup\limits_{pq} {\bigcup\limits_{i'j'}{\Phi _{i'j'}^{pq} (t^\ast )} } $ and a $\mathcal{X}_{mn} (t^\ast )$. Since $\mathcal{X}_{mn}(t^\ast )$ locates in sub-square$(m,n)$, which is $M$ cells away from the border of super-subSquare$(m,n)$, the distance between a disjoint subset of $\bigcup\limits_{pq} {\bigcup\limits_{i'j'} {\Phi_{i'j'}^{pq} (t^\ast )} } $ and a $\mathcal{X}_{mn} (t^\ast )$ is still no less than $M$ cells. Comprehensively, $S(t^\ast )$ consists of disjoint subsets
which are separated by at least $M$ cells.

Note that a disjoint subset of $S(t^\ast )$ does not equalize to a $S_{ij} (t^\ast )$ since a $S_{i'j'} (t^\ast -1)$ may be reserved completely in different super-subSquares at time slot $t^\ast $. Here we denote the disjoint subset by $\psi _i (t^\ast )$, and $S(t^\ast )=\bigcup {\psi _i (t^\ast )} $.

By Lemma 2, for each link $l\in \psi _i (t^\ast )$, where $\psi _i (t^\ast)$ comes from
the former part of equation (9), we have
{\small{$ I_{S(t^\ast )}^l  \le {I_{\max }^l}_{.}$}}
Meanwhile, for each $l\in \psi _i (t^\ast )$, where $l\in \psi _i (t^\ast )$ comes
from
the later part of (9), it holds that
{\small{ $I_{S(t^\ast )}^l  \le {I_{\max }^l}_{.}$ }}
Then we have
\begin{equation}
I_{S(t^\ast )}^l  \le I_{\max}^l, \forall l\in S(t^\ast ),
\end{equation}
indicating that $S(t^\ast )$ is an independent set.

Next we consider situations at time slot $t^\ast +1$. Similarly, $S(t^\ast +1)$ composes of disjoint subsets separated by no less than  $M$ cells. Using the same technique as at time slot $t^\ast $, we can get that $S(t^\ast +1)$ is still an independent set.

By induction we can infer that $S(t)$ is a union of disjoint activated subsets separated by  $M$ cells at least, thus it is an independent set. Herein we finish the first phrase of the proof.

\textbf{Phase II}: We derive the approximation ratio by the pigeonhole principle, Lemma 4, and Proposition 1.

 Let $D(t)$ denote the link set of the removed strips at Partition$(K,a_t,b_t)$. $D^\ast (t)$ represents a subset of $S^\ast (t)$, links of which fall inside $D(t)$, \emph{i.e.}, $D^\ast (t)=D(t)\cap S^\ast (t)$.  Recalling that there are $K^2$ different partitions for a plane totally.
  If we tried all these partitions in a single slot, each $\mbox{cell}(i,j)$ would appear in the ``removed" strips at most $2KM$ times. Thus the weight of all $D^\ast (t)$ in the $K^2$ partitions should be $2KM W(S^{\ast}(t))$, \emph{i.e.},
    \begin{equation} \sum\limits_{\mbox{i=0}}^{K^2-1} {W(D_i^\ast (t))} \leq {2KM}W(S^\ast (t)). \end{equation}
 Since actually we only experience one partition at time slot $t$,  by the pigeonhole principle the corresponding  Partition$(K,a_t,b_t)$ has a probability at least $1/K^2$ to be an \emph{optimal partition}, the weight of whose removed links in $D^\ast (t)$ is no greater than that of other partitions. Instantly we have the following with probability of $1/K^2$ at least,

{\small{\begin{equation}
W(D_i^\ast (t))\le \frac{1}{K^2}\cdot \sum\limits_{\mbox{i=0}}^{K^2-1}
{W(D_i^\ast (t))} \leq \frac{2M}{K}W(S^\ast (t)),
\end{equation}}}
and then,
{\small{
\begin{equation}
    W(\cup S_{ij}^\ast (t))= W(S^\ast (t)\backslash D^\ast (t))\ge (1-\frac{2M}{K})W(S^\ast (t)).
\end{equation}
}}

Following Lemma 4, the following holds with a probability of $1/K^2$,
{\small{\begin{equation}
(1-\frac{2M}{K})W(S^\ast (t))\le  \frac{4}{(1-\varepsilon )^2}W(\cup \mathcal{X}_{ij} (t)).
\end{equation}}}
So by Proposition 1 we get that the approximation ratio for the optimal is
{\small{${\frac{4}{(1-2\cdot\frac{M}{K})(1-\varepsilon )^2}}_{.}$}}
\end{IEEEproof}

\subsection{Time and communication complexity}
We then give a brief analysis on time complexity and communication complexity.

 We define the \emph{time complexity} as the time units required by the running of Algorithm 1 inside each local area in the worst case \cite{peleg87}. It includes time units for local information collection and local computation time at the center node.
  At every time slot, the coordinators shall collect queue information and last scheduling status of all links inside the super-subSquare. After computation, the coordinators broadcast the scheduling results throughout the super-subSquares.
  It may require multihop propagation to collect and broadcast the needed information inside super-subSquares.  We implement it as follows.

   To avoid collision, the coordinator can first compute a tree that determines the sequence of transmissions at each node in the super-subSquare based on topology information. As the root node it broadcasts the results to its children (or child). The children acquire the order of transmissions and relay this message as the specified order. The message is relayed until it reaches every node.
      Then the required information (or the scheduling results) can be collected (or broadcast) according to the sequence. We suppose that the the parent node can packet the children's messages into its own packet for collection.
      In the worst case each node has to transmit one by one at different mini slot, then it causes $O(n_{ij})$ time complexity where $n_{ij}$ is the number of nodes inside super-subSquare$(i,j)$. Herein we just give a basic scheme, the time complexity may be further reduced with the help of other techniques.

   After all required information gathered, the local computation complexity at the center node is $O(2^{|L_{ij}|})$, which can be ignored comparing to the time for information collection. Thus the total time complexity is $O(n_{ij})$.
 The \emph{ communication complexity} is the total number of basic messages transmitted during each scheduling in the worst case \cite{peleg87}. The communication complexity is $O(|V|)$, and the number of messages transmitted in each local area is $O(n_{ij})$.

\section{The algorithm in the Uniform Power Setting}
We now extend the framework to the uniform power setting.
Instead of enumeration, we compute a MWISL of the candidate links for each sub-square by adopting the method
proposed in\cite{S:phy9}. The reader may wonder why we do not use enumeration directly as we do in the linear power setting. The difficulty lies behind the fact that the cardinality of the optimal MWISL in each sub-square is no longer bounded by a constant in the uniform power setting, which drives $M$ no longer bounded by a constant and  simple enumeration time consuming.
Except for the difference, the general structure of the algorithm is the same with that of the linear power setting. The following we will give a detailed analysis, and derive a logarithmic bound on throughput capacity.

We first describe the main idea of Algorithm $1$ in \cite{S:phy9} for computation of MWISL inside each sub-square, and show that the cardinality of the resulted link set is upper bounded by a constant. Given a set of links and weights associated with the links, the algorithm works as follows:

\textbf{Phase I:} Remove every link whose associated weight is at most $\frac{w_{\max}}{n}$ where $w_{\max}$ denotes the maximum weight among all links and $n$ is the number of all given links. Let $w_{\min}$ denote the minimum weight from the remaining links.

\textbf{Phase II:}  Partition the remaining links into $\log{\frac{w_{\max}}{w_{\min}}}$ groups such that  the links of the $i-$th group $G_i$ have weights within $[2^i w_{\min}, 2^{i+1} w_{\min}]$. For each group $G_i$ of links, it finds an independent set of links among it by adopting the method in \cite{S:phy4}. Totally it will get $\log{\frac{w_{\max}}{w_{\min}}}$  ISLs, one for each group.
Then it chooses the one with the maximum weight among the $\log{\frac{w_{\max}}{w_{\min}}}$ ISLs as the final solution.

The remove operation of Phase I simply excludes those links with smaller weight so that we can get a smaller number of groups when partition is done according to weight. In that way, it is more likely to get a better approximation ratio to the optimal in practical, though the theoretical guarantee is not explicitly improved. We also notice that the resulted link scheduling  set is an ISL by the method in \cite{S:phy4}.  With the help of the following lemma presented in \cite{S:phy4}, we will get that the the resulted link set is indeed with limited number of links.
\begin{lemma}(\cite{S:phy4})
Consider a link $l=(u,v)$ and a set $N$ of nodes other than $u$ whose distance from $u$ is at most $\rho\|uv\|$. If link $l$ succeeds in the presence of the interference from $N$, then
\[
    |N| \le \frac{(\rho+1)^{\kappa}}{\sigma} \biggl[1 - (\frac{\|uv\|}{R})^{\kappa} \biggl]
\]
\end{lemma}

The corollary below asserts a upper bound of  the number of successfully transmitting links in a local link set with size $JR \times JR$ when utilizing Algorithm $1$ in \cite{S:phy9}.
\newtheorem{corollary}{Corollary}
\begin{corollary}
The cardinality of the resulted link scheduling set by Algorithm $1$ in  \cite{S:phy9} is upper bounded, \emph{i.e.},
\[
    |\mathcal{X}_{ij}(t)| \le \frac{(\sqrt{2}J \frac{R}{r}+1)^{\kappa}}{\sigma} \biggl[ 1 - (\frac{r}{R})^{\kappa} \biggl]
\]
\end{corollary}
\begin{IEEEproof}
In the method presented in \cite{S:phy4}, it adds firstly the shortest link among all the candidates to the scheduling set. And the distance between any pair of nodes will be no greater than $\sqrt{2}JR$. Therefore, let $r$  be the shortest link length and then  $\rho=\frac{\sqrt{2}JR}{r}$,  we derive the upper bound.
\end{IEEEproof}

We let $ |\mathcal{X}_{ij} (t)|_{ub}$ denote a upper bound of the size of the local computed ISL for $Z_i$ by Algorithm 1 in \cite{S:phy9}. Then we present the lemma below to show that in the uniform power setting, if any two disjoint local link sets of a given network are separated by a certain constant distance away, independent scheduling inside each local link set without consideration of interference outside is possible.
\begin{lemma}
In a given network $(V,E)$ under the physical interference model in the uniform power setting, if the Euclidean distance between any two disjoint local link sets is at least $M\times R $ , then activated links in each local set suffer negligible cumulative interference from all other activated link sets, \emph{i.e.}, for each activated link $l$ in local link set $Z_i $,
\vspace*{-0.25\baselineskip}
\[
 I_{out}^l \le \varepsilon \cdot I_{\max} \mbox{, } 0< \varepsilon <1  \mbox{,}
 \vspace*{-0.25\baselineskip}
\]
where $I_{\max}$ is the maximum interference that the longest links in $E$ can tolerant during a successful transmission,  $M$ is a constant, satisfying
{\small{$
    M \ge \biggl[ \frac{2 \pi \eta P \cdot |\mathcal{X}_{ij} (t)|_{ub}}{(\kappa-2) \varepsilon  I_{\max}R^{\kappa}} \biggr]^{\frac{1}{\kappa}}_{\mbox{~.}}
$}}
\label{lemma5}
\end{lemma}
\begin{IEEEproof}
The proof is available in Appendix C.
\end{IEEEproof}

In light of Lemma $6$, the partition strategy to enable distributed implementation will remain effective  in the uniform power setting.

Next, we reveal the relationship between the newly computed scheduling set  for sub-square$(i,j)$ and the corresponding $S^*_{ij}(t)$. It will help us to derive an approximation ratio of the global solution generated by our algorithm later.
\begin{lemma}
The weight of the newly computed link scheduling  set inside each sub-square$(i,j)$,
\emph{i.e.}, $W(\mathcal{X}_{ij} (t))$,  has a  approximation ratio of $\frac{4 \mu}{(1-\varepsilon)^2}$ to
the weight of the intersection set by the corresponding local link set $L_{ij} $ and the global optimal MWISL $S^\ast (t)$.
\label{lemma6}
\end{lemma}
\begin{IEEEproof}
 The newly computed local scheduling set for each sub-square should be a $(1-\varepsilon)$-signal set, so its weight is at least $\frac{(1-\varepsilon)^2}{4}$ times the corresponding $1$-signal set. Since the solution for MWISL with uniform power assignment proposed in \cite{S:phy9} finds an ISL with affectness no greater than $1$, it is obvious that
\begin{equation}
W(S^*_{ij}(t)) \le W(OPT_{ij}(t)) \le \frac{4\mu}{(1-\varepsilon)^2}W( \mathcal{X}_{ij} (t) ),
\end{equation}
where $\mu = \log |V|  $ is the approximation bound of the algorithm in \cite{S:phy9}.
\end{IEEEproof}

Then, we will state an exact bound on the throughput performance of our algorithm in Theorem $2$.
\begin{theorem}
\label{theorem2}
The union of the local computed scheduling link sets, \emph{i.e.},
$S(t)=\bigcup S_{ij}(t)$,  is feasible under the uniform power setting. The weight of $S(t)$ achieves a fraction of $O(\log|V| ) $ times the optimal solution.
\end{theorem}
\begin{IEEEproof}
We first show that the resulted scheduling link set by our algorithm is an independent link set with uniform power at every time slot.
 Using the same technique in the proof of Theorem $1$, we can inductively conclude that each global scheduling set $S(t)$ is
 composed by disjoint link sets which are kept at least $M$ cells away from each other. Therefore, by Lemma $6$, we can get that
 for each link $l \in S(t)$, the total interference it receives satisfies that,
 \begin{equation}
    {I_{S(t)}^l \le (1-\varepsilon) \cdot I_{\max}^l + \varepsilon \cdot I_{\max} \le I_{\max}^l}_{\mbox{.}}
 \end{equation}
It then follows that $S(t)$ is an independent link set.

We next prove the approximation bound on the performance of our algorithm. Remember that a Partition$(K, a_t, b_t)$ has a probability of $1/K^2$ to be a optimal partition. In that case, with the help of Lemma $6$ it follows that
{\small{\begin{eqnarray}
 W(\cup \mathcal{X}_{ij} (t) ) &\ge&  \frac{(1-\varepsilon)^2 (K-2M)}{4K\mu} W( \cup{ S_{ij}^*(t) } ) \\
       &\ge&  \frac{(1-\varepsilon)^2 (K-2M)}{4K\mu}W(S^*(t))
\end{eqnarray}}}
Hence, we get
{\small{
\begin{equation}
    \mathbb{P}\big( W(S(t)) \ge  \frac{(1-\varepsilon)^2 (K-2M)}{4K\mu} S^*(t) \big) \ge \frac{1}{K^2}.
\end{equation}}}

Since $\mu = \log |V|$ , it is clear that Algorithm $1$ can achieve $O(\frac{1}{\log |V|})$ of the optimal capacity region in the uniform power setting by Proposition $1$.
\end{IEEEproof}

The time complexity and communication complexity is the same as the corresponding of the linear power assignment.
\section{PERFORMANCE EVALUATION}

We do simulation experiments to verify the correctness, and evaluate the throughput performance of our proposed algorithms.
The general setting of our experiments is as follows. We consider a network with $500$ nodes, half of which as senders randomly located on a plane with size $200\times 200$ units, the other half as receivers positioned uniformly at random
inside disks of radius $R$ around each of the senders. Packets arrive at each link independently in a Poisson process \footnote{The results hold under any arrival process satisfying the strong law of large numbers, using the fluid model approach \cite{sha07} \cite{S:pick3}.} with the same average arrival rate $\lambda $. Initially, we assign each link $k$ packets where $k$ is randomly chosen from $[100,300]$. The path loss exponent is set to be $3$.

We  conduct three series of experiments to fully evaluate our algorithms. In the first series, we verify the correctness of our algorithms, that is, whether each activated link receives bounded interference from all other links outside the local link set it belongs to. The other two series of experiments focus on the evaluation of average throughput performance in terms of total backlog (the total number of unscheduled packets).
The total backlog fluctuates slightly in a region if the arrival rate vector lies in the achievable capacity region of a link scheduling  algorithm. Or it increases dramatically if the arrival rate vector exceeds the achievable capacity region. If the total backlog increases unboundedly to infinity, the network will become unstable. Thus we can roughly approximate an average supportable arrival rate through the changes of backlog.

In the second series, we study how some related variables affect the performance of the algorithms.
In the third series, we further study the performance efficiency of the proposed algorithms by comparisons with two centralized algorithms:
\begin{enumerate}
  \item Greedy Maximal Schedule (GMS), which is observed to perform throughput optimally in simulations \cite{S:GMS}. It greedily adds to the scheduling set the most weighted link that is independent with the previously added links.
  \item Randomized Algorithms, where each time slot links are randomly picked, and added to current scheduling set if they do not conflicts with previous selected links. The process is repeatedly performed until no new links can be added.
\end{enumerate}

The reasons why we do not compare with distributed implementation of the two centralized algorithms are as follows.
Undoubtedly, these two are  distributed in nature according to some locality-sensitive approaches \cite{peleg87}, if  considering binary interference models. And previous works on binary interference models indeed implement them in a distributed manner, \emph{i.e.}, \cite{S:LGS} proposing a local greedy link scheduling algorithm and \cite{S:constant3} making scheduling decisions by probability. The key point is that the effect of this category of interference is localized and binary. It is easy to keep feasibility of the final global scheduling set by inactivating all other links in the interference range of an activated link surely or \emph{w.h.p.}. However, it is not so obvious and easy to guarantee under the physical interference model.
It is hard to localize these approaches because the interference effect is global and additive. Simply inactivating all other links in some neighborhood of an activated link will not help much to keep the final scheduling set feasible. Considering these difficulties, we focus on comparisons with the centralized visions.

\subsection{Under Linear Power Setting}
In this subsection we conduct the three series of experiments  for the linear power assignment.

\begin{figure}[t]
    \centering
 \begin{minipage}{0.49\textwidth}
    \centering
        \subfigure[Interference vs. links]{
        \includegraphics[height=1.2in,width=1.6in]{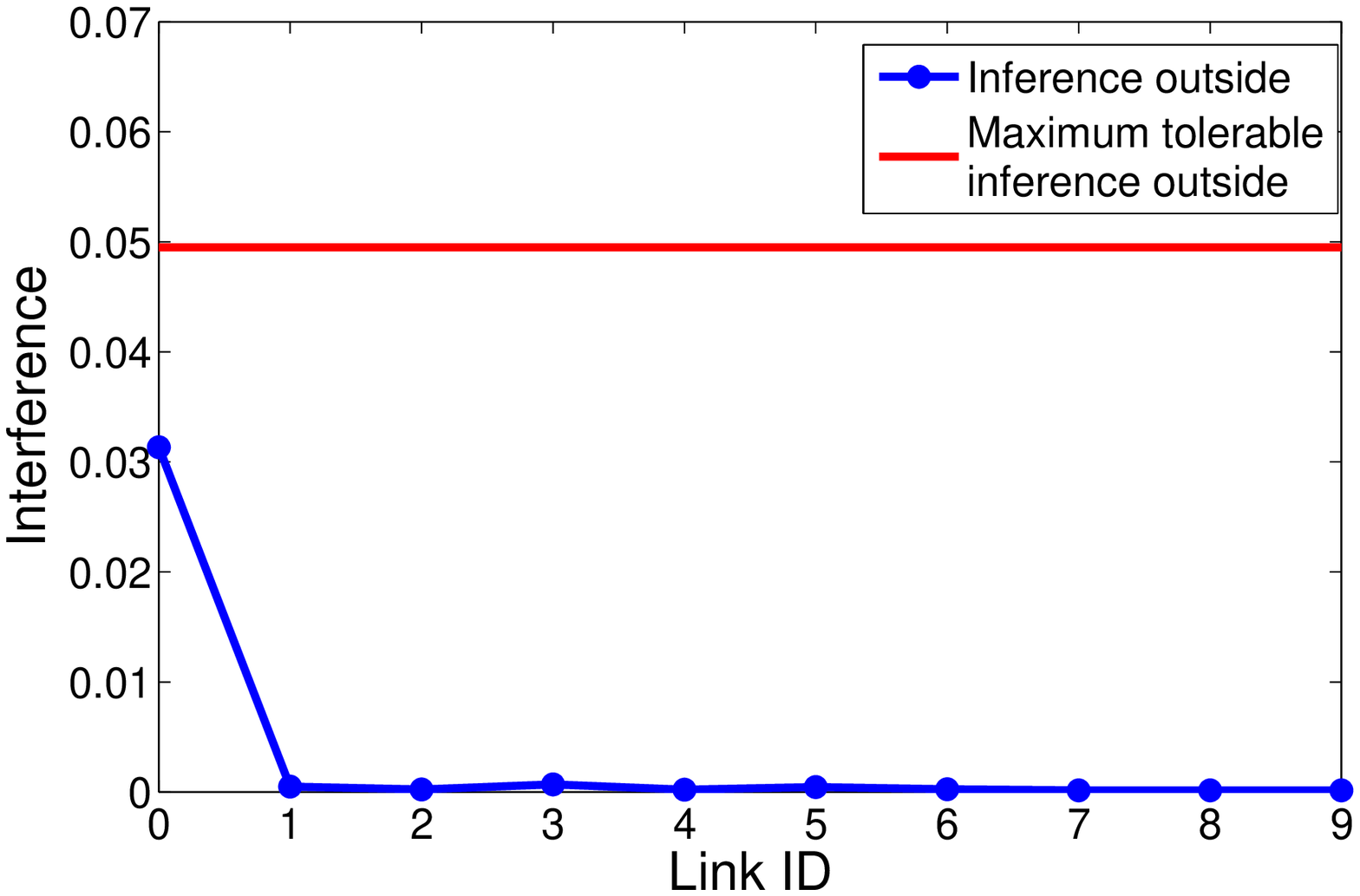}}
    \hspace{0cm}
        \subfigure[Affectness vs. links]{
        \includegraphics[height=1.2in,width=1.6in]{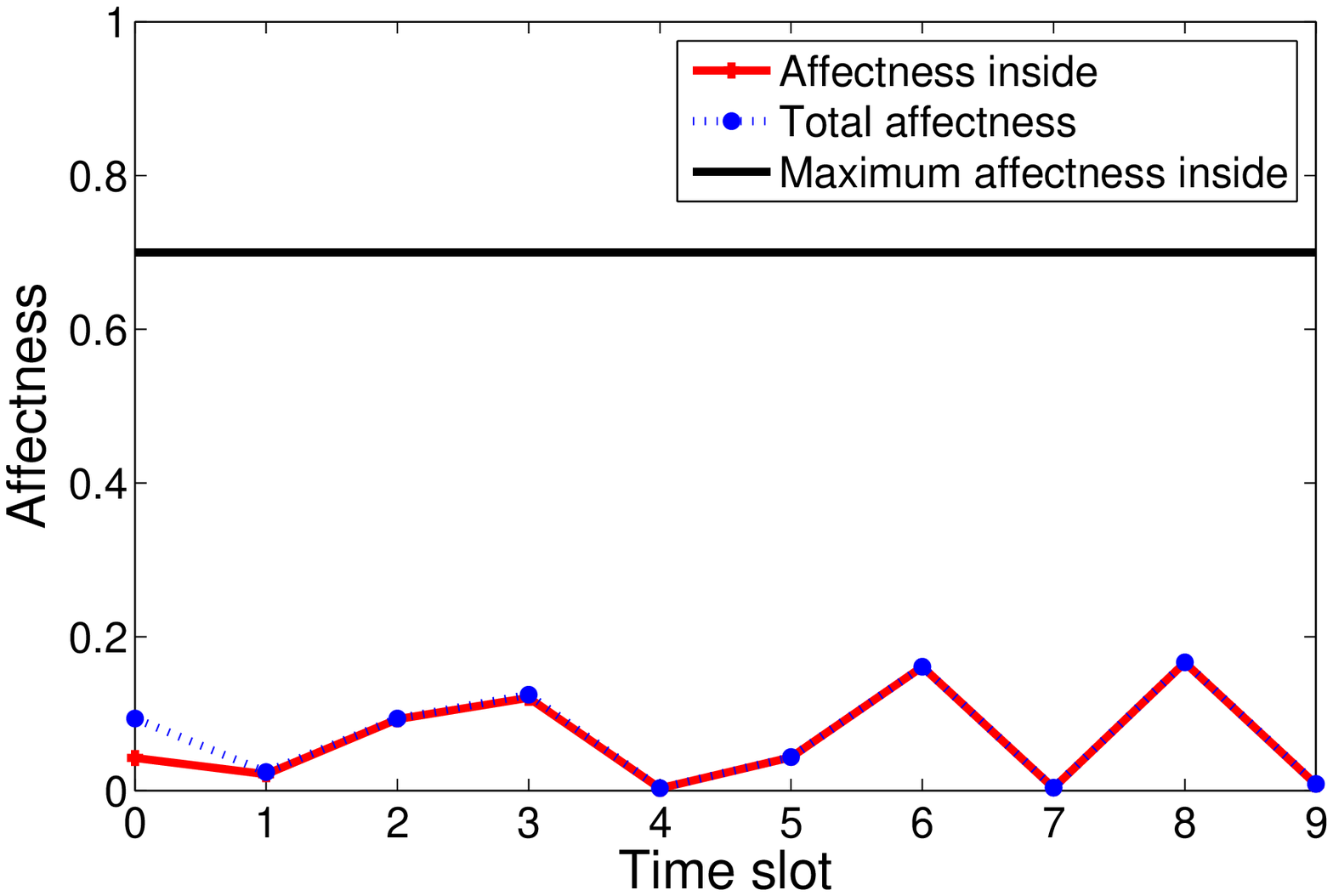}}
     \caption{{\small{The interference/affectness each randomly chosen transmitting link receives with the linear power assignment. (a) shows the outside interference each link receives. (b) shows the inside affectness for each link, and the total affectness each receives from all other simultaneously transmitting links.}}}
  \end{minipage}
  \hspace{0cm}
  \hfill
 \begin{minipage}{0.49\textwidth}
    \centering
     \subfigure[Average interference vs. time slot]{
        \includegraphics[height=1.2in,width=1.60in]{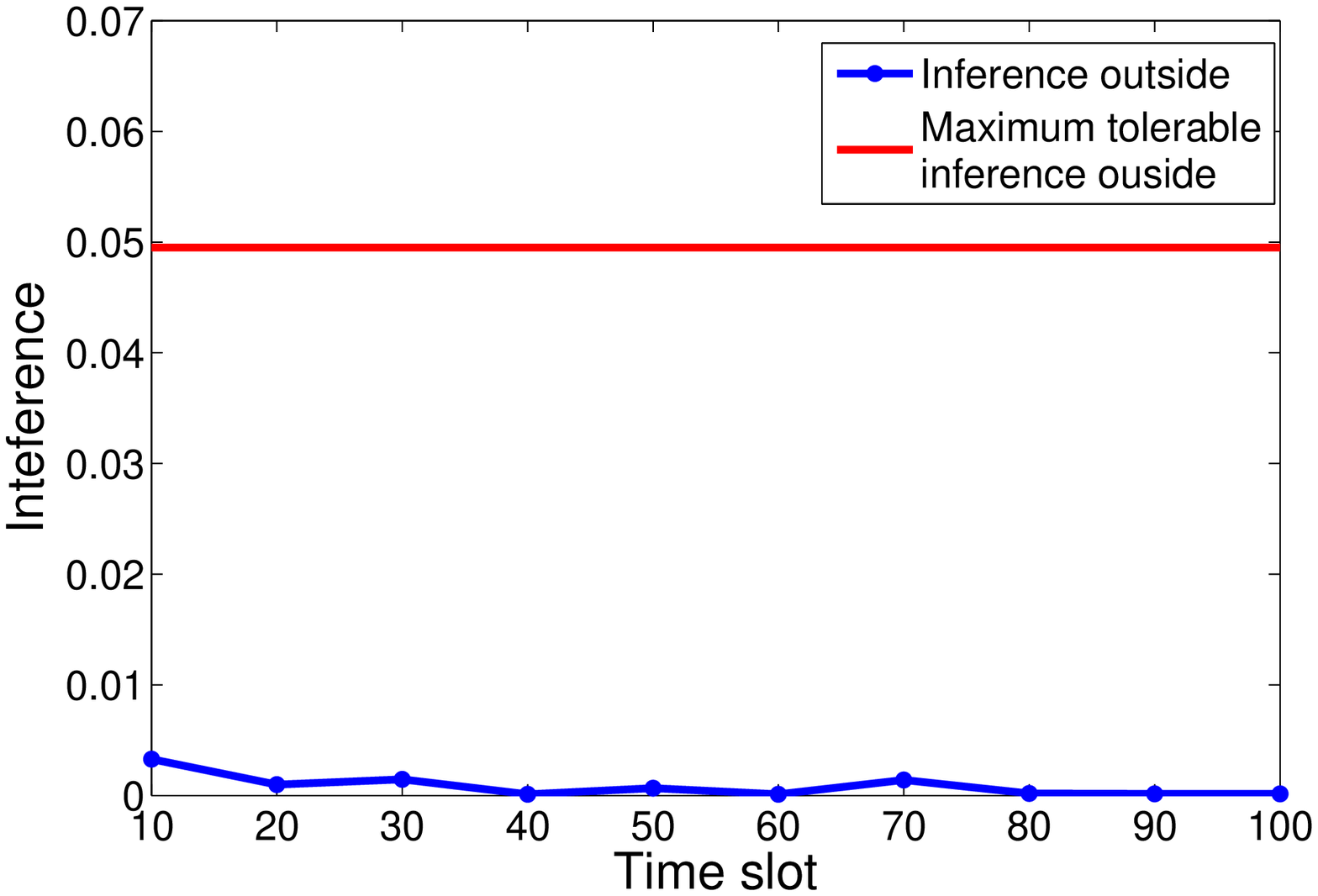}}
  \hspace{0cm}
    \subfigure[Average affectness vs. time slot]{
        \includegraphics[height=1.2in,width=1.6in]{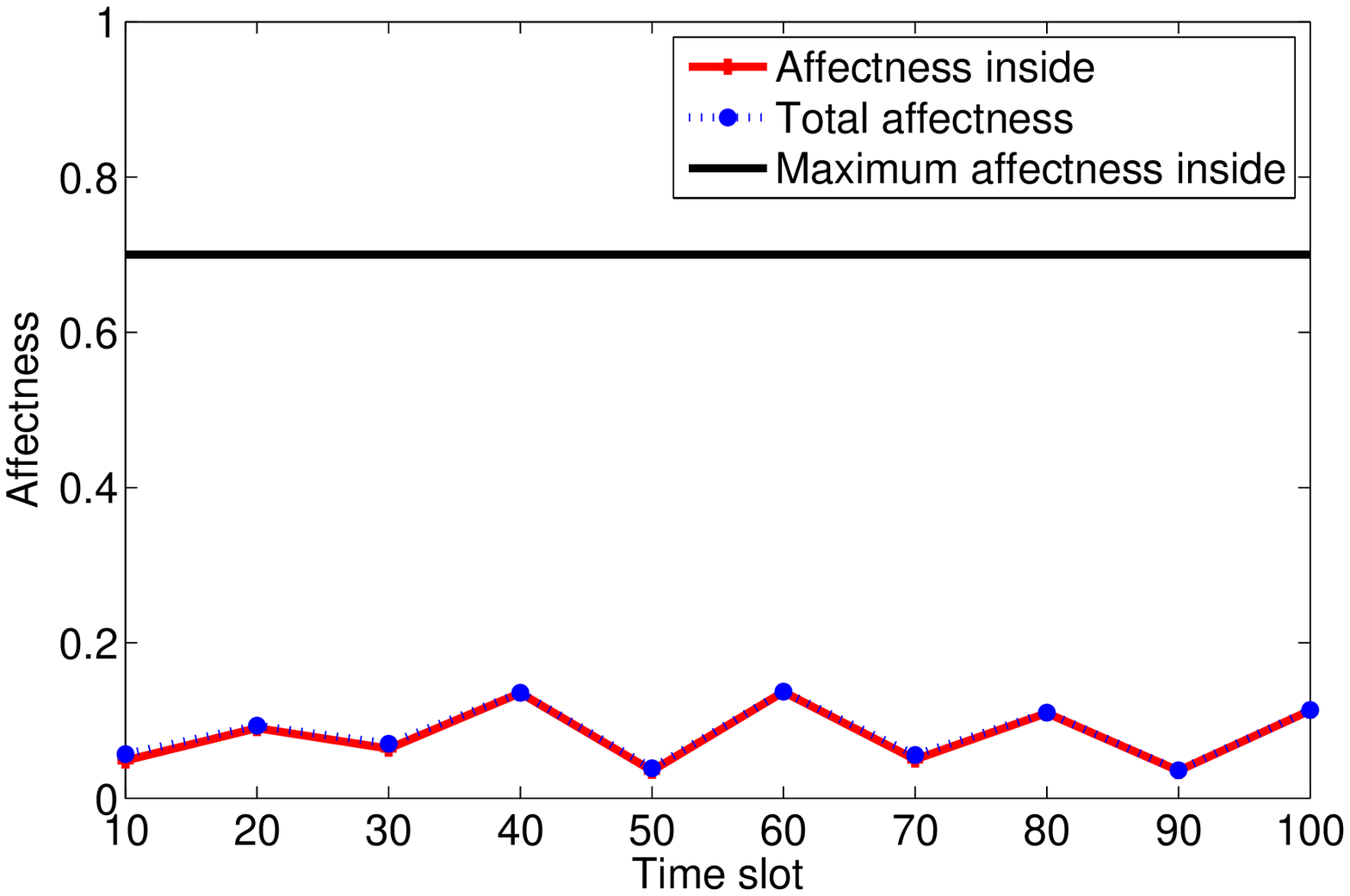}}
     \caption{{\small{Average interference per transmitting link receives at different time slots with the linear power assignment. (a) shows the average outside interference  per transmitting link receives. (b) shows the average inside affectness, and the average total affectness per transmitting link receives from all other simultaneously transmitting links.} }}
 \end{minipage}
\end{figure}
We first plot the numerical results for the first series of experiments in Fig. 3, 4. For convenience, we call the interference/affectness each link receives from all other simultaneously transmitting links outside/inside the corresponding super-subSquare it belongs to as outside interference/inside affectness. Since the maximum tolerable interference of each link for successful transmission is different, the corresponding inside and total interference each link receives is shown as the value of affectness for better illustration. Recall that, for a successful transmission, the inside and total affectness shall be less than $1-\varepsilon$ and $1$ respectively.

In this set of experiments, we set $\varepsilon$ a small value of $0.3$.  For Fig. 3, we randomly choose a time slot and $10$ transmitting links. As shown in Fig. 3(a), the outside interference each chosen link receives is much smaller than the constant bound $\varepsilon I_{max}$. From Fig. 3(b), we can further confirm that the transmitting links suffer negligible interference outside. For Fig. 4, we show the average outside interference per transmitting link receives at different time slots in Fig. 4(a), and the total and outside average affectness  per transmitting link in Fig. 4(b). These results cooperate to show that the outside interference a transmitting receives is indeed bounded by $\varepsilon I_{max}$, and indicate the correctness of Algorithm $1$.

Next we set another set of experiments to study the performance of Algorithm $1$ alone under different values of variables that may impact the average throughput performance actually. The variables $\frac{K}{M} \mbox{~and~} \varepsilon$ dominate the theoretical bound of Algorithm $1$. When $\varepsilon$ is fixed, a larger $\frac{K}{M}$ implies a bigger fraction of the optimal capacity region, but with a smaller probability of $\frac{1}{K^2}$ to achieve it.  $\varepsilon$ denotes a weighting factor between the interference a activated link suffers inside and outside the sub-square. A bigger value of $\varepsilon$ indicates a smaller value of $M$ theoretically. Therefore, though under the fixed value of $\frac{K}{M}$ a smaller  $\varepsilon$ leads to a bigger fractional capacity region, the probability to achieve this region gets smaller because of the resulted bigger $M$. Since the running time of the simulation is much shorter than the time for the algorithm to achieve the theoretical value, the probability will impact the actual average throughput in our experiments. Typically, a small probability of $\frac{1}{K^2}$ may cause poorer throughput performance. An  experiment study of the two variables are illustrated in Fig. 5 and Fig. 6.

\begin{figure*}[htpb]
    \centering
    \subfigure[$\varepsilon=0.2$]{
        \includegraphics[height=1.4in,width=2in]{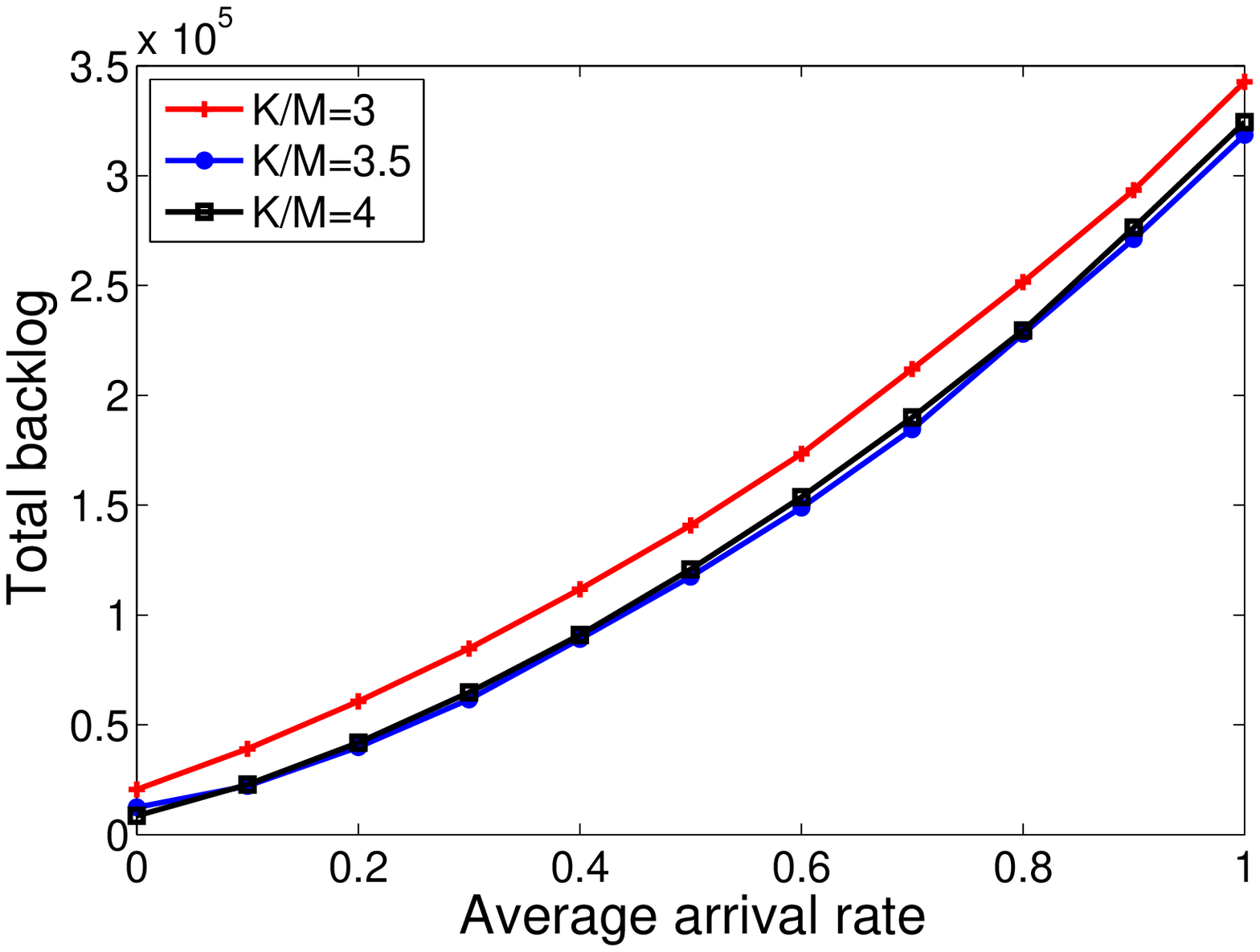}}
    \hspace{0cm}
    \subfigure[$\varepsilon=0.4$]{
        \includegraphics[height=1.4in,width=2in]{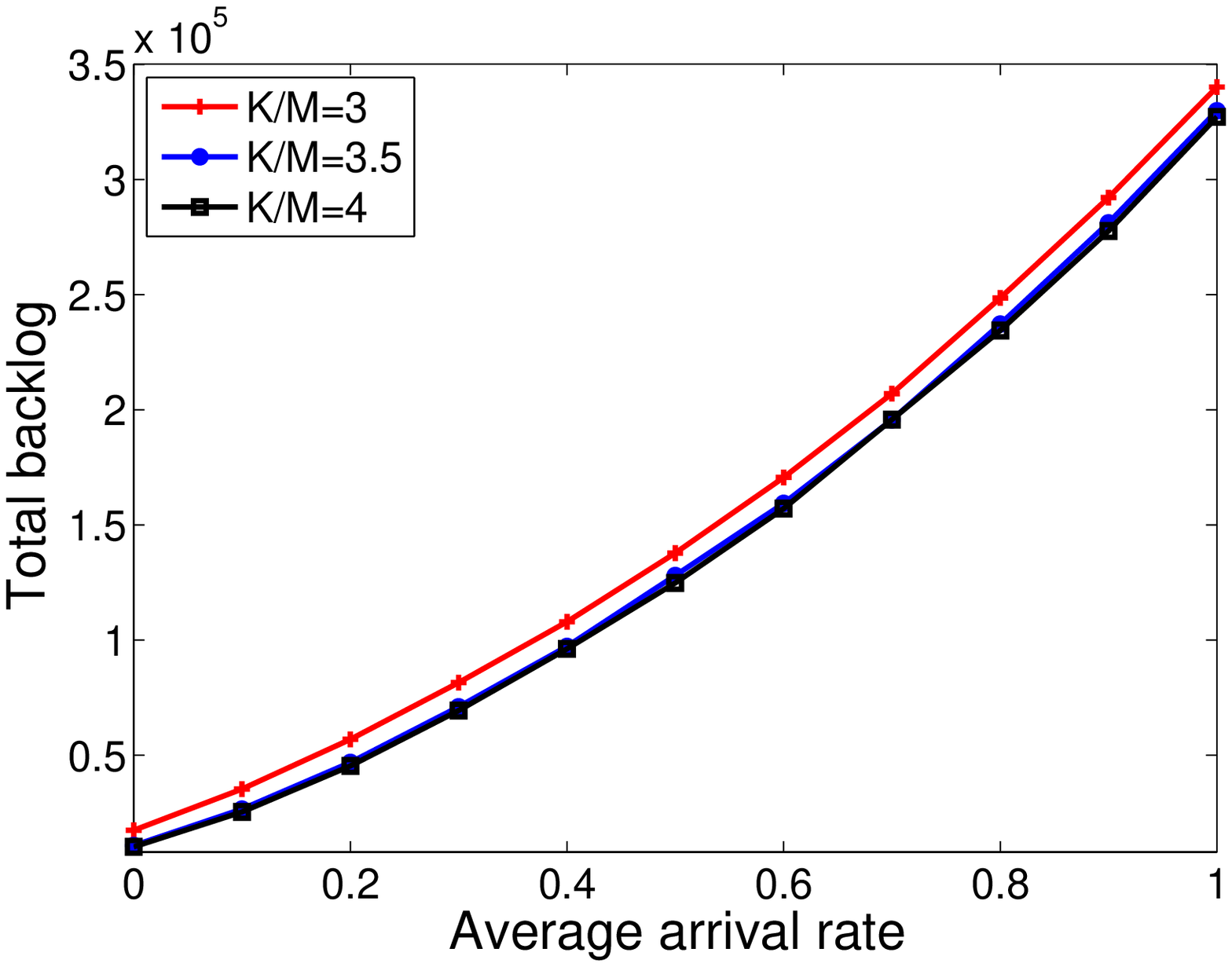}}
    \hspace{0cm}
    \subfigure[$\varepsilon=0.8$]{
        \includegraphics[height=1.4in,width=2in]{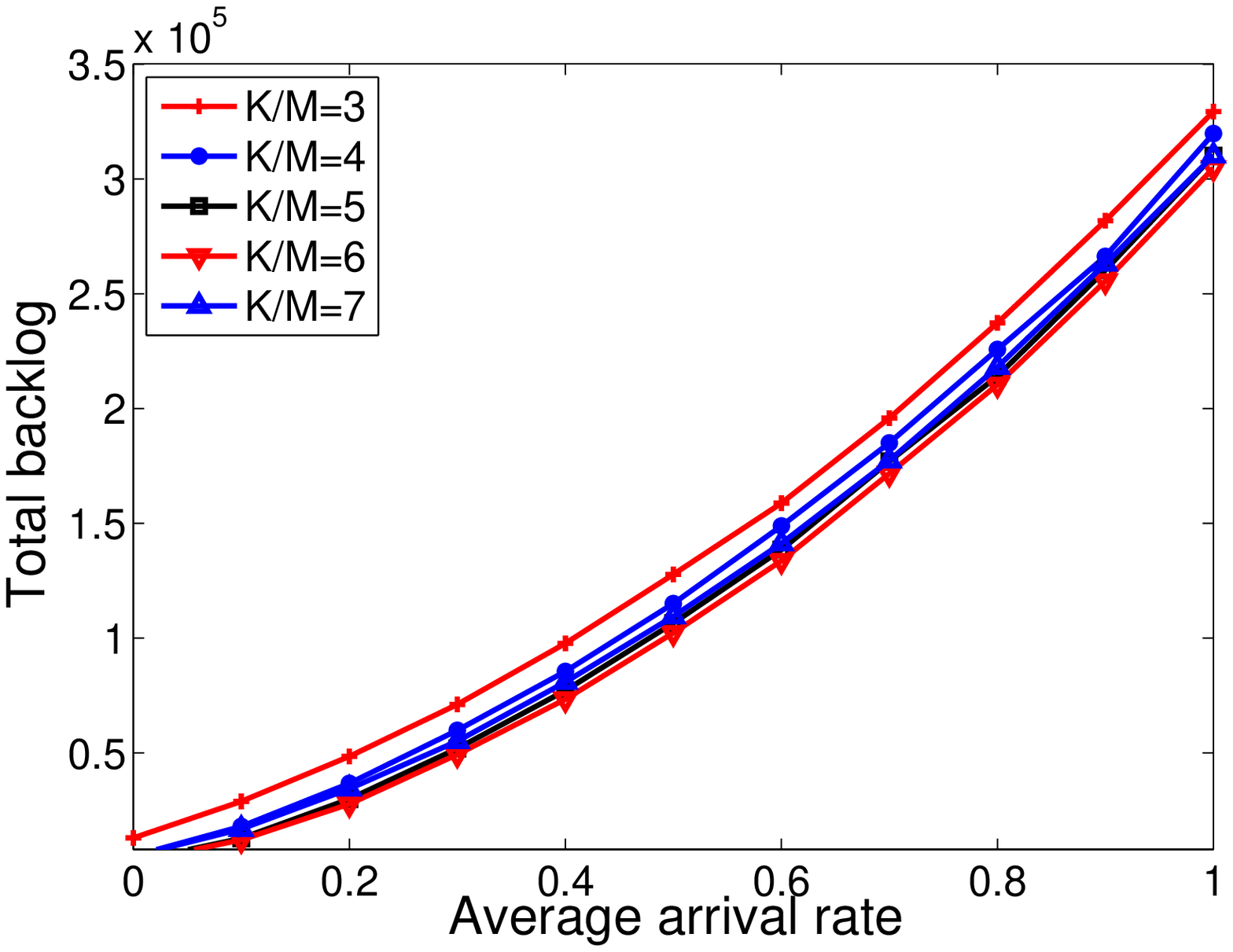}}
    \captionsetup{justification=centering}
     \caption{{\small{Total backlog vs. average arrival rate  vs. different values of $\frac{K}{M}$  at time slot $1000$ in the linear power setting}}}
\vspace*{-1\baselineskip}
\end{figure*}

In Fig. 5 we compare the total backlog by increasing the feasible values of $\frac{K}{M}$ where $\varepsilon$ serves as an constant. Fig. 5(a), 5(b), 5(c) respectively shows the comparisons of total backlog at time slot $1000$ with increasing arrival rate when $\varepsilon=0.2$, $\varepsilon=0.4$, $\varepsilon=0.8$. Note that the value of $\frac{K}{M}$ must be greater than $2$, or the size of the sub-squares will be $0$. It shall also be noticed that the feasible values of $\frac{K}{M}$ are different when $\varepsilon$ varies, since $\varepsilon$ affects the value of $M$. It then explains why we set different values for $\frac{K}{M}$ for varying $\varepsilon$.
All the three figures show that the throughput performance generally improves as $\frac{K}{M}$ increases, which coincides with the theoretical results we derive. Though the relevant probability shall become smaller as $\frac{K}{M}$ increases, there is no obvious sign shown in Fig. 5. (a), 5. (b). We can see an obvious impact in Fig. 5(c) where $\frac{K}{M}$ can be set bigger values. It shows the average throughput becomes a little worse at $\frac{K}{M}=7$ than $\frac{K}{M}=6$.

\begin{figure}[h]
    \centering
    \subfigure[$\frac{K}{M}=3$]{
        \includegraphics[height=1.2in,width=1.65in]{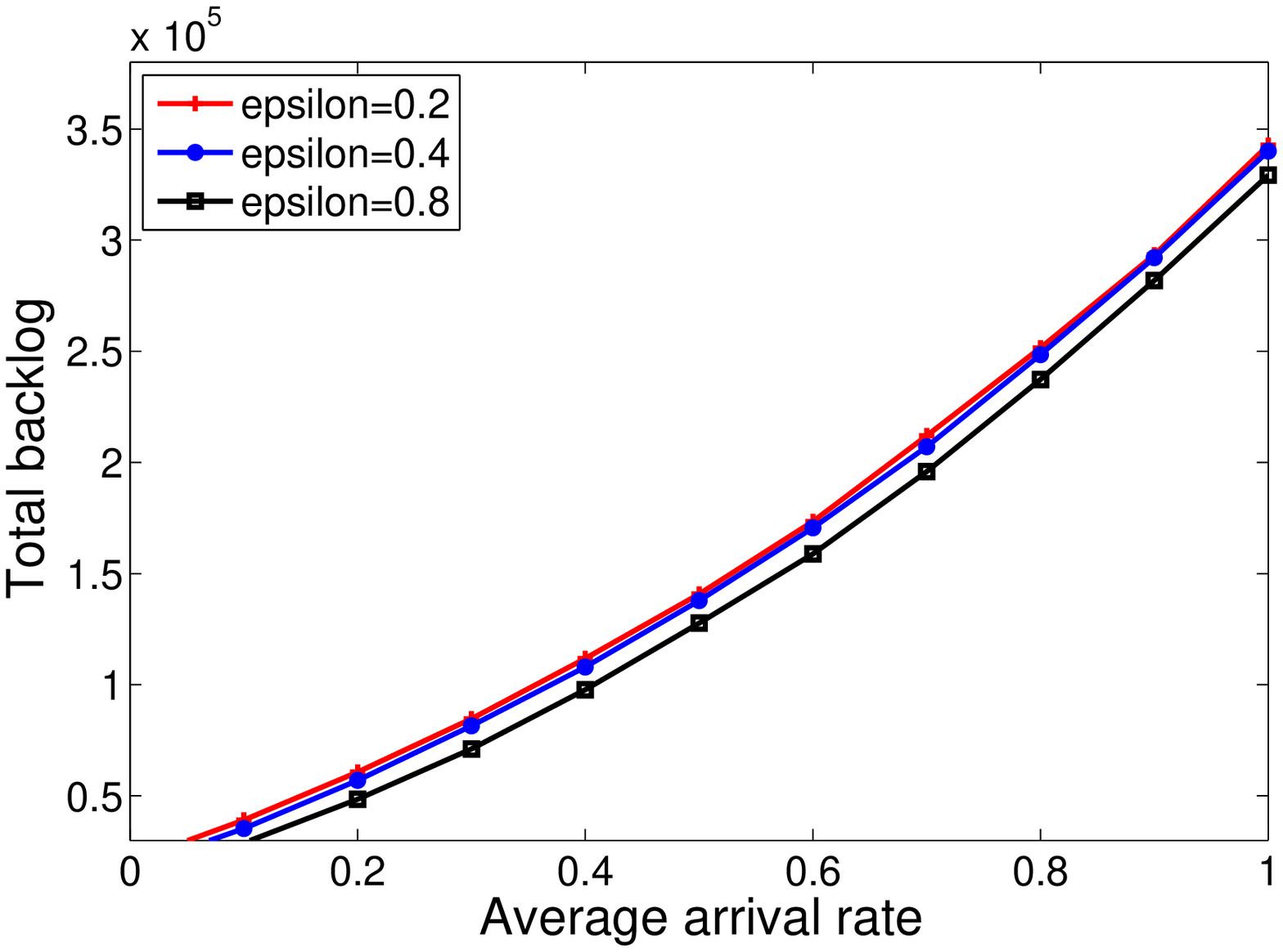}}
  \hspace{0cm}
    \subfigure[$\frac{K}{M}=4$]{
        \includegraphics[height=1.2in,width=1.65in]{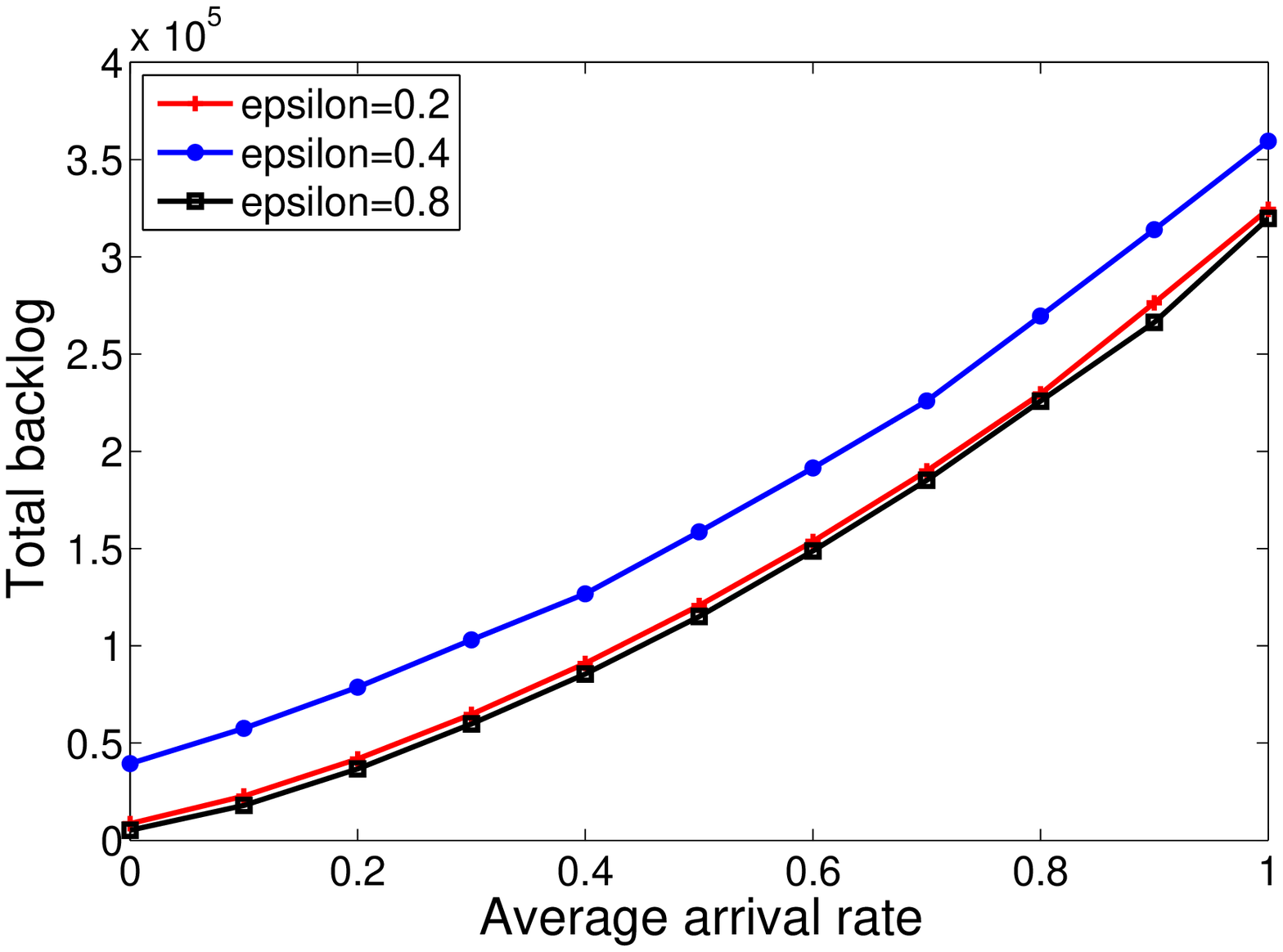}}
   \caption{{\small{Total backlog vs. average arrival rate vs. different values of $\varepsilon$ at time slot $1000$ in the linear power setting}}}
\end{figure}
We briefly illustrate the impact of $\varepsilon$ at fixed $\frac{K}{M}$ in Fig. 6. Though the theoretically achievable capacity region shall have been greater with a smaller $\varepsilon$, it seems to show contradicted results in Fig. 6(a). The apparent contradiction lies behind the probability of $\frac{1}{K^2}$ which becomes quite small because of a much bigger $M$ caused by a smaller $\varepsilon$.   Fig. 6(b) shows a similar consequence caused by the crucial impact of $\varepsilon$ both on the achievable capacity region and the relevant probability.

We then focus on comparisons with the two centralized algorithms in Fig. 7, 8. We set $\varepsilon=0.8 \mbox{, } \frac{K}{M}=6$ to conduct the following simulations. These simulation results indicate that our distributed scheduling algorithm(denoted by DS in figures) achieves comparable performance with the centralized GMS and the heuristic RA.
\begin{figure}[htpb]
    \centering
    \subfigure[\hspace*{-0.5\baselineskip} Total backlog vs. average arrival rate]{
        \includegraphics[height=1.2in,width=1.7in]{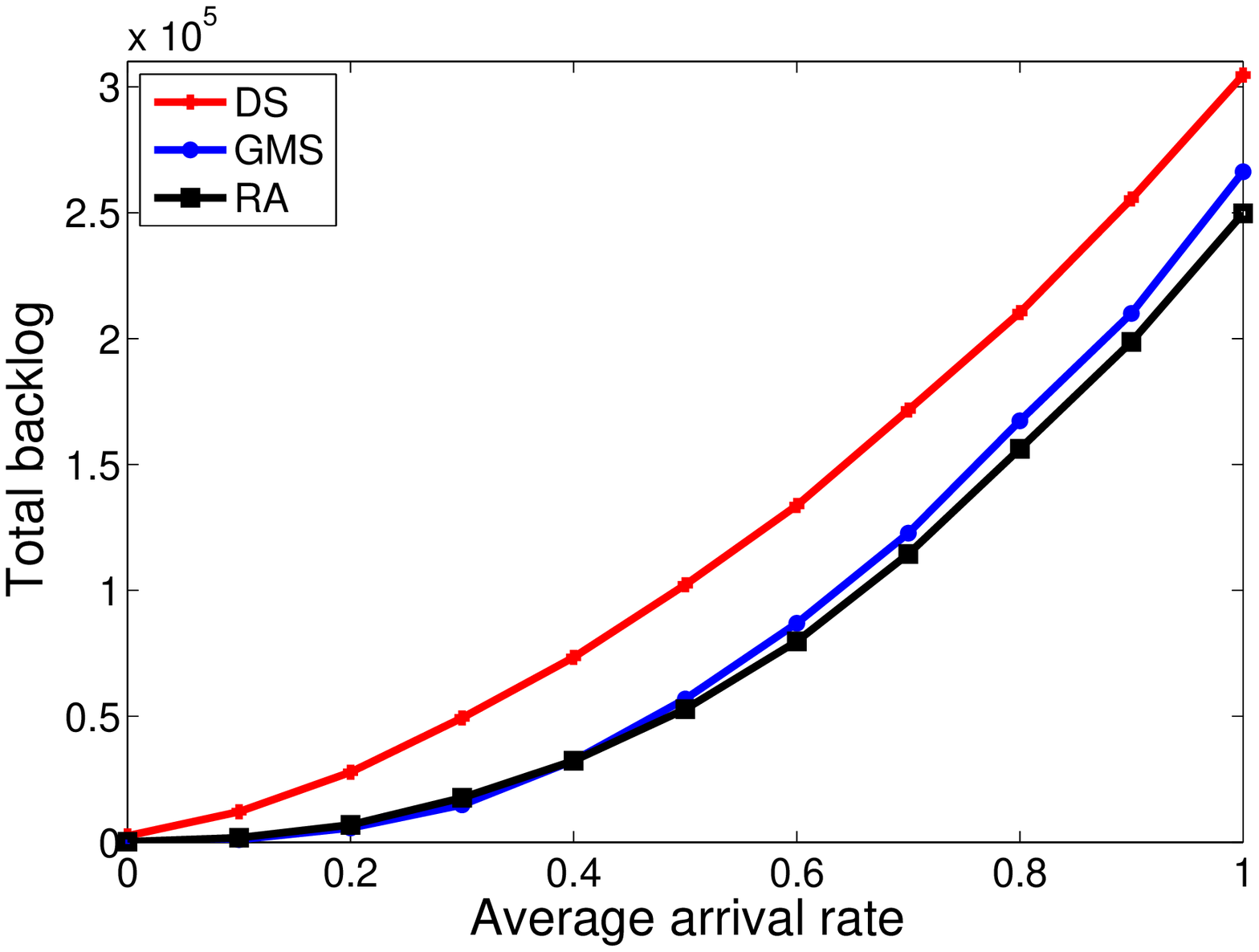}}
  \hspace{0cm}
    \subfigure[Zoom in of (a)]{
        \includegraphics[height=1.2in,width=1.6in]{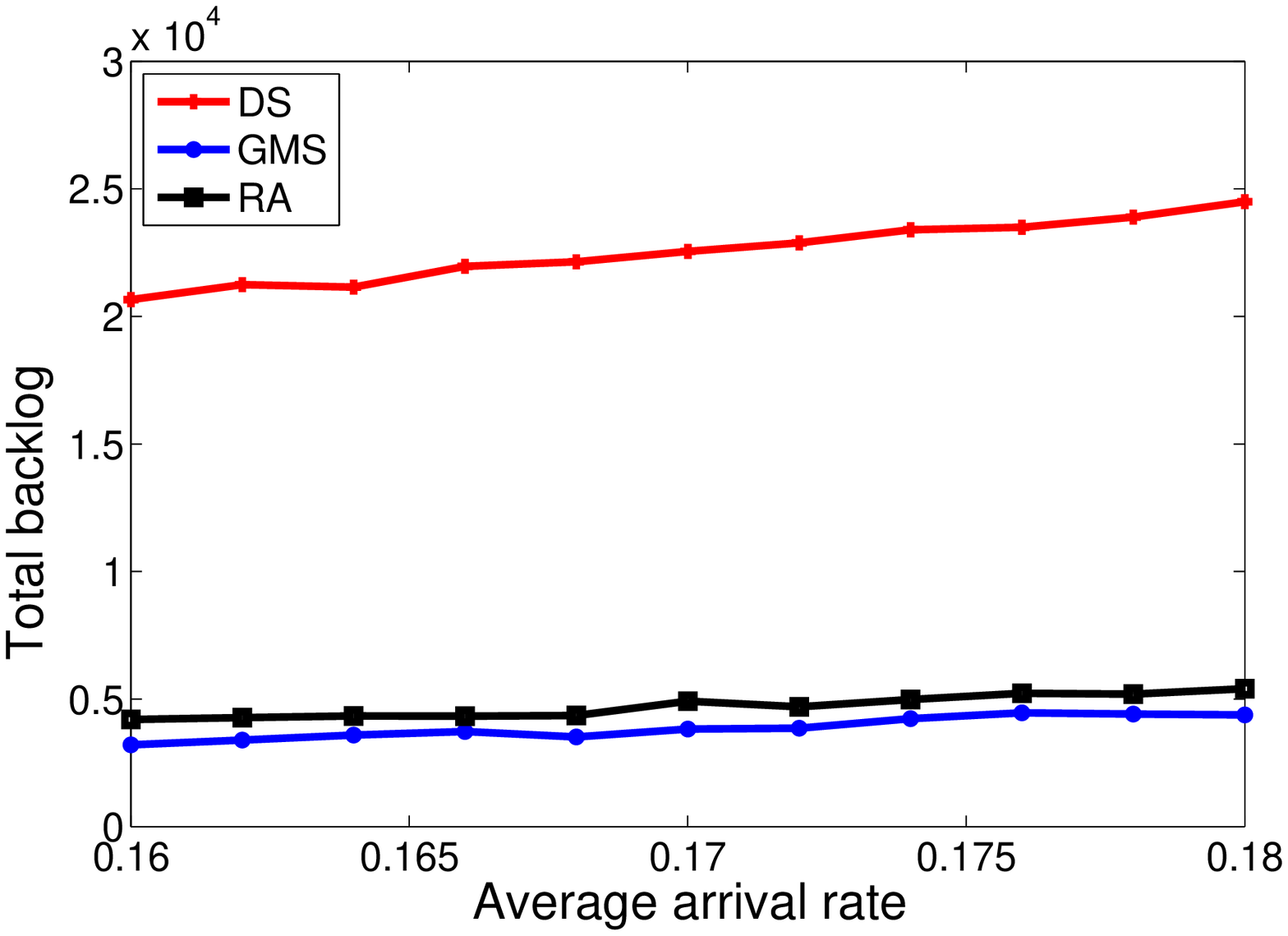}}
     \caption{{\small{Total backlog vs. average arrival rate  vs. different algorithms at time slot $1000$ in the linear power setting} }}
\end{figure}

In Fig. 7 we compare the total backlog changes of the three algorithms as average arrival rate increases. It shows that our algorithm may approximately support a maximum average rate no greater than $0.2$, and the other two support  no greater than $0.3$. We zoom in a subgraph of the Fig. 7(a) in the Fig. 7(b), where the average arrival rate is in $[0.16, 0.18]$. It shows that our algorithm can keep the total backlog stable in the region.
Fig. 8 then illustrates detailed comparisons of achievable capacity region for the three algorithms. It shows that our algorithm can support a comparable traffic arrival rate vector with the two centralized algorithms.  From the three subfigures we can see that our proposed algorithm can keep the total backlog stable at an arrival rate no greater than $0.18$, while the counterpart of the greedy algorithm and random algorithm is $0.28$ and $0.26$.

\begin{figure*}
    \centering
    \subfigure[DS]{
        \includegraphics[height=1.4in,width=2.0in]{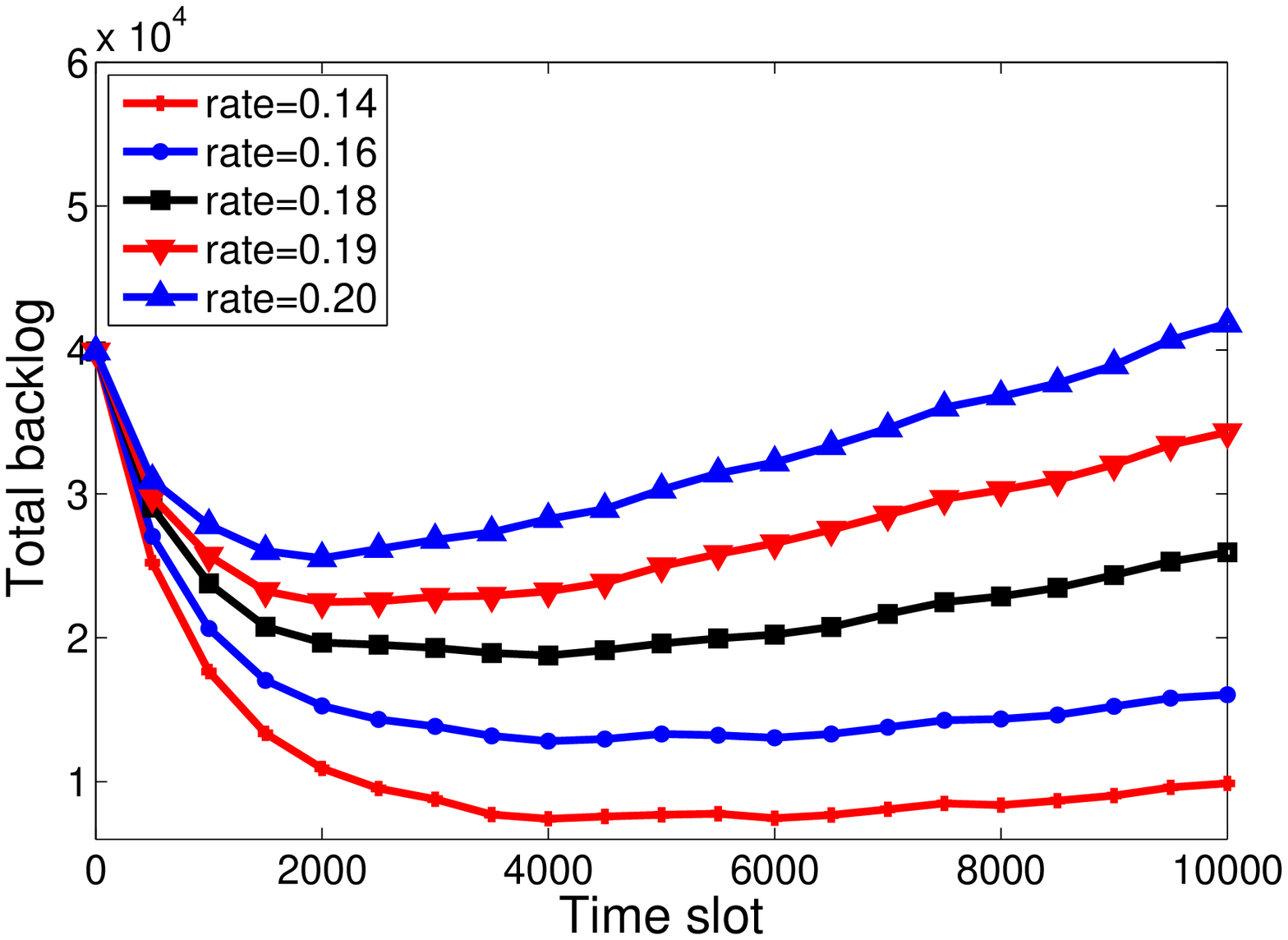}}
  \hspace{0cm}
    \subfigure[GMS]{
        \includegraphics[height=1.4in,width=2.0in]{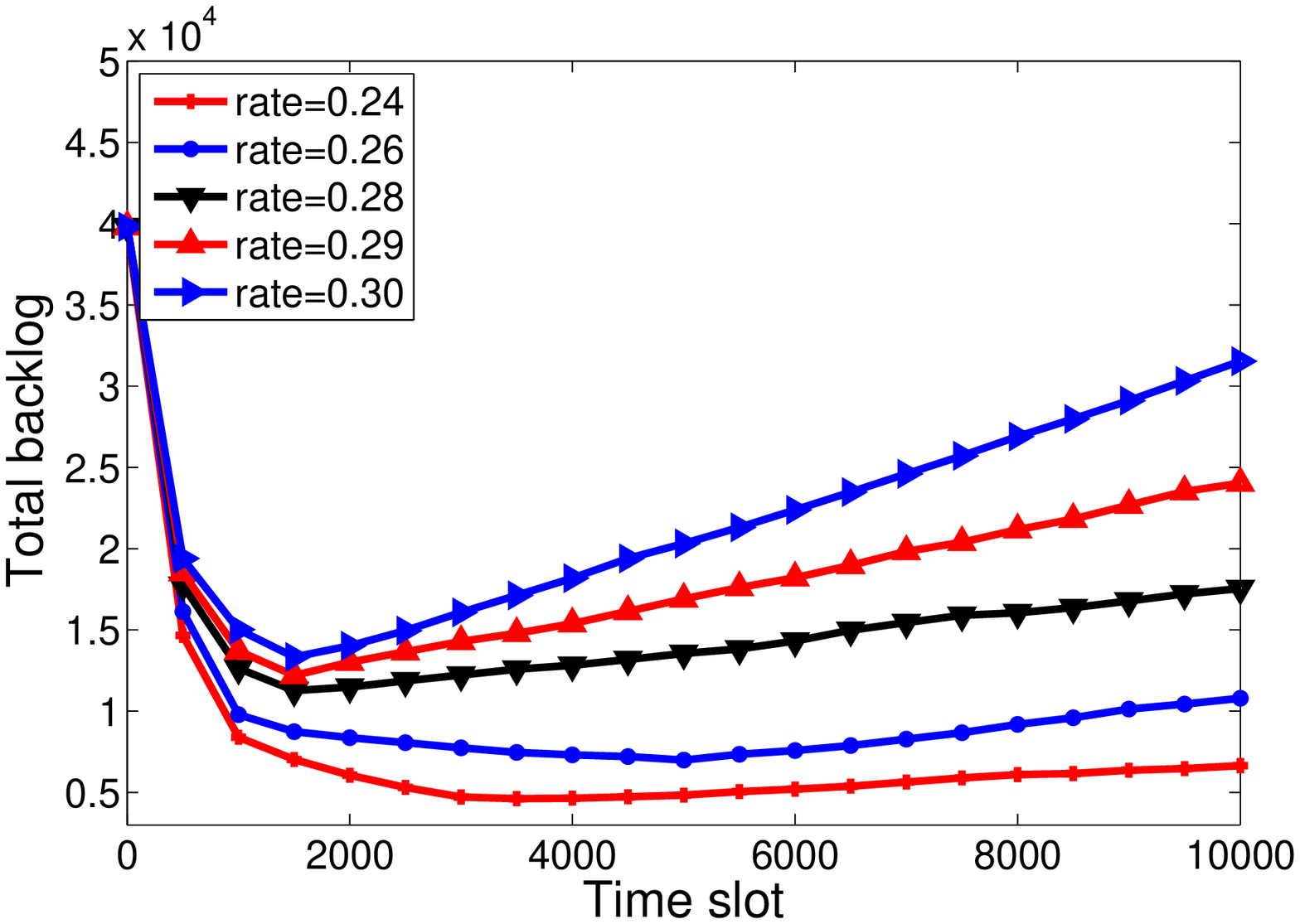}}
  \hspace{0cm}
    \subfigure[RA]{
        \includegraphics[height=1.4in,width=2.0in]{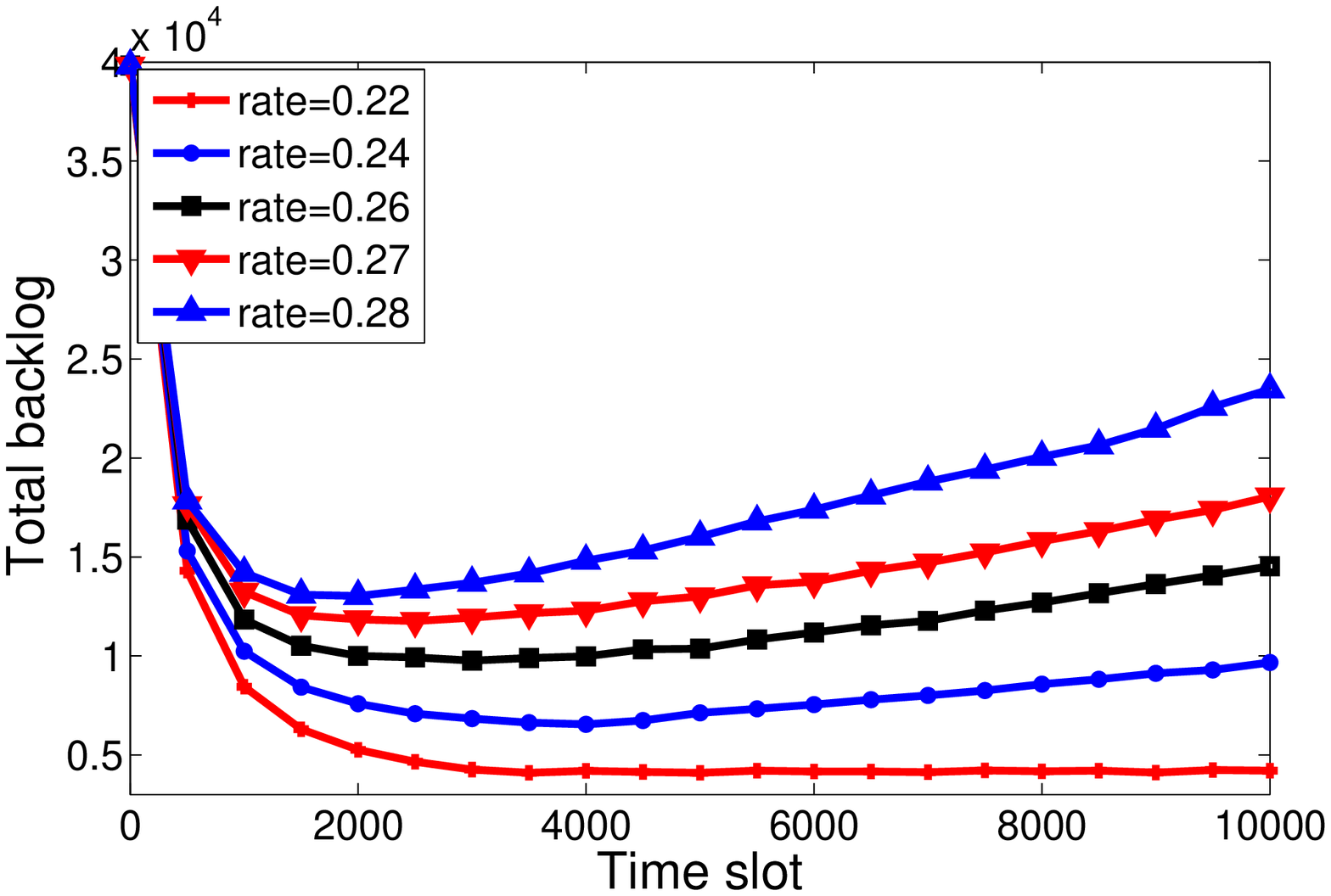}}
   \caption{{\small{Achievable capacity region of different algorithms in the linear power setting}}}
\end{figure*}

\subsection{Under Uniform Power Setting}
Now we conduct the same three series of experiments for the uniform power assignment.


The setting of the first series of experiments is the same as the linear power setting. The corresponding numerical results are plotted in Fig. 9, 10. From the two figures it concludes that the transmitting links receive much smaller interference outside than the theoretical bound. Therefore, it indicates that our algorithm is correctly applicable to the uniform power setting as well.

Then we then conduct another set of experiments to study how $\frac{K}{M}$  and $\varepsilon$ affect the performance.

We show the effect of $\frac{K}{M}$ at different values of $\varepsilon$ where $\varepsilon=0.2,$  $ 0.4,$  $ 0.8$ respectively in Fig. 11. Fig. 11(a), 11(b) and 11(c) plot the total backlog changes as increasing average arrival rate under different values of $\frac{K}{M}$. The trends are a little different from those in the linear power setting. From the three figures we can see that the average throughput performance gets better as increasing $\frac{K}{M}$ under a fixed $\varepsilon$. The decreasing probability of $\frac{1}{K^2}$ shows no obvious influence on the results. It may be partly caused by the algorithm for computing new schedulings inside sub-squares since it selects candidate links based on their distance with previous selected links. Thus a larger area implies more links get scheduled. This improvement remits the influence of the decreasing probability of  $\frac{1}{K^2}_.$

\begin{figure}[htpb]
 \centering
    \centering
    \subfigure[Interference vs. links]{
        \includegraphics[height=1.2in,width=1.6in]{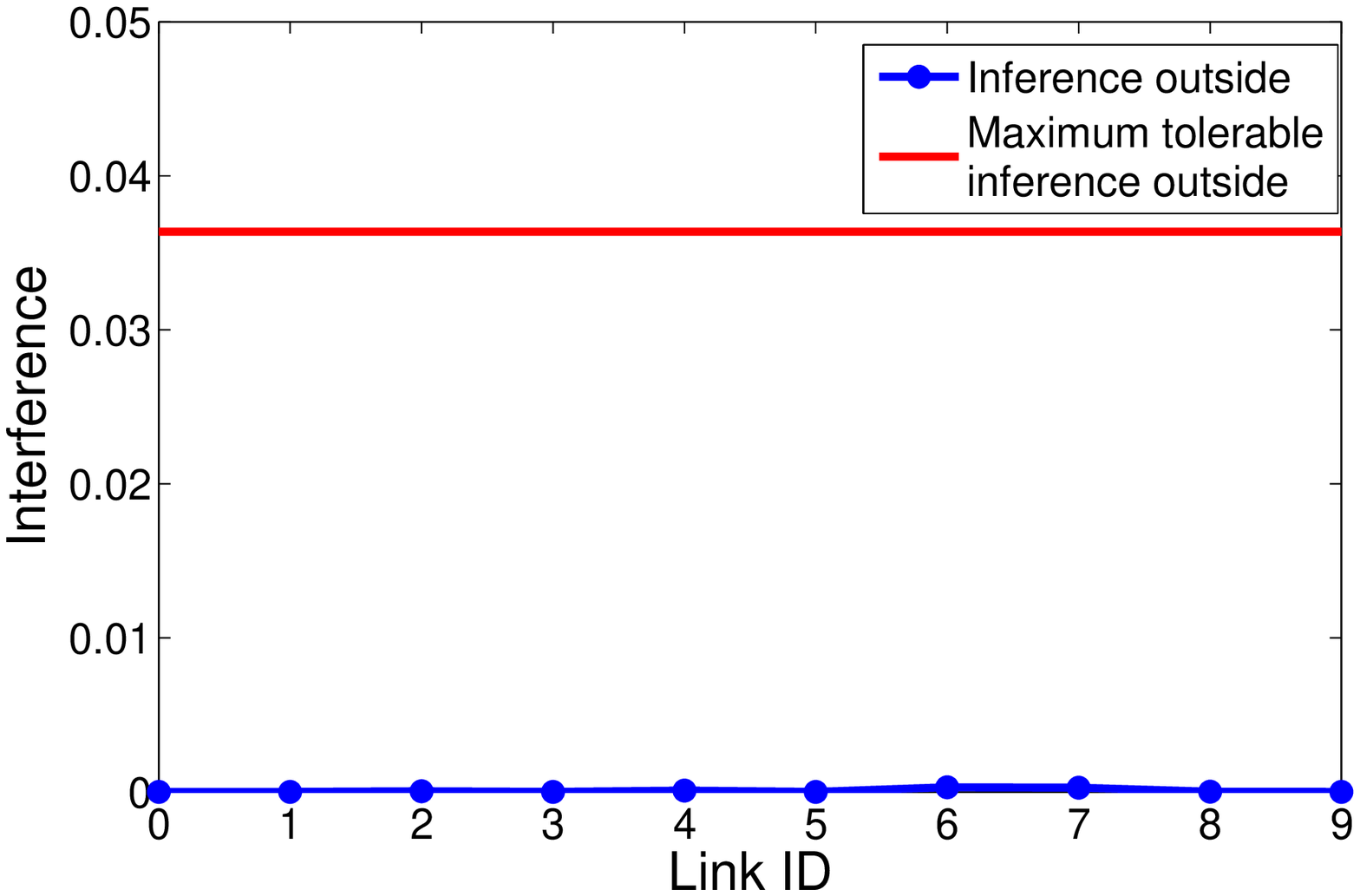}}
  \hspace{0cm}
 \hfill
    \subfigure[Affectness vs. links]{
        \includegraphics[height=1.2in,width=1.6in]{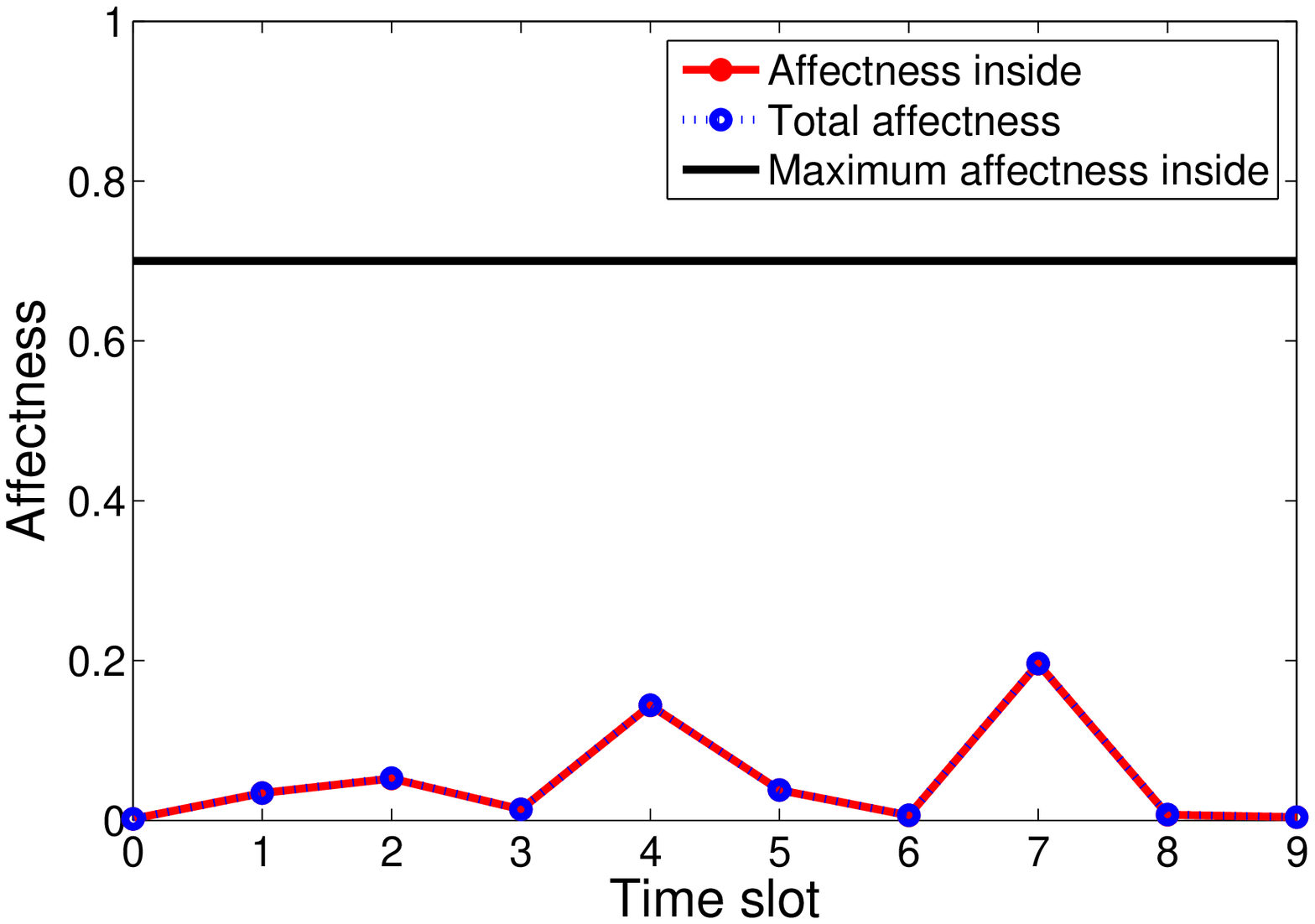}}
     \caption{{\small{The interference/affectness each randomly chosen transmitting link receives with the uniform power assignment. (a) shows the outside interference each link receives. (b) shows the inside affectness for each link, and the total affectness each receives from all other simultaneously transmitting links.}}}
 \end{figure}

\begin{figure}
    \centering
     \subfigure[Average interference vs. time slot]{
        \includegraphics[height=1.2in,width=1.60in]{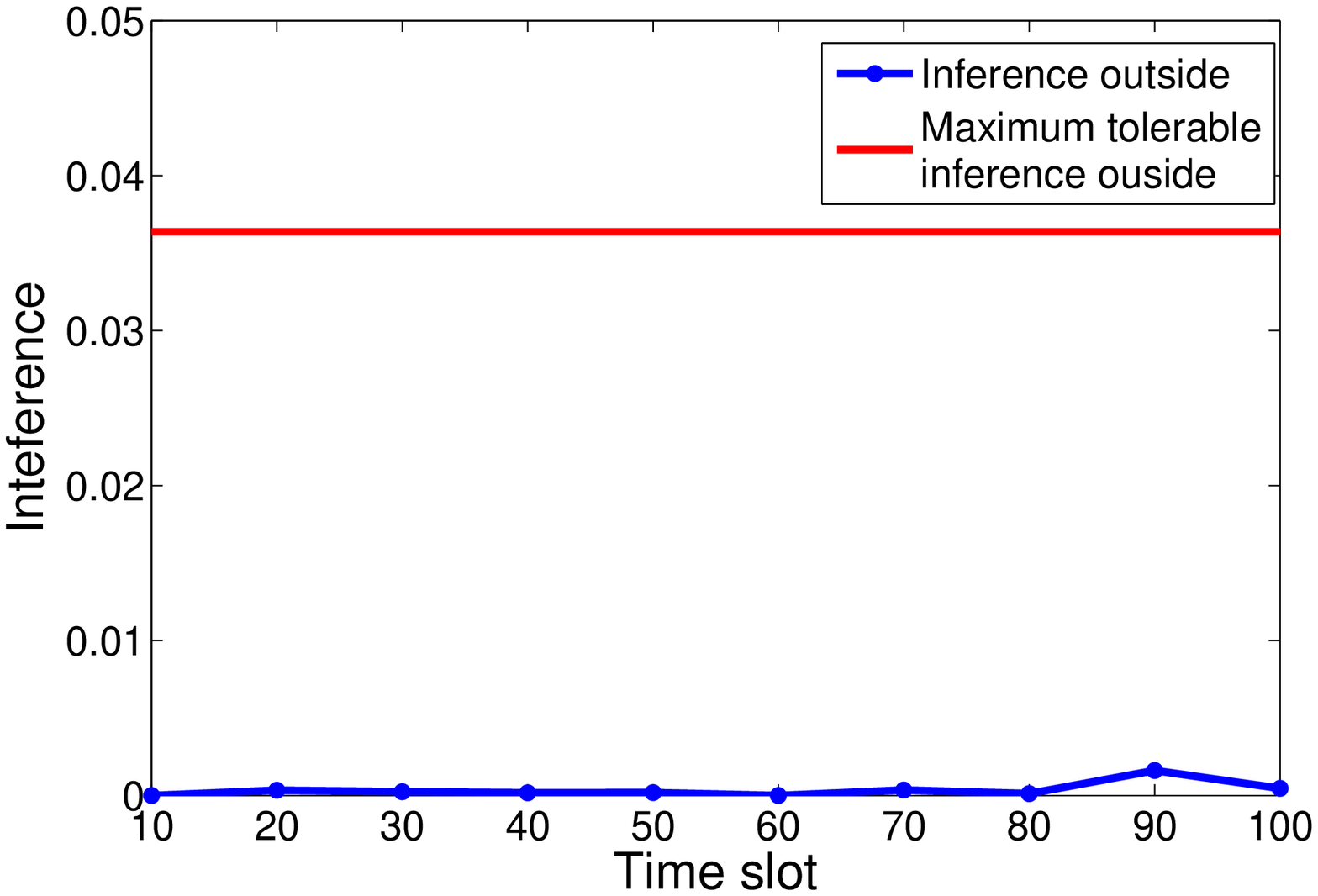}}
  \hspace{0cm}
  \hfill
    \subfigure[Average affectness vs. time slot]{
        \includegraphics[height=1.2in,width=1.60in]{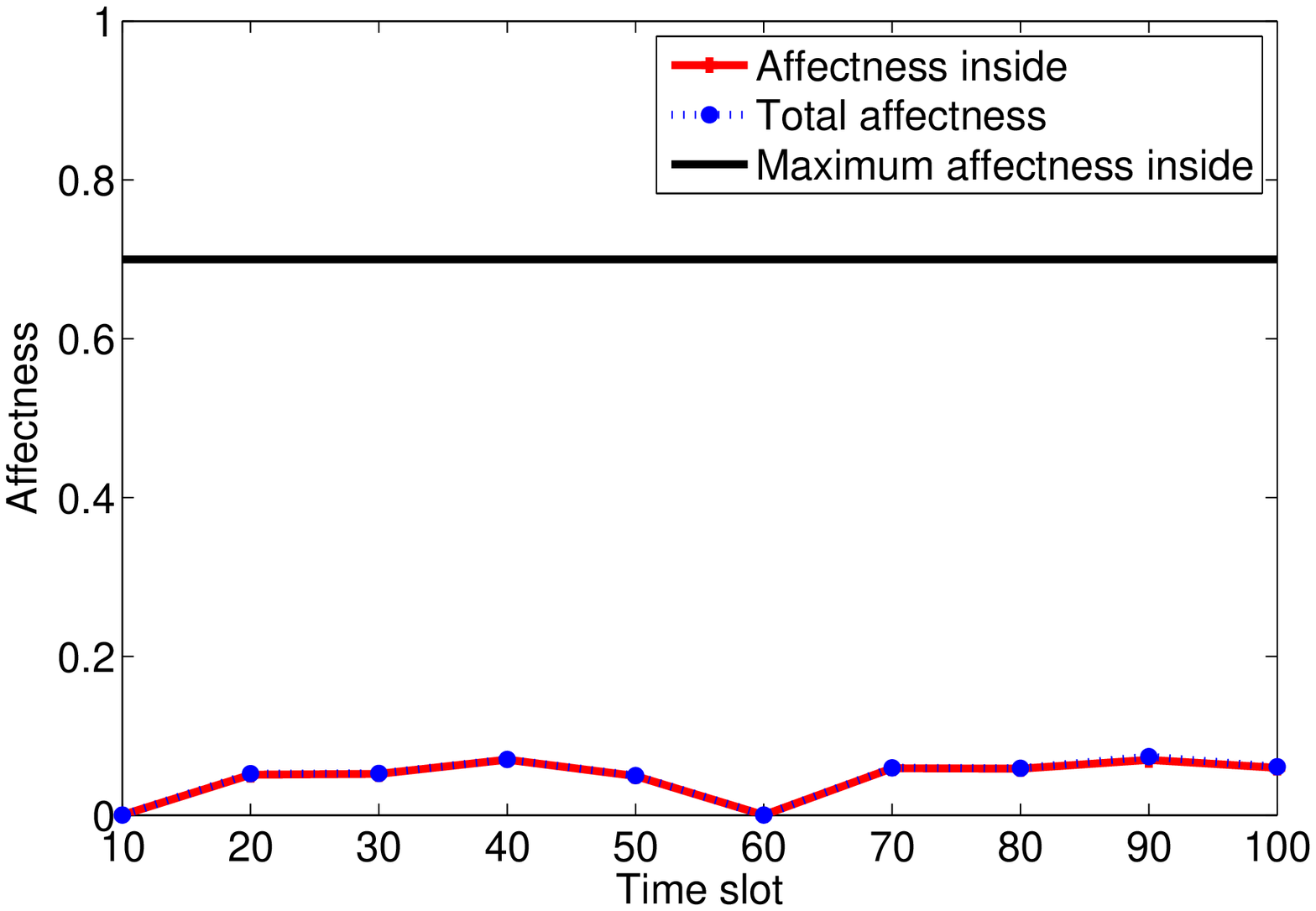}}
     \caption{{\small{Average interference per transmitting link receives at different time slots with the uniform power assignment. (a) shows the average outside interference  per transmitting link receives. (b) shows the average inside affectness, and the average total affectness per transmitting link receives from all other simultaneously transmitting links.} }}
\end{figure}

\begin{figure*}[htpb]
    \centering
    \subfigure[$\varepsilon=0.2$]{
        \includegraphics[height=1.4in,width=2.0in]{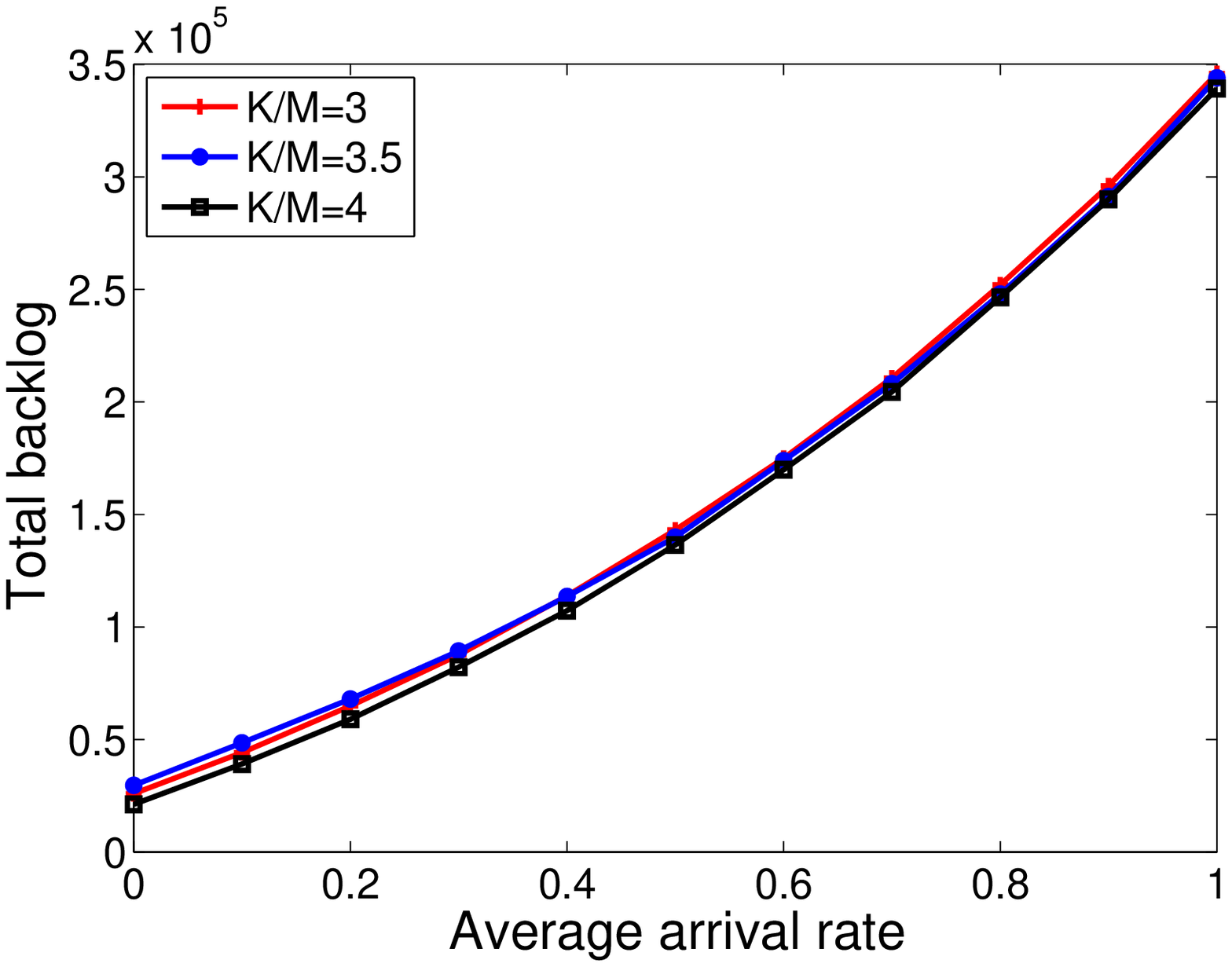}}
  \hspace{0cm}
    \subfigure[$\varepsilon=0.4$]{
        \includegraphics[height=1.4in,width=2.0in]{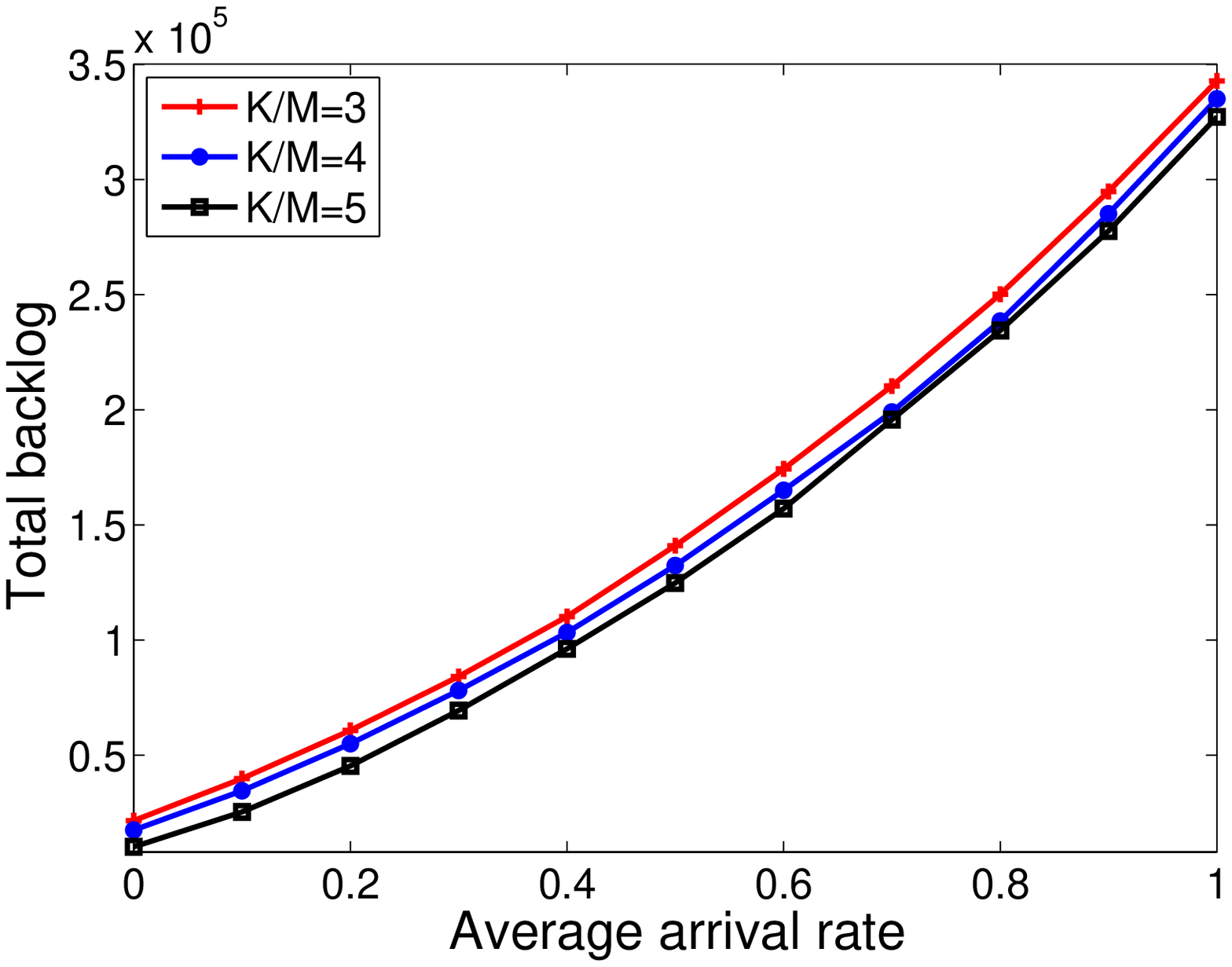}}
  \hspace{0cm}
    \subfigure[$\varepsilon=0.8$]{
        \includegraphics[height=1.4in,width=2.0in]{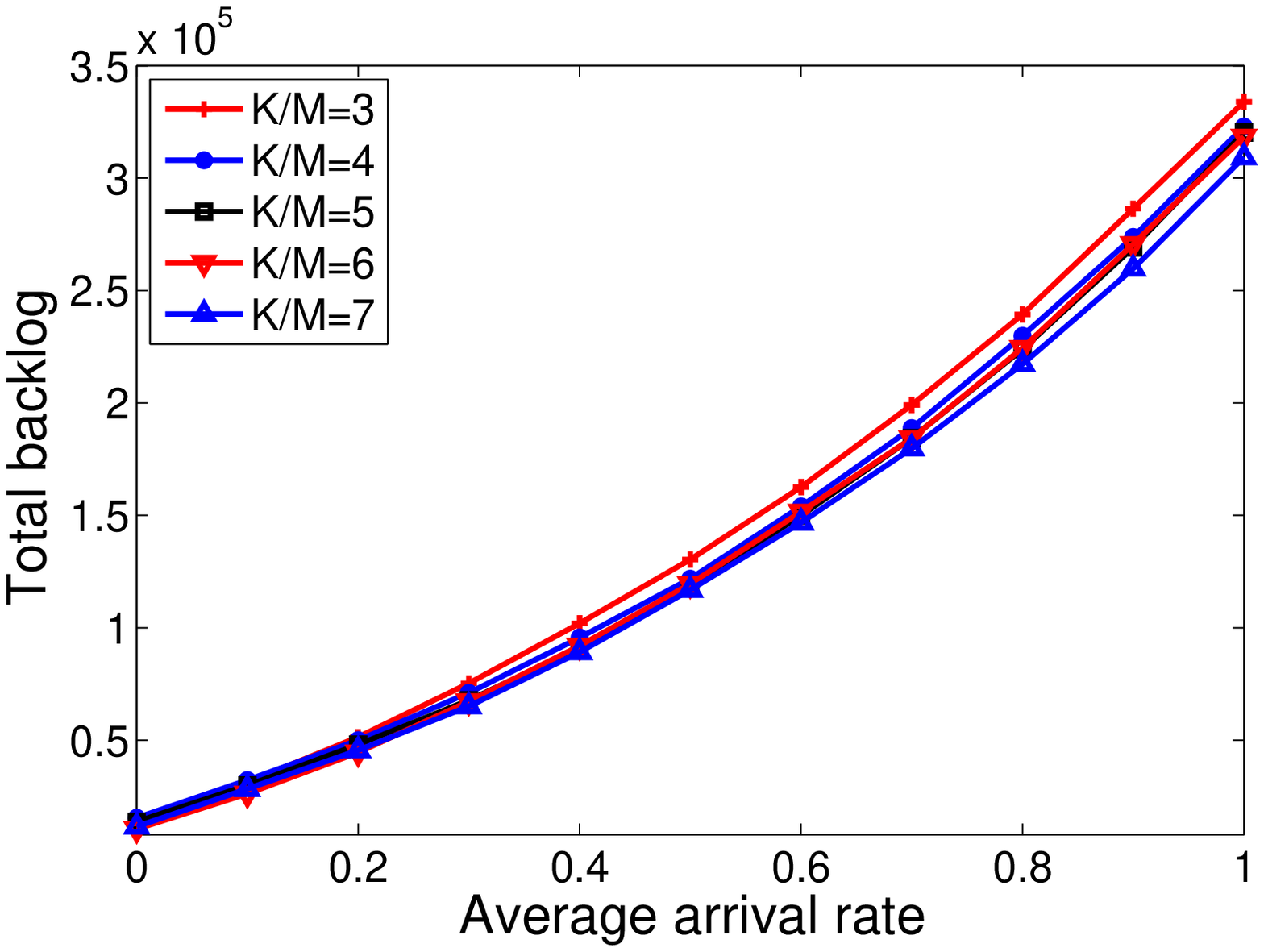}}
   \caption{{\small{Total backlog vs. average arrival rate  vs. different values of $\frac{K}{M}$  at time slot $1000$ in the uniform power setting}}}
\end{figure*}

The similar trend occurs when $\varepsilon$ increases at different values of $\frac{K}{M}_.$ We give a brief illustration in Fig. 12. Both the figures show that a larger $\varepsilon$ generates better performance at fixed $\frac{K}{M}_. $
Since the affectness of local ISLs computed by the algorithm of \cite{S:phy9} inside each sub-square may be much smaller than $1-\varepsilon$, it explains why $\varepsilon$ has little influence on the theoretical bound actually. But $\varepsilon$ has much more influence on the value of $M$ and the corresponding probability. Therefore, in the uniform power setting, the algorithm achieves better performance with a larger $\varepsilon$.

\begin{figure}[htpb]
    \centering
    \subfigure[$\frac{K}{M}=3$]{
        \includegraphics[height=1.2in,width=1.65in]{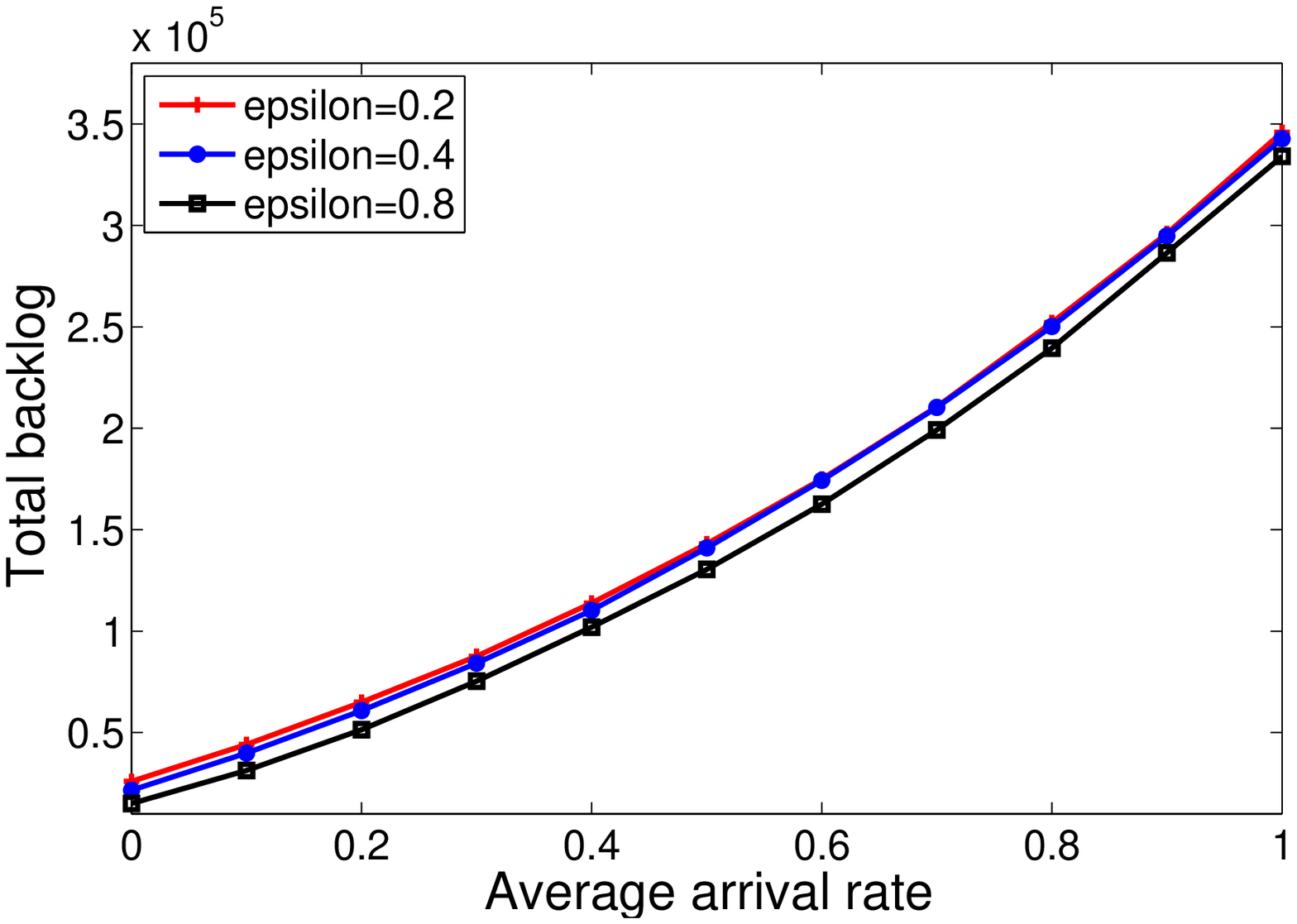}}
  \hspace{0cm}
    \subfigure[$\frac{K}{M}=4$]{
        \includegraphics[height=1.2in,width=1.65in]{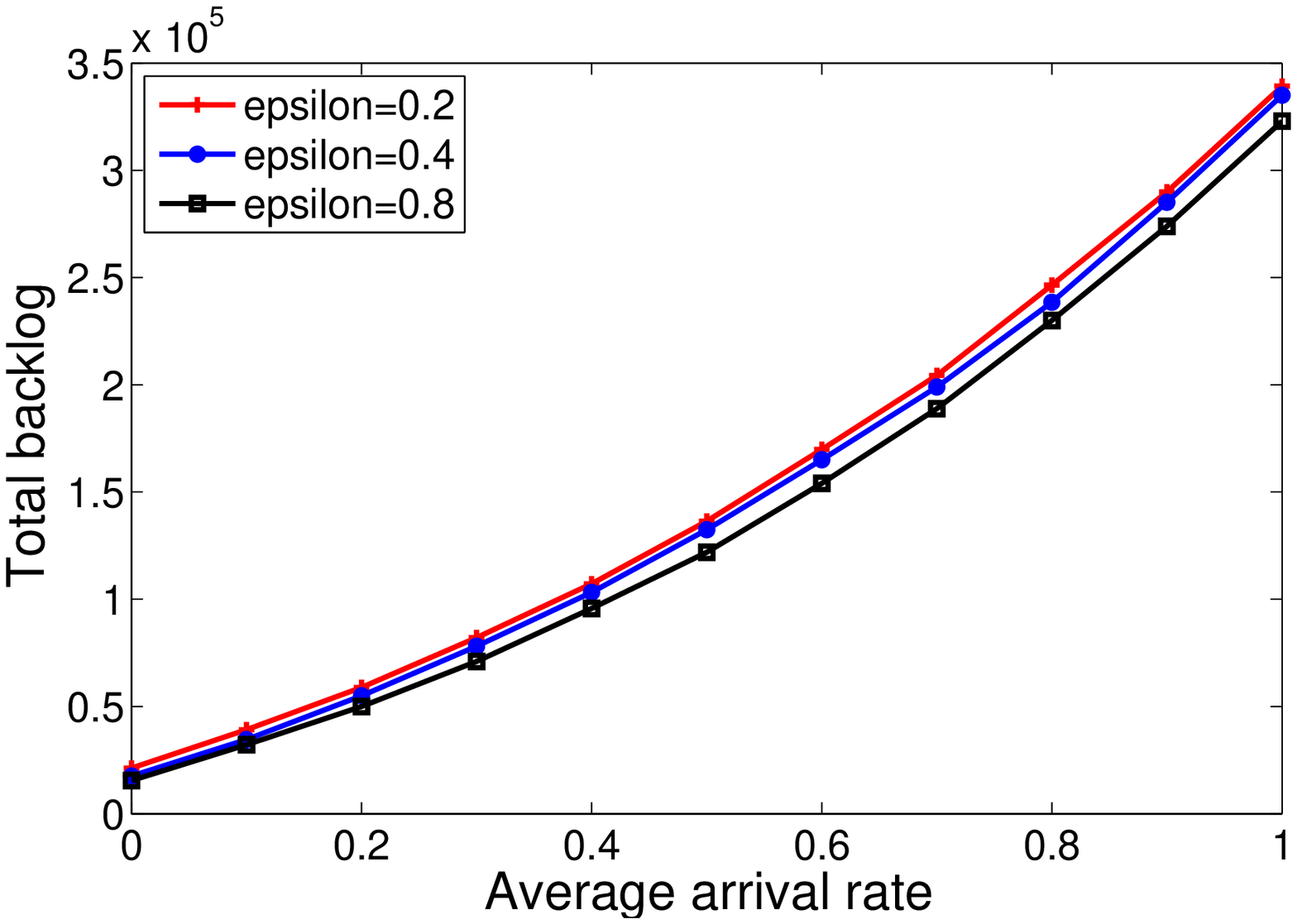}}
   \caption{{\small{Total backlog vs. average arrival rate vs. different values of $\varepsilon$ at time slot $1000$ in the uniform power setting}}}
\end{figure}

We  conduct the third set of experiments to compare with the two centralized algorithms on average throughput where we set $\varepsilon=0.9$, $\frac{K}{M}=9$. The results in Fig. 13, 14  convey the same kind of information as those in Fig. 7, 8. These graphs jointly show that Algorithm $1$ still keep comparable throughput performance with the two centralized in terms of total backlog. For example, the maximum supportable average arrival rate of our algorithm is around $0.12$, nearly a half of $0.24$ and $0.22$ achieved by the two centralized algorithm.

\begin{figure}[htpb]
    \centering
    \subfigure[\hspace*{-0.6\baselineskip} Total backlog vs. average arrival rate]{
        \includegraphics[height=1.2in,width=1.7in]{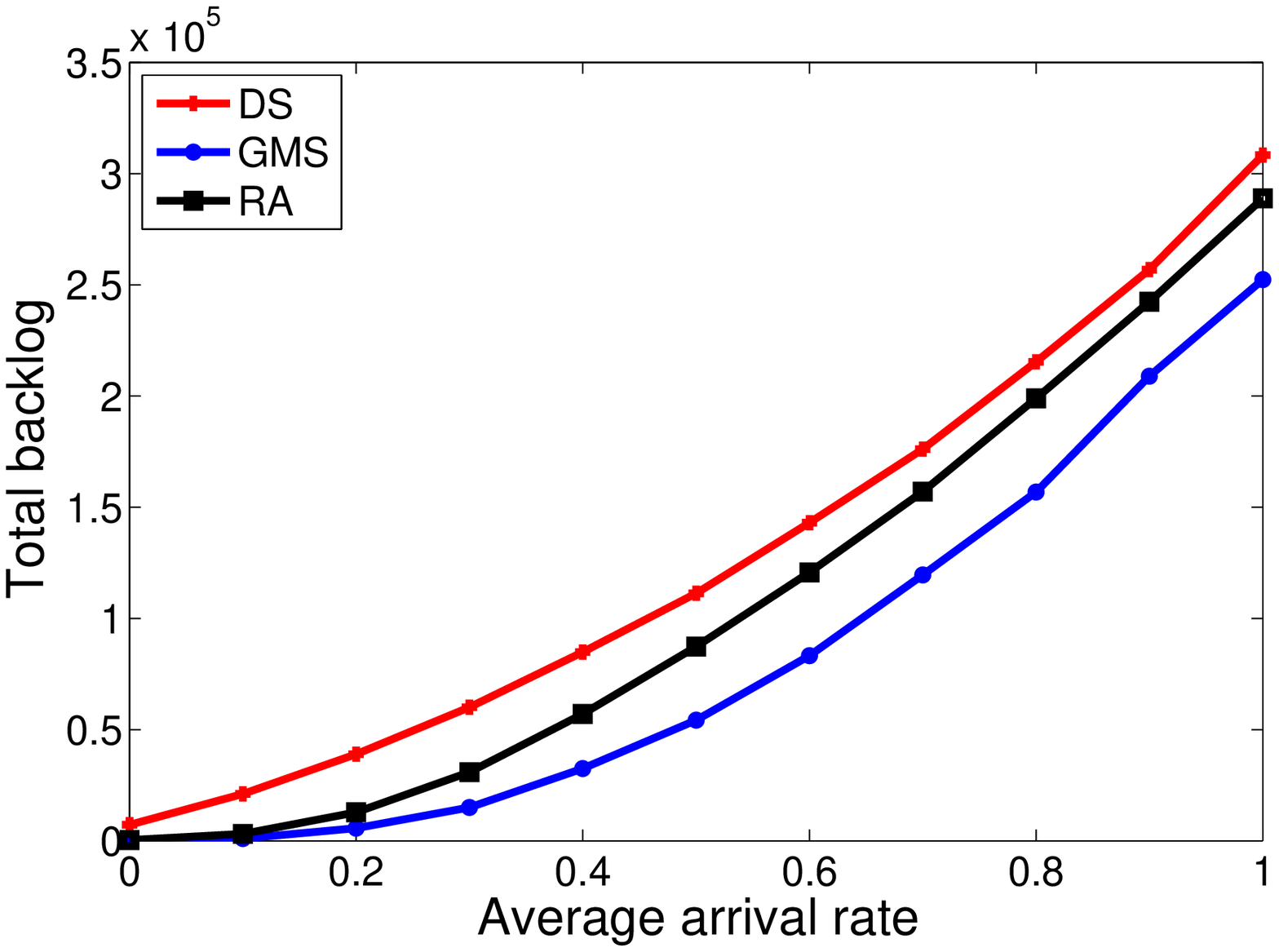}}
  \hspace{0cm}
    \subfigure[Zoom in of (a)]{
        \includegraphics[height=1.2in,width=1.6in]{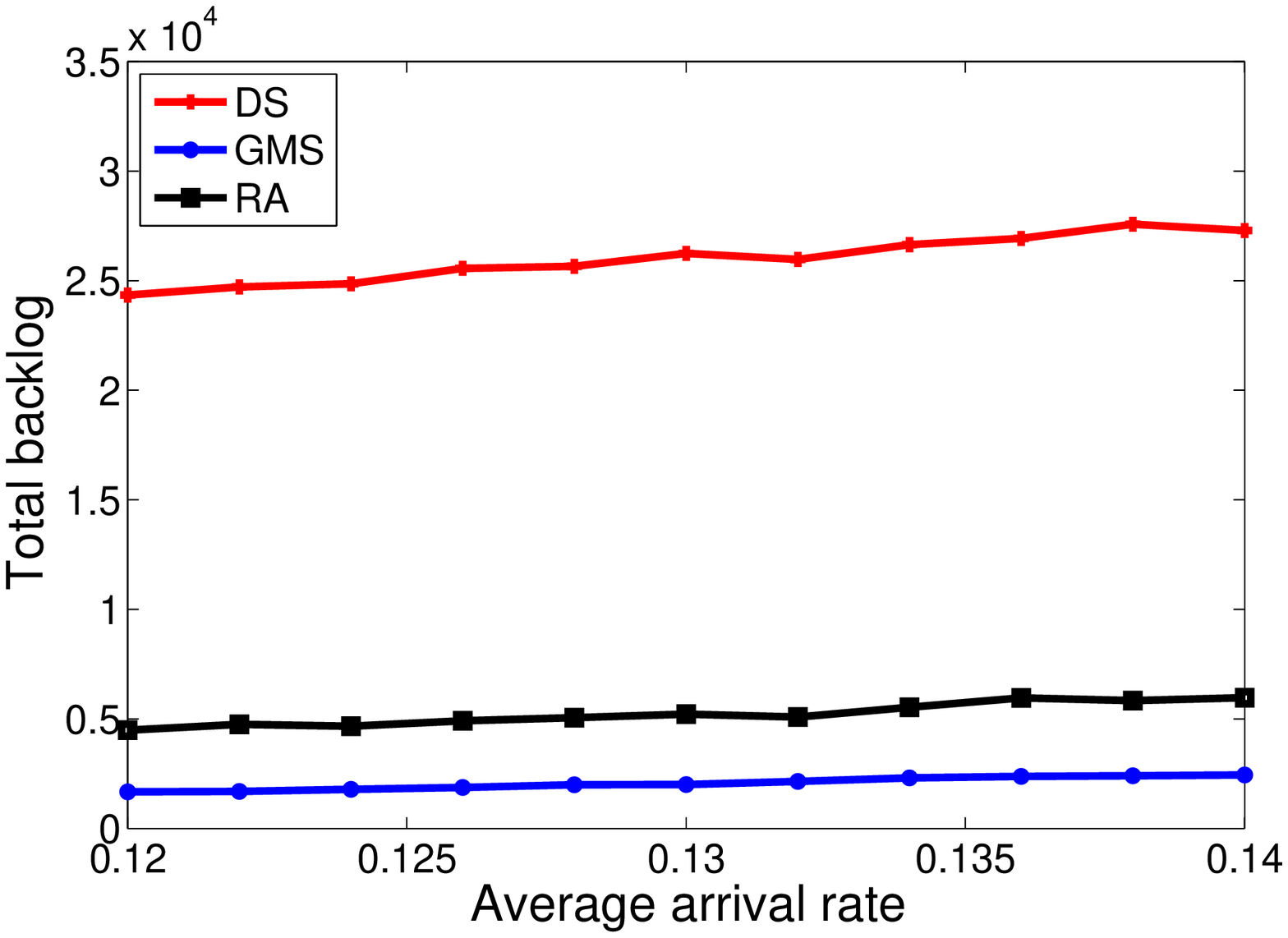}}
     \caption{{\small{Total backlog vs. average arrival rate  vs. different algorithms at time slot $1000$ in the uniform power setting} }}
\end{figure}

\begin{figure*}[htpb]
    \centering
    \subfigure[DS]{
        \includegraphics[height=1.4in,width=2.0in]{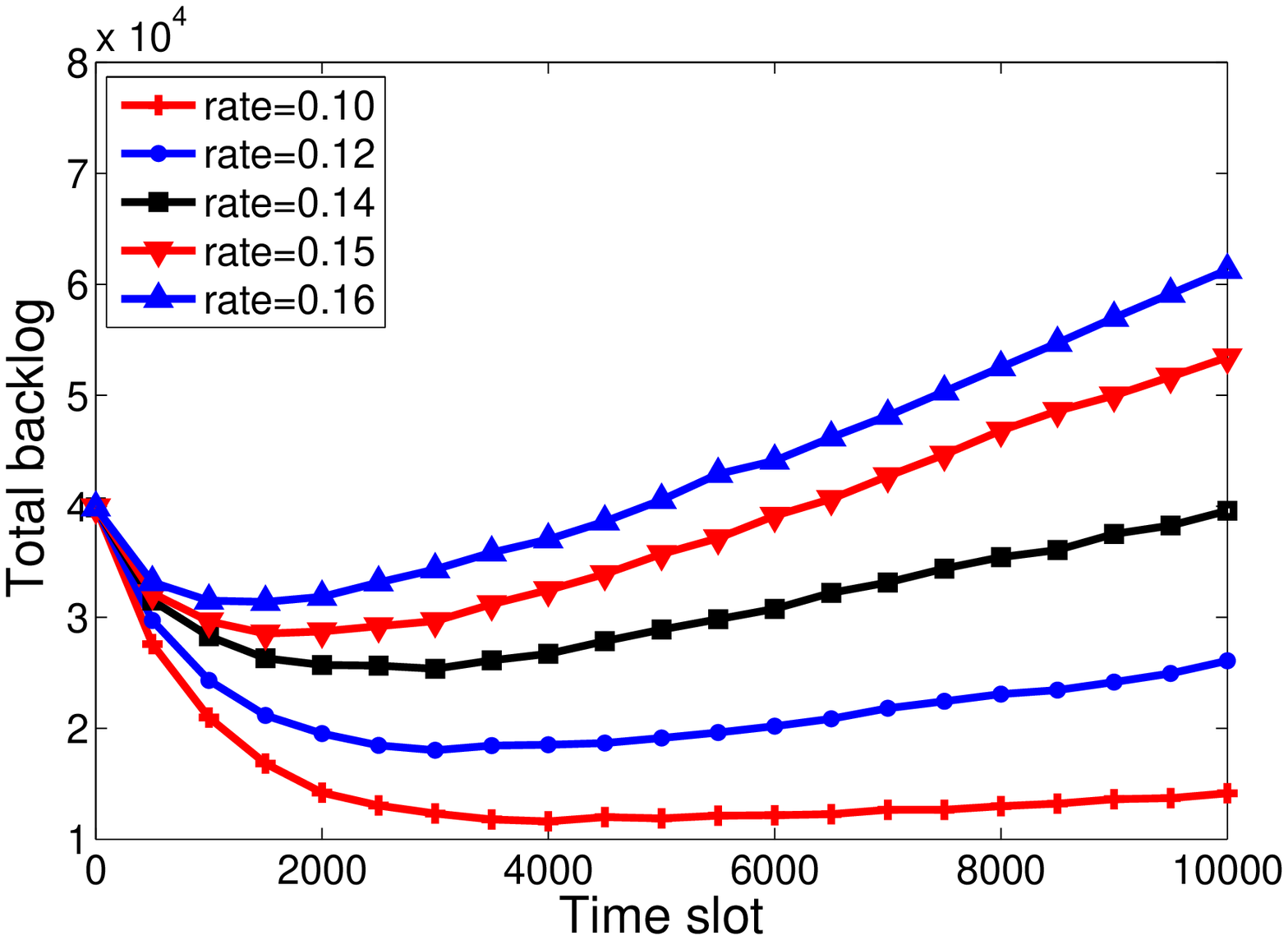}}
  \hspace{0cm}
    \subfigure[GMS]{
        \includegraphics[height=1.4in,width=2.0in]{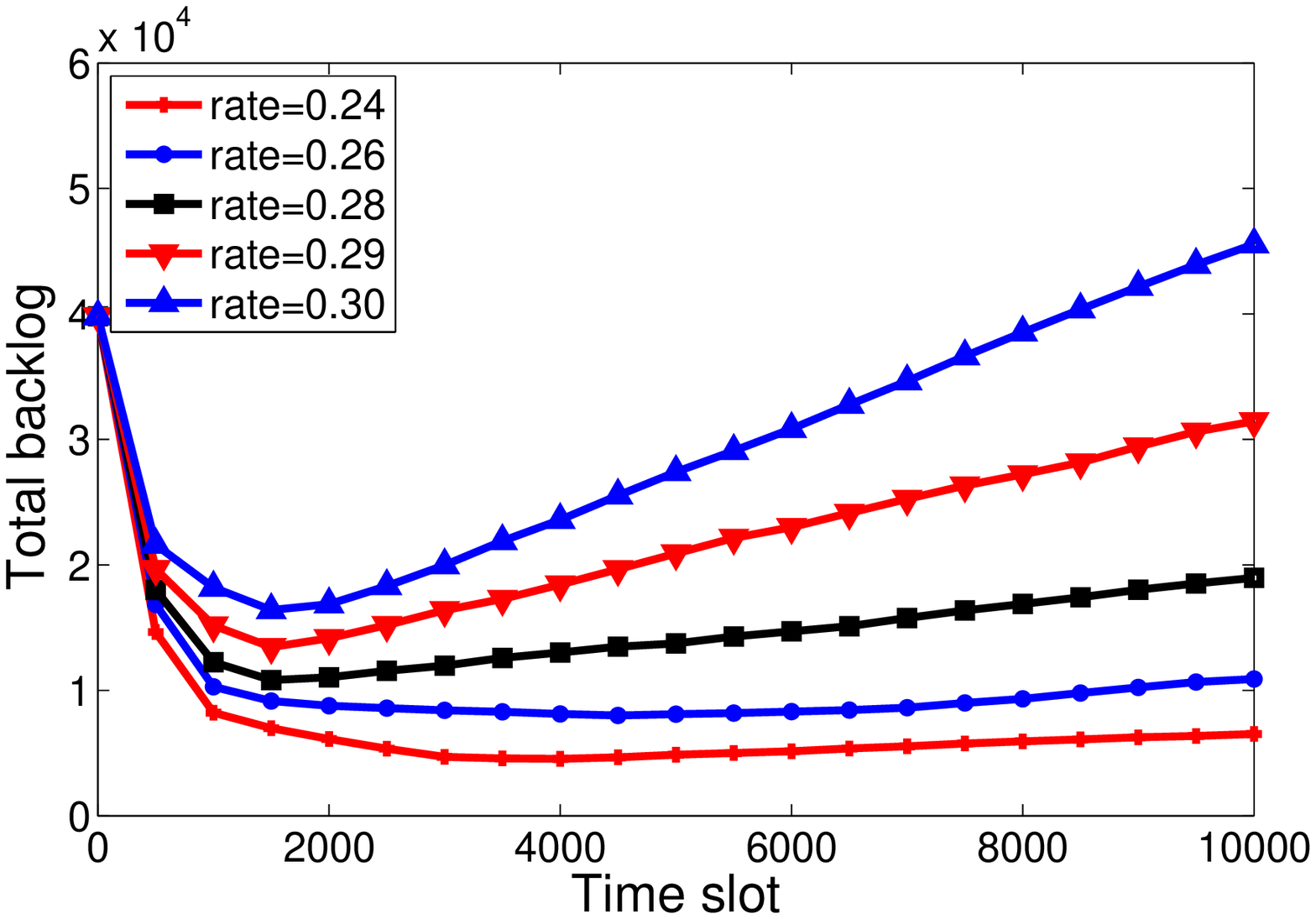}}
  \hspace{0cm}
    \subfigure[RA]{
        \includegraphics[height=1.4in,width=2.0in]{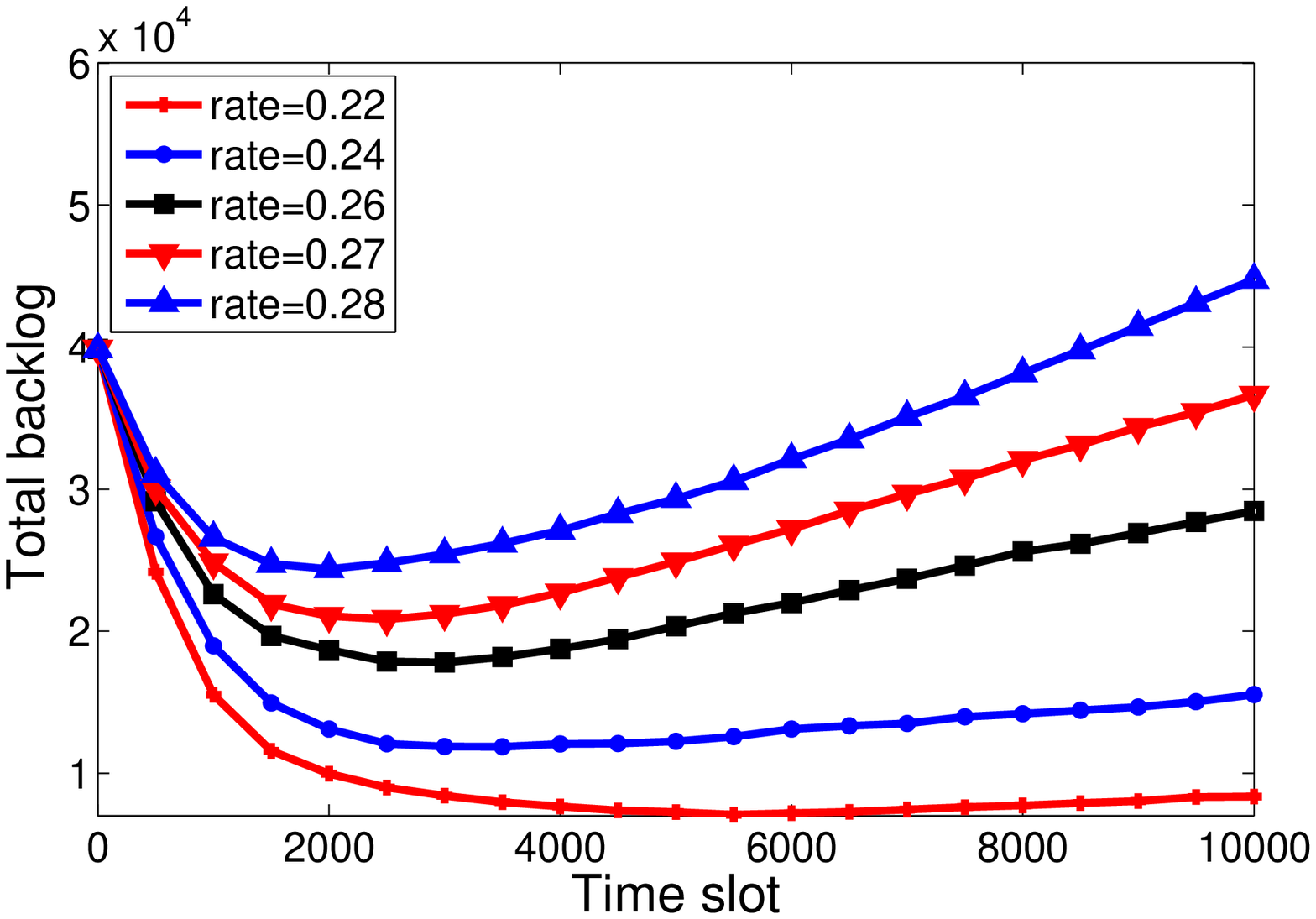}}
   \caption{{\small{Achievable capacity region of different algorithms in the uniform power setting}}}
   \vspace*{-1.0\baselineskip}
\end{figure*}

\subsection{Discussions }
Compared with experiment results with the linear power assignment, results under the uniform power setting are a little poorer in terms of total backlog under the same average arrival rate vector. It is probably because local computation of MWISL in the uniform power setting can just provide a logarithmic efficiency ratio, while it provides a constant efficiency ratio in the linear power setting.

In both settings, our proposed distributed algorithms achieve almost half of the maximum supportable average arrival rate by the two centralized algorithms. These comparable results show that our distributed algorithms perform efficiently without much sacrifice of throughput performance. The lost throughput performance is mainly caused by these silent links outside the local scheduling sets.  We have computed that the theoretical distance  between these independent local scheduling sets is at least $Md$. In our simulation settings links are uniformly dispersed in the network, which means nearly $1-(2M/(K+2M))^2$ fractional links are excluded to ensure the theoretical distance for independent local scheduling. In our simulation results, it also shows that the theoretical distance is too large for practice since the actual interference outside the local scheduling sets are far smaller than maximum tolerable interference.

From fluctuations of total backlog as time, we can roughly get  convergence time of scheduling algorithms. In both settings, the total backlog begins to convergence to a stable region around time slot $2000$ under supportable arrival rate vectors, e.g. Fig. 8, 14. The figures show that our localized algorithms have almost the same time of convergence as the two centralized algorithms.

\section{Related work}
As a fundamental problem in multihop wireless networks, link scheduling has been extensively studied. Existing works in the literature on wireless scheduling consider different optimization measures and assume different interference models. Though we mainly work on scheduling for throughput maximization under the physical interference model, we give a complete review on scheduling with distinct optimization measures and interference models.\\
\indent Scheduling for throughput maximization has been well understood under binary interference models in literature ranging from capacity analysis  to  algorithm designs
\cite{S:pick1}, \cite{S:pick2}, \cite{S:pick3}, \cite{S:constant1}-\cite{S:constant4}, \cite{S:GMS}, \cite{S:MWM2}, \cite{S:MS}. In their seminal work \cite{S:MWM1}, Tassiulas and Ephremides prove that the celebrated maximum weighted scheduling (MWS) achieves optimal throughput region. Since finding a MWS is NP-hard in general interference models, they further propose a linear-complexity centralized algorithm using an approach called pick-and-compare \cite{S:MWM2}. Subsequently, a variety of simpler and/or suboptimal scheduling algorithms are proposed to achieve full or fractional optimal throughput region under binary models. However, scheduling on a more practical physical interference model is only at the beginning.\\
\indent Distributed link scheduling subject to binary interference mainly approximates maximal scheduling algorithms locally, or uses the pick-and-compare approach.\\
\indent For the former, the works provide fractional throughput guarantee are presented in \cite{S:constant1}-\cite{S:constant4}, \cite{S:MS}. Using queue-length based random access approach, Lin and Rasool \cite{S:constant1} propose a constant overhead algorithm with $1/3-\epsilon$  of the capacity region under the primary interference model and   $\frac{1}{1+\Delta}-\epsilon$ of the capacity region under the $2$-hop interference model. Their works are further refined and studied in \cite{S:constant2}-\cite{S:constant4} with throughput capacity guarantee no more than $1/2$  of the optimal under the general interference models.\\
\indent The pick-and-compare is extended to distributed scheduling firstly by Modiano \emph{et al.} \cite{S:pick3} in the primary interference model. However, it is difficult to be applied to the more general $k$-hop interference models and omits consideration of scheduling overheads. Sanghavi \emph{et al.} \cite{S:pick1} further present a distributed algorithm with constant overhead, which can gain arbitrary fraction of capacity region by utilizing local augmentation technique and the pick-and-compare. Their work is limited to the primary interference model as well. A novel distributed scheduling  algorithm using graph partition under the pick-and-compare framework is proposed by Tang \emph{et al.} \cite{S:pick2} under more mature but still binary interference models.\\
\indent Until recently, research works begin to cast attentions on link scheduling under the physical interference model, and achieve some progress.\\
\indent Goussevskaia \emph{et al.} \cite{S:phy1} firstly  present the NP-completeness proofs for the scheduling problem  under the physical interference model. It firstly proposes algorithms with logarithmic  approximation ratio $O(g(|E|))$ under the uniform power setting without noise, where  $O(g(|E|))$  presents the link diversity among all links.  Goussevskaia \emph{et al.} further \cite{S:phy2} attempt to develop a constant approximation-ratio algorithm for the problem  of maximum independent set of links (MISL), a special case of MWISL with uniform weight. However, it is valid only without noise as pointed in \cite{S:phy3}. Wan \emph{et al.} finally succeed in developing a constant approximation-ratio algorithm for MISL with the existence of noise in  \cite{S:phy4}.\\
\indent Under the linear power setting, by utilizing partition and shifting strategies, Xu \emph{et al.}  get a constant approximation algorithm for MWISL subject to physical interferences \cite{S:phy5}. Very recently Xu \emph{et al.} \cite{S:phy9} propose another  constant approximation algorithm for the same problem based on the solution for the maximum weighted independent set of disks problem in \cite{S:ptas}. They also develop a logarithmic approximation algorithm for the uniform power setting. Chafekar \emph{et al.}  \cite{S:phy12} provides  algorithms for maximizing throughput with logarithmic approximation ratio  in the two power settings as well. However, the attained bound is not relative to the original optimal throughput capacity, but to the optimal value by using slightly smaller power levels.

Despite the main concern of this paper is on link scheduling for throughput maximization, we also make a review on the closely related problem of minimum length scheduling, which seeks a link schedule of  minimum length that satisfies all link demands. A very recent work in \cite{S:phy6} gives an overall analysis on link scheduling problems under the physical interference model  from an algorithm view. It reveals that the algorithmic reduction from the minimum length scheduling to the throughput maximization scheduling is approximation-preserving.

The NP-completeness proofs of the  minimum length scheduling problem, and a logarithmic approximation algorithm without noise in the uniform power setting, is  available in \cite{S:phy1}. With noise taken into consideration, Moscibroda \emph{et al.} present a scheduling algorithm for the problem with power control, but without provable guarantee in \cite{S:phy13}. They subsequently get a linear approximation bound in \cite{S:phy14}. An attempt on a constant approximation bound with uniform power setting fails in \cite{S:phy8}, pointed out by Wan. \emph{et al.} \cite{S:phy15}. They \cite{S:phy15} then propose a logarithmic approximation algorithm for the minimum length scheduling problem with power control.

Blough \emph{et al.} \cite{S:phy17} claim that the so-called black links, with length exactly at the maximum transmission range of the sender, hinder a tighter approximation bounds for the minimum length scheduling problem. They try to get a constant approximation ratio by limiting scheduling of such kind of links. Thus they revise the algorithm GOW proposed in \cite{S:phy1} by partitioning links according to the SNR diversity, instead of the length diversity. However, their revised algorithm GOW* can only guarantee a constant approximation ratio when the number of black links is bounded by a constant. They have also recognized the necessity of distributed implementation for the scheduling algorithm. Some discussions on the suitability of the algorithm for distributed execution are then presented.

A distributed implementation of the centralized algorithm \cite{Brar-mobicom06} is presented in \cite{brar-icdcs08}. The algorithm assumes a variation of the physical interference model we study in this paper. Nodes rely on the carrier sensing mechanism to detect collisions, \emph{i.e.}, a node can successfully detect a carrier even if multiple nodes are concurrently transmitting. In each slot a node with unsatisfied demand is selected as the controller. At least one link associated with the controller is scheduled in the slot. Other candidate links are added to the slot in iteratively. Whenever active nodes are added that cause previously allocated links to lose packets, this is detected through the carrier sensing mechanism and the SCREAM approach is used to notify all active nodes that the last scheduled links should be removed from the slot. Hence, this approach has to propagate the notification of SCREAM throughout the network for a feasible schedule.
When nodes are distributed uniformly at random in a square of unit area, and that radio signal propagation obeys the log-distance path model with path loss exponent $\kappa > 2$, this resulted schedule length is the same as the one computed by the centralized algorithm of \cite{Brar-mobicom06}. In comparison, our proposed localized algorithms  do not need capability of carrier sensing and the theoretical results hold for arbitrary wireless networks.

\indent A recent work \cite{S:phy16} proposes a column-generation-based solution for an optimal solution for the minimum length scheduling problem. They firstly formulate the problem into a linear programming problem, and then use the column-generation-based decomposition techniques to solve the problem iteratively in fact. Though the column generation avoids an enumeration of all possible schedules, it still grows exponentially with the number of the links. The method with so high computing complexity, like the optimal primal-dual solution for binary interference models, is unrealistic for wireless networks in nature, though optimal.\\
\indent A closely related work \cite{S:phy7} analyzes the performance of longest queue first (LQF) algorithm under the physical model with the uniform power assignment, and employs a technique named ``interference localization" to prevent the performance of LQF vanishing, and enable distributed implementation of LQF. The implementation of the distributed LQF algorithm actually relies on the radius of interference neighborhood.  However, the trivial procedure for determination of interference neighborhood is centralized, and iterated at every link to ensure feasibility of the scheduling. Obviously it is impractical in large networks since number of links is unbounded. Further, the achievable throughput capacity is $\frac{\gamma_{LQF}}{2}$ fraction of the original throughput where $\gamma_{LQF}$ is efficiency ratio of LQF under interference localization constraints. Despite the value of  $\gamma_{LQF}$ is bounded away from zero, but still uncertain. In contrast, the efficiency ratio of our algorithm under the uniform power setting is logarithmic. Meanwhile, our algorithm is much simpler and more practical. It implements distributed scheduling by compulsory separation of local link sets where the separation distance is predefined.  We do not need globally collected information  in the scheduling process, but just some neighborhood information in constant distance.
 Moreover, we consider links transmits at heterogeneous power and aim at distributed throughput-optimum scheduling, but not improvement of a specified scheduling algorithm LQF.

 Another closely related work in \cite{S:dphy1} proposes a CSMA-type distributed link scheduling approach, where a link that is not active at time slot $t-1$ randomly waits for a backoff time to be active if it does make links in global scheduling $S(t-1)$ violate the SINR threshold. In order to compute a feasible schedule, this random access based approach has to consume much time on the backoff process. Moreover, it induces high communication overhead since whenever a candidate link is to be activated, the corresponding notification should reach all previously added links for verification of the SINR requirement. Conversely, our algorithm computes local results parallelly with local information within constant hops.

\section{Discussions}

Note that our proposed algorithms have assumed a  static wireless network where every node knows its own location and locations of neighbors within a distance of $R$. It is necessary in the partition process and in the linear power assignment. To get all these required location information, we only need to perform some localization techniques with high accuracy once for a static wireless network, \emph{e.g.}, storing all these information at each node when deploying  networks. Thus we do not necessarily need an extra location device, \emph{e.g.}, GPS device, embedded in every node. As to the impact of some inaccurate locations on the performance of our algorithms, we argue that
\begin{enumerate}
  \item It has little influence on the correctness of the algorithms. The distance we set for separation of each local area is safe enough to ensure local independent scheduling, since we have considered the worst case (with a maximum power and a highest node density that a feasible schedule can tolerate) when computing the distance. The distance can be set even larger if considering inaccurate locations.  An inaccurate location may influence the decision that whether a node shall participate in a local scheduling, affecting the local scheduling set but not the correctness of the algorithms.
  \item Inaccurate locations will impact local scheduling sets, and thus the throughput performance. An inaccurate location may lead to an inaccurate link length. Since the link length is bounded by  a minimum length and a maximum length both previously known, the error link length shall also lies in this region. The assigned powers in the linear power setting remain in the original region, though some links have a power bigger than they should have and some links have a smaller power.
     Consequently, the size and feasibility of local scheduling sets will be partly affected. To ensure the feasibility of local scheduling set, we can deliberately assume a smaller maximum tolerable inside interference  for each link.  The size and then weight of the resulted local scheduling sets might be smaller than the optimal local scheduling sets. From simulation results, the actual inside interference suffered by each activated link is much smaller than the theoretical value. Thus we infer that the influence won't be too serious if just some inaccurate locations occurs.
\end{enumerate}

 We have to predefine $\varepsilon$, $K$ and $M$ to initiate the algorithm for the partition and shifting process. But we do not need other extra information to determine the values, except the topology information. We just need to set $\varepsilon$ among $(0,1)$, and a greater value leads to better performance. We shall set $M$ as some constant no less than the lower bound and $K/M$ greater than $2$. The exact values of $K$ and $M$ also depend on the number of squares we want to partition the network. Combining our simulations, it is better to set $K/M$ around $6$ for better throughput performance, but not necessarily.
As to the value of longest and shortest link length, these values are also easy to get as long as the topology is determined. We shall notice that once the longest link length $R$ is determined as a constant, $I_{\max}$ is also determined as a constant, even for the case  when $I_{\max}$ is close to $0$. In such a case, the optimal capacity region of the network itself is also constrained by the small value of  $I_{\max}$. Thus we do not have to take extra care of long links. The passive impact of long links on capacity region instead suggests that we shall avoid long links in network deployment if possible.

Despite we only consider  single hop link scheduling at each link, our work is  extendible to  the joint routing and scheduling problem using the back-pressure mechanism \cite{sha07}. The difference is that the weight of each link becomes the difference of unscheduled packets destined to the same node at  two neighboring nodes.

Throughout the paper we have mainly focused on link scheduling for optimal throughput. Another important optimization objective for scheduling is delay. It is intuitive that MWM-type scheduling which tries to minimize total queue length may suffer serious delay and long convergence time, detailed analysis seen in \cite{bui-info09} \cite{lei-info08} and their references.
 In general MWM-type scheduling,  factors such as  time complexity of  scheduling algorithms, duration of data transmission and infrequent schedule updates impact delay. Some theoretical analysis and engineering implications is available \cite{yi-mobihoc09}.
Many techniques such as ``shadow queues''\cite{bui-info09}, or cross-layer methods\cite{ruogu-ton11}, are proposed to improve delay performance. We leave it as future works to study efficient   localized scheduling algorithms for different delay and throughput demands.

\section{Conclusion}
We tackle  the problem of throughput-optimum localized link scheduling in multihop wireless networks subject to physical interference constraints.
We successfully address the primary challenge of global interference constraint in the physical interference model that hinders the development of localized scheduling algorithms. We then utilize the graph partition and shifting techniques to design a localized scheduling method with a constant approximation ratio to the optimal solution under the linear power setting. We further extend the solutions to the uniform power setting, and develop a localized algorithm with a logarithmic approximation ratio. Our extensive simulation results show that under both power settings the achieved capacity regions of our algorithms are nearly half of these achieved by counterpart centralized algorithms, which further demonstrates performance efficiency of our algorithms.  We believe that our work can find applications in some time-slotted wireless networks, \emph{e.g.}, time-slotted wireless sensor networks or wireless mesh networks. Knowing these factors that determine the throughput of a network also helps to better deploy a multihop wireless network and enhance the overall throughput performance.

For future work, there remain many interesting open questions.
The first is to further improve the efficiency of localized link scheduling algorithms. The second is to improve the approximation ratio of a method in uniform power setting. Up to now there have been only logarithmic approximation algorithms with the uniform power assignment. In addition, our results as well as many other results, assume the whole topology and geometry location of the network known by each node. This information is not always available and accurate in some cases. Some simple distributed methods without using explicit global and location information may be more interesting and practical.



\bibliography{mylib}

\begin{thebibliography}{10}
\providecommand{\url}[1]{#1}
\csname url@samestyle\endcsname
\providecommand{\newblock}{\relax}
\providecommand{\bibinfo}[2]{#2}
\providecommand{\BIBentrySTDinterwordspacing}{\spaceskip=0pt\relax}
\providecommand{\BIBentryALTinterwordstretchfactor}{4}
\providecommand{\BIBentryALTinterwordspacing}{\spaceskip=\fontdimen2\font plus
\BIBentryALTinterwordstretchfactor\fontdimen3\font minus
  \fontdimen4\font\relax}
\providecommand{\BIBforeignlanguage}[2]{{%
\expandafter\ifx\csname l@#1\endcsname\relax
\typeout{** WARNING: IEEEtran.bst: No hyphenation pattern has been}%
\typeout{** loaded for the language `#1'. Using the pattern for}%
\typeout{** the default language instead.}%
\else
\language=\csname l@#1\endcsname
\fi
#2}}
\providecommand{\BIBdecl}{\relax}
\BIBdecl

\bibitem{li2009underground}
M.~Li and Y.~Liu, ``Underground coal mine monitoring with wireless sensor
  networks,'' \emph{ACM Transactions on Sensor Networks}, vol.~5, no.~2, p.~10,
  2009.

\bibitem{liu2010passive}
Y.~Liu, K.~Liu, and M.~Li, ``Passive diagnosis for wireless sensor networks,''
  \emph{IEEE/ACM Transactions on Networking}, vol.~18, no.~4, pp. 1132--1144,
  2010.

\bibitem{S:effcient}
X.-H. Xu, X.-Y. Li, P.-J. Wan, and S.-J. Tang, ``Efficient scheduling for
  periodic aggregation queries in multihop sensor networks,'' \emph{IEEE/ACM
  Transactions on networking}, Aug. 2011.

\bibitem{S:GMS}
C.~Joo, X.~Lin, and N.~B. Shroff, ``Understanding the capacity region of the
  greedy maximal scheduling algorithm in multi-hop wireless networks,'' in
  \emph{Proc. IEEE Infocom}, 2008, pp. 1103--1111.

\bibitem{S:MWM2}
L.~Tassiulas and A.~Ephremides, ``Linear complexity algorithms for maximum
  throughput in radio networks and input queued switches,'' in \emph{Proc. IEEE
  Infocom}, 1998, pp. 533--539.

\bibitem{S:constant1}
X.~Lin and S.~B. Rasool, ``Constant-time distributed scheduling policies for ad
  hoc wireless networks,'' in \emph{Proc. IEEE CDC}, 2006, pp. 1258--1263.

\bibitem{S:constant2}
------, ``Constant-time distributed scheduling policies for ad hoc wireless
  networks,'' \emph{IEEE/ACM Transactions on Automatic Control}, vol. $54$, pp.
  231--242, 2009.

\bibitem{S:constant3}
A.~Gupta, X.~Lin, and R.~Srikant, ``Low-complexity distributed scheduling
  algorithms for wireless networks,'' \emph{IEEE/ACM Transactions on
  Networking}, vol. $17$, pp. 1846--1859, 2009.

\bibitem{S:constant4}
C.~Joo and N.~B. Shroff, ``Performance of random access scheduling schemes in
  multi-hop wireless networks,'' \emph{IEEE/ACM Transactions on Networking},
  vol. $17$, pp. 1481--1493, 2009.

\bibitem{S:pick1}
S.~Sanghavi, L.~Bui, and R.~Srikant, ``Distributed link scheduling with
  constant overhead,'' in \emph{Proc. ACM SIGMETRICS}, 2007, pp. 313--324.

\bibitem{S:pick2}
S.-J. Tang, X.-Y. Li, X.~Wu, Y.~Wu, X.~Mao, P.~Xu, and G.~Chen, ``Low
  complexity stable link scheduling for maximizing throughput in wireless
  networks,'' in \emph{Proc. IEEE SECON}, 2009, pp. 1--9.

\bibitem{S:pick3}
E.~Modiano, D.~Shah, and G.~Zussman, ``Maximizing throughput in wireless
  networks via gossiping,'' in \emph{Proc. ACM SIGMETRICS}, 2006, pp. 27--38.

\bibitem{S:MS}
P.~Chaporkar, K.~Kar, and S.~Sarkar, ``Throughput and fairness guarantees
  through maximal scheduling in wireless networks,'' \emph{IEEE/ACM
  Transactions on Information Theory}, vol. $54$, pp. 572--594, 2008.

\bibitem{S:phy6}
P.-J. Wan, O.~Frieder, X.-H. Jia, F.~Yao, X.-H. Xu, and S.-J. Tang, ``Wireless
  link scheduling under physical interference model,'' in \emph{Proc. IEEE
  Infocom}, 2011, pp. 838--845.

\bibitem{S:phy13}
T.~Moscibroda and R.~Wattenhofer, ``The complexity of connectivity in wireless
  networks,'' in \emph{Proc. IEEE Infocom}, 2006, pp. 1--13.

\bibitem{S:phy14}
------, ``Topology control meets {SINR}: the scheduling complexity of arbitrary
  topologies,'' in \emph{Proc. IEEE Mobihoc}, 2006, pp. 310--321.

\bibitem{S:phy1}
O.~Goussevskaia, Y.~Oswald, and R.~Wattenhofer, ``Complexity in geometric
  {SINR},'' in \emph{Proc. ACM Mobihoc}, 2007, pp. 100--109.

\bibitem{S:phy2}
O.~Goussevskaia, R.~Wattenhofer, M.~M. Halldorsson, and E.~Welzl, ``Capacity of
  arbitrary wireless networks,'' in \emph{Proc. IEEE Inforcom}, 2009, pp.
  1872--1880.

\bibitem{multicast}
X.-Y. Li, ``Multicast capacity of wireless ad hoc networks,'' \emph{IEEE/ACM
  Transactions on networking}, vol. $17$, pp. 950--961, 2009.

\bibitem{S:phy5}
X.-H. Xu, S.-J. Tang, and P.-J. Wan, ``Maximum weighted independent set of
  links under physical interference model,'' in \emph{LNCS}, vol. $6221$, 2010,
  pp. 68--74.

\bibitem{S:phy8}
M.~Halldorsson and R.~Wattenhofer, ``Wireless communication is in {APX},'' in
  \emph{Proc. 36th International Colloquium on Automata, Languages and
  Programming}, 2009, pp. 525--536.

\bibitem{S:phy12}
D.~Chafekar, V.~Kumar, M.~Marathe, S.~Parthasarathy, and A.~Srinivasan,
  ``Arrpoximation algorithms for computing capacity of wireless networks with
  {SINR} constraints,'' in \emph{Proc. IEEE Infocom}, 2008, pp. 1166--1174.

\bibitem{S:phy15}
P.-J. Wan, X.-H. Xu, and O.~Frieder, ``Shortest link scheduling with power
  control under physical interference model,'' in \emph{Proc. IEEE MSN}, 2010,
  pp. 74--78.

\bibitem{S:phy9}
X.~H. Xu, S.~J. Tang, and X.-Y. Li, \emph{Stable Wireless Link Scheduling
  Subject to Physical Interferences With Power Control}, 2011, manuscript.

\bibitem{S:phy17}
D.~M. blough, G.~Resta, and P.~Santi, ``Approximation algorithms for wireless
  link scheduling with {SINR-Based} inteference,'' \emph{IEEE/ACM Transactions
  on Networking}, vol. $18$, pp. 1701--1712, 2010.

\bibitem{S:phy7}
L.-B. Le, E.~Modiano, C.~Joo, and N.~B. Shroff, ``Longest-queue-first
  scheduling under {SINR} interference model,'' in \emph{Proc. ACM Mobihoc},
  2010, pp. 41--50.

\bibitem{brar-icdcs08}
G.~Brar, D.~Blough, and P.~Santi, ``The {SCREAM} approach for efficient
  distributed scheduling with physical interference in wireless mesh
  networks,'' in \emph{Proc. IEEE ICDCS}, 2008, pp. 214 --224.

\bibitem{peleg87}
D.~Peleg, \emph{Distributed computing{:} a locality-sensitive approach}, ser.
  {SIAM} Monographs on Discrete Mathematics and Applications.\hskip 1em plus
  0.5em minus 0.4em\relax SIAM, 1987, vol.~5.

\bibitem{S:MWM1}
L.~Tassiulas and A.~Ephremides, ``Stability properties of constrained queueing
  systems and scheduling policies for maximum throughput in multihop radio
  networks,'' \emph{IEEE/ACM Transactions on Automatic Control}, vol. $37$, pp.
  1936--1948, 1992.

\bibitem{sharma2006complexity}
G.~Sharma, R.~Mazumdar, and N.~Shroff, ``On the complexity of scheduling in
  wireless networks,'' in \emph{Proc. ACM MobiCom}, 2006, pp. 227--238.

\bibitem{lin2005info}
X.~Lin and N.~B. Shroff, ``The impact of imperfect scheduling on cross-layer
  rate control in multihop wireless networks,'' in \emph{Proc. IEEE Infocom},
  2005, pp. 1804--1814.

\bibitem{S:phy4}
P.-J. Wan, X.-H. Jia, and F.~Yao, ``Maximum independent set of links under
  physical interference model,'' in \emph{WASA}, 2009, pp. 169--178.

\bibitem{sha07}
S.~Shakkottai and R.~Srikant, ``Network optimization and control,''
  \emph{Foundations and Trends in Networking}, vol.~2, no.~3, pp. 271--379,
  2007.

\bibitem{S:LGS}
C.~Joo, ``A local greedy scheduling scheme with provable performance
  guarantee,'' in \emph{Proc. ACM Moihoc}, 2008, pp. 111--120.

\bibitem{S:phy3}
X.-H. Xu and S.-J. Tang, ``A constant approximation algorithm for link
  scheduling in arbitrary networks under physical interference model,'' in
  \emph{Proc. ACM FOWANC}, 2009, pp. 13--20.

\bibitem{S:ptas}
X.-Y. Li and Y.~Wang, ``Simple approximation algorithms and {PTASs} for various
  problems in wireless ad hoc networks,'' \emph{Journal of Parallel and
  Distributed Computing}, vol. $66$, pp. 515--530, 2006.

\bibitem{Brar-mobicom06}
G.~Brar, D.~M. Blough, and P.~Santi, ``Computationally efficient scheduling
  with the physical interference model for throughput improvement in wireless
  mesh networks,'' in \emph{Proc. ACM MobiCom}, 2006, pp. 2--13.

\bibitem{S:phy16}
S.~Kompella, J.~E. Wieselthier, and A.~Ephremides, ``On optimal {SINR-Based}
  scheduling in multihop wireless networks,'' \emph{IEEE/ACM Transactions on
  Networking}, vol. $18$, pp. 1713--1724, 2010.

\bibitem{S:dphy1}
J.~Ryu, C.~Joo, T.~T. Kwon, N.~B. Shroff, and Y.~Choi, ``Distributed {SINR}
  based scheduling algorithm for multi-hop wireless networks,'' in \emph{Proc.
  ACM MSWIM}, 2010, pp. 376--380.

\bibitem{bui-info09}
L.~Bui, R.~Srikant, and A.~Stolyar, ``Novel architectures and algorithms for
  delay reduction in back-pressure scheduling and routing,'' in \emph{Proc.
  IEEE Infocom}, april 2009, pp. 2936 --2940.

\bibitem{lei-info08}
L.~Ying, R.~Srikant, and D.~Towsley, ``Cluster-based back-pressure routing
  algorithm,'' in \emph{Proc. IEEE Infocom}, april 2008, pp. 484 --492.

\bibitem{yi-mobihoc09}
Y.~Yi, J.~Zhang, and M.~Chiang, ``Delay and effective throughput of wireless
  scheduling in heavy traffic regimes: Vacation model for complexity,'' in
  \emph{Proc. ACM Mobihoc}, 2009.

\bibitem{ruogu-ton11}
R.~Li, A.~Eryilmaz, L.~Ying, and N.~Shroff, ``A unified approach to optimizing
  performance in networks serving heterogeneous flows,'' \emph{Networking,
  IEEE/ACM Transactions on}, vol.~19, no.~1, pp. 223 --236, feb. 2011.

\end{thebibliography}

\appendices
\section{Proof of Lemma 1}
\begin{IEEEproof}
Suppose $l^\ast $ is the shortest link in $OPT_{i}$,
let $l$ denote any other link different from $l^\ast $ in $OPT_{i}$.
Assuming $l^\ast =(u^\ast, v^\ast )$, $l=(u, v)$, we have
{\small{$ \left\| {{uv}^\ast } \right\| \le \sqrt 2 JR.$}}
The relative interference $l^\ast$ received from $OPT_{i}$ is
shown below.

\begin{eqnarray*}
 r_{OPT_{i}} (l^\ast )&=& \frac{\sum\limits_{l\in \{OPT_{i}\backslash
\{l^\ast \}\}} {c\cdot \left\| l \right\|^\beta \cdot \eta \cdot \left\|
{{uv}^\ast } \right\|^{-\kappa }} }{c\cdot \left\| {l^\ast }
\right\|^\beta \cdot \eta \cdot \left\| {l^\ast } \right\|^{-\kappa }} \\
&\ge& \sum\limits_{l\in \{OPT_{i}\backslash \{l^\ast \}\}} {(\sqrt 2
JR)^{-\kappa }} \frac{\left\| l \right\|^\beta }{\left\| {l^\ast }
\right\|^\beta }\\
&\ge&  {\left| {OPT_{i}-1} \right|\cdot (\sqrt 2 JR)^{-\kappa }}_.
 \end{eqnarray*}

Since $l^\ast \in OPT_{i} $, so it should satisfy the SINR constraint that {\small{$\frac{c\cdot \left\| {l^\ast } \right\|^\beta \cdot \eta \cdot
\left\| {l^\ast } \right\|^{-\kappa }}{\xi +(1-\varepsilon )I_{\max }^{l^*} }\ge
\sigma \mbox{ ,}$}}
then we have
\[r_{_{OPT_{i}} }^{max } (l^\ast )\le \frac{1}{1-\varepsilon
}\left[ {\frac{1}{\sigma }-\frac{\xi \cdot \left\| {l^\ast }
\right\|^{\kappa -\beta }}{c\eta }} \right]_. \]
Therefore we derive

\[ \left| {OPT_{i}} \right| \le  \frac{(\sqrt 2 JR)^\kappa }{1-\varepsilon
}\left[ {\frac{1}{\sigma }-\frac{\xi \cdot r^{\kappa -\beta}}{c\eta }} \right]+1.\]
\end{IEEEproof}

\section{Proof of Lemma 2}
\begin{IEEEproof}
For any two links $l^*$ and $ l$ which do not belong to the same local link set $Z_i$, assuming $l^*=(u^*, v^*)$,
$l=(u,v)$, it always holds that
$    \|uv^*\| \ge M \times R \mbox{~.}$

The total interference $l^*$ suffered form all concurrent transmitting links located in other sub-squares is

\begin{eqnarray*}
    I_{out}^ {l^\ast } &=& \sum\limits_{l\in \{\cup Z_i \backslash Z'\},l^\ast \in Z'}
                            c \times {\left\| l \right\|}^\beta \times \eta \times \left\| {uv^\ast }\right\|^{-\kappa } \\
    & \le & c \times \eta \times R^{\beta} \times \sum\limits_{l\in \{\cup Z_i \backslash Z'\},l^\ast \in Z'}{\|uv^*\|}^{-\kappa} \\
    & \le & c \times \eta \times R^{\beta} \times \int_1^{\infty}{ \frac{2 \pi x M R}{M R} \times |OPT_{i}|_{ub} \times
            (x M R)^{-\kappa} dx}  \\
    & =   & c \times \eta \times R^{\beta-\kappa} \times 2\pi \times |OPT_{i}|_{ub} \times M^{-\kappa} \times \frac{1}{\kappa-2}
\end{eqnarray*}

since $\kappa$ is a constant greater than $2$ typically.

To assure {\small{$I_{out} ^ {l^\ast } \le \varepsilon \cdot I_{\max}$}}, we derive that
\[
    M \ge \biggl[ \frac{2 \pi c \eta R^{\beta-\kappa} \cdot |OPT_{i}|_{ub}}{(\kappa-2) \varepsilon  I_{\max}} \biggr]^{\frac{1}{\kappa}}
\]
where {\small{$I_{\max} = \frac{c \eta R^{\beta-\kappa}}{\sigma} - \xi$ }} is a constant for a given network, $|OPT_{i}|_{ub}$ is a constant upper bound of $|OPT_{i}|$.
\end{IEEEproof}

\section{Proof of Lemma 6}
\begin{IEEEproof}
For any two links $l^*$ and $ l$ which do not belong to the same local link set $Z_i$, assuming $l^*=(u^*, v^*)$,
$l=(u,v)$, it always holds that
   $\|uv^*\| \ge M \times R \mbox{~.}$
The total interference $l^*$ suffered form all concurrent transmitting links located in other sub-squares is

\vspace*{-0.5\baselineskip}
\begin{eqnarray*}
    I_{out}^ {l^\ast }
    &  =  & \sum\limits_{l\in \{\cup Z_i \backslash Z'\},l^\ast \in Z'}
                            P \times \eta \times \left\| {uv^\ast }\right\|^{-\kappa } \\
    & \le & P \times \eta \times \int_1^{\infty}{ \frac{2 \pi x M R}{M R} \times |\mathcal{X}_{ij} (t)|_{ub} \times
            (x M R)^{-\kappa} dx}  \\
     \vspace*{-0.5\baselineskip}
       & =   & P \times \eta \times R^{-\kappa} \times 2\pi \times |\mathcal{X}_{ij} (t)|_{ub} \times M^{-\kappa} \times \frac{1}{\kappa-2}
\end{eqnarray*}

since $\kappa$ is a constant greater than $2$ typically.

To assure $I_{out} ^ {l^\ast } \le \varepsilon \cdot I_{\max}$, we derive
{\small{$
    M \ge \biggl[ \frac{2 \pi \eta P \cdot |\mathcal{X}_{ij} (t)|_{ub}}{(\kappa-2) \varepsilon  I_{\max}R^{\kappa}} \biggr]^{\frac{1}{\kappa}}_{\mbox{~,}}$}}
where $I_{\max} = \frac{P \eta R^{-\kappa}}{\sigma} - \xi$ is a constant, and $|\mathcal{X}_{ij} (t)|_{ub} $ is a constant upper bound of $|\mathcal{X}_{ij} (t)|$.
\vspace*{-0.5\baselineskip}
\end{IEEEproof}




\end{document}